\documentclass[10pt,aps,prd,onecolumn,showpacs,amsmath,amssymb,nofootinbib,eqsecnum,preprintnumbers,superscriptaddress]{revtex4-2}

\usepackage{mathtools}					
\usepackage[colorlinks=true,
            citecolor=red,
            linkcolor=blue,
            urlcolor=violet,
            filecolor=cyan,
            backref=false]{hyperref}
\usepackage{xcolor}
\usepackage{nicefrac}
\usepackage{empheq}
\usepackage{stmaryrd}
\SetSymbolFont{stmry}{bold}{U}{stmry}{m}{n}

\usepackage[shortlabels]{enumitem}
\usepackage{centernot}

\usepackage{pst-node}

\usepackage{tikz}

\usepackage{longtable}
\usepackage{array}

\usepackage[section]{placeins}

\usepackage{pifont}

\newcommand{\cmark}{\ding{51}}%
\newcommand{\xmark}{\ding{55}}%

\newcommand{\descstyle}[1]{\underline{#1}}

\newcommand{\abbrev}[1]{\textsf{\underline{\smash{#1}}}}

\newcommand{\appsection}[1]{\section{\MakeUppercase{#1}}}
\newcommand\bs[1]{\boldsymbol{#1}} 
\newcommand\feq{\mathrel{\phantom{=}}} 

\renewcommand{\cosh}{\operatorname{ch}}
\renewcommand{\sinh}{\operatorname{sh}}
\renewcommand{\tanh}{\operatorname{th}}
\renewcommand{\coth}{\operatorname{cth}}
\newcommand{\csch}{\operatorname{csch}}
\newcommand{\sech}{\operatorname{sech}}

\newcommand{\sgn}{\operatorname{sgn}}

\newcommand{\lie}{\pounds}

\begin{document}
\title{
Symmetry reduction of gravitational Lagrangians
}

\author{Guillermo Frausto}
\email{guillermo.frausto@usu.edu}
\affiliation{Department of Physics, Utah State University, Logan, Utah 84322, USA}

\author{Ivan Kol\'a\v{r}}
\email{ivan.kolar@matfyz.cuni.cz}
\affiliation{Institute of Theoretical Physics, Faculty of Mathematics and Physics, Charles University, V Hole\v{s}ovi\v{c}k\'ach 2, Prague 180 00, Czech Republic}

\author{Tom\'a\v{s} M\'alek}
\email{malek@math.cas.cz}
\affiliation{Institute of Mathematics of the Czech Academy of Sciences, \v{Z}itn\'a 25, 115 67 Prague 1, Czech Republic}

\author{Charles Torre}
\email{charles.torre@usu.edu}
\affiliation{Department of Physics, Utah State University, Logan, Utah 84322, USA}

\date{\today}

\begin{abstract}
We analyze all possible symmetry reductions of Lagrangians that yield fully equivalent field equations for any 4-dimensional metric theory of gravity. Specifically, we present a complete list of infinitesimal group actions obeying the principle of symmetric criticality, identify the corresponding invariant metrics (and $l$-chains), discuss relations among them, and analyze simplifications allowed by the residual diffeomorphism group and Noether identities before and after variation of the reduced Lagrangian. We employ rigorous treatment of the symmetry reduction by Fels and Torre and use the Hicks classification of infinitesimal group actions by means of Lie algebras pairs of isometries and their isotropy subalgebras. The classification is recast in the coordinates in which the well-known symmetries and geometries are easily recognizable and relatable to each other. The paper is accompanied by a code for the symmetry reduction of Lagrangians that we developed in \texttt{xAct} package of \textsc{Mathematica}.
\end{abstract}

\maketitle


\section{Introduction}

\textit{Spacetime symmetries} play a crucial role in our understanding of gravitational theories. This is apparent already by looking at the large number of exact solutions admitting several isometries \cite{Stephani2009-dt,Griffiths:2009dfa}. The idea of \textit{reduction of Lagrangians} by specified symmetry groups is attractive for several reasons: i) When dealing with complicated theories of gravity (e.g., the higher-derivative gravity models), one can bypass the variation of the full Lagrangian and the subsequent substitution of the metric ansatz into the field equations and replace it by a simpler variation of the reduced Lagrangian evaluated directly on this ansatz, which typically yields the reduced field equations much faster. This shortcut to symmetric solutions of gravitational theories is called the \textit{Weyl trick} as it was first used by Weyl in \cite{Weyl:1917gp}. Its advantages and drawbacks have been emphasized in various papers focusing mainly on spherical symmetry (with stationarity\footnote{Although the stationarity implies staticity in the spherically symmetric spacetimes, we use the former terminology as our analysis is purely local.}) \cite{Lovelock1973SphericallySM,Deser:2003up,Deser:2004gi,Deser:2005gr,sym11070845} or homogeneous cosmology \cite{Hawking:1968zw,Maccallum:1972er}. ii) The reduced Lagrangians themselves can be viewed as interesting mini/midi-superspace models that may be used in applications that cannot easily deal with the full phase space, such as the quantization, etc. \cite{DeWitt:1967,Misner:1969,Kuchar:1971,Torre:1998dy}. iii) One can scan for special theories that are tuned to desired symmetries, e.g., so that they possess a reduced order of derivatives in the field equations. Examples of these theories are the (generalized) quasi-topological gravities \cite{Oliva:2010eb, Myers:2010ru,
Dehghani:2011vu, Cisterna:2017umf, Bueno:2019ycr, Bueno:2022res}, which attracted considerable attention for their interesting properties, and more recent cosmological gravities \cite{Moreno:2023arp}.

In a large number of theoretical physics papers, the symmetry reduction is often used heuristically and even considered a somewhat dubious trick (though frequently applied), as it may not always give the set of equivalent equations depending on the chosen symmetry group action and other important details of the symmetry reduction. If not checked against the full set of reduced field equations, one may easily end up with incorrect solutions of the theory. Naturally, such a check (if complex) partially defies the purpose of the Weyl trick as a shortcut for finding solutions effortlessly. Also, the mini/midi-superspace models or the symmetry-fine-tuned theories based on such symmetry reductions are not reliable. Apart from the use of the symmetry reduction in mathematically unjustified cases of symmetries, the common heuristic often involves ill-defined integration over the unbounded orbits rather than properly reducing the volume elements to the lower-dimensional one. Another issue comes with simplifying the form of the metric ansatz that is substituted into Lagrangians by fixing metric functions using diffeomorphism invariance. If it is not the most generic form of the metric invariant under the group action, the symmetry reduction of Lagrangians (in contrast to the symmetry reduction of the field equations) may still fail despite the simpler metric form can be related to the generic one just by diffeomorphisms. Finally, the procedure of symmetry reduction itself, although completely covariant in nature, may often seem coordinate dependent when not formulated appropriately; this could unjustly discourage some from its use.

The reason for the imprecise/incorrect use in the literature is because the rigorous mathematical formulation is not well-spread among researchers. As it was shown in \cite{Fels:2001rv} (following previous studies \cite{Anderson1997,Anderson:1999cn,Anderson:1999cm}, see also \cite{Torre:2010xa}), the Weyl trick, when done rigorously, is justified by the \textit{principle of symmetric criticality (PSC)} provided that two conditions on the infinitesimal symmetry group action are satisfied (independent of the gravitational theory in question). PSC then allows one to canonically symmetry-reduce any Lagrangian and guarantees that its field equations are fully equivalent to the reduced field equations. Naively, one would think that the restricted variation may only lead to insufficient number of equations resulting in extra wrong solutions apart from the correct ones; this happens if one of the two conditions is not satisfied. Unfortunately, the situation can be even worse if the other condition is not satisfied. Then, one may obtain the field equation that are completely wrong, and hence, miss the correct solution whatsoever. This was noticed in cosmological models with Bianchi class B already in \cite{Hawking:1968zw,Maccallum:1972er}. The rigorous treatment by \cite{Fels:2001rv} also clarified issues with the seeming coordinate dependence of reduction of the volume element. On the other hand, it did not consider fixing the metric functions by means of the diffeomorphism invariance and Noether identities, which was only partially addressed in \cite{Lovelock1973SphericallySM,Anderson:1999cn}. On top of that, the appropriate treatment due to \cite{Fels:2001rv} involves rather advanced knowledge of differential geometry and Lie algebras; hence, it may be hard to follow.

In order to analyze every possibility of PSC being satisfied, one also needs an exhaustive classification of all infinitesimal group actions in four spacetime dimensions. There exists the \textit{Petrov classification}\footnote{It should not be confused with the more famous classification by Petrov according to the algebraic properties of the curvature.} of spacetimes according to their isometries \cite{Petrov}, which is quite old, contains a number of mistakes, and is incomplete. A better, more modern classification that makes use of Lorentzian Lie algebra-subalgebra (isometry-isotropy) pairs was elaborated in the thesis \cite{Hicks:thesis} (by completing previous classification of abstract Lorentzian Lie algebra-subalgebra pairs \cite{Fels_Renner_2006,Bowers:2012,Snobl2014-te,Rozum2015-lp} and associating vector fields to them); we refer to it as the \textit{Hicks classification}. It was built on \texttt{DifferentialGeometry} package \cite{diffgeo} of \textsc{Maple} \cite{maple} and implemented as a digital database therein. Although much better (still containing a number of typos in the text version), the Hicks classification is not given in natural coordinates and vector-field bases. The exact solutions of general relativity (some of which have been classified in \cite{Hicks:thesis,Hwang:thesis}) are not easily recognizable in the original Hicks coordinates. Furthermore, the relations among individual infinitesimal group actions are not transparent, which makes the orientation in the Hicks classification rather difficult.

The main objectives of this paper are as follows: i) Identification of all infinitesimal group actions in Hicks classification that are compatible with PSC, which is needed as only a handful of examples is known from \cite{Fels:2001rv,Torre:2010xa}. ii) Providing the essential ingredients for a successful symmetry reduction, i.e., the complete list of $\Gamma$-invariant metrics and $l$-chains. (Although the Hicks classification \cite{Hicks:thesis} includes the $\Gamma$-invariant metrics, it does not account for the $\Gamma$-invariant $l$-chains and Noether identities, as it does address the symmetry reduction.) iii) Recasting these in more natural coordinates and vector-field bases that highlight mutual relations among individual symmetries and allow simple recognition of well known geometries. iv) Analysis of possible gauge fixing of the $\Gamma$-invariant metrics by means of the residual diffeomorphisms and Noether identities, which was to this day studied only in a single example of static spherically symmetric spacetime in \cite{Anderson:1999cn}. v) Building a new digital database with all necessary invariant objects in the \texttt{xAct} package \cite{xact} of \textsc{Mathematica} \cite{Mathematica} and developing the first code for automating the rigorous symmetry reduction of gravitational Lagrangians.

The main body of the paper is organized as follows:
\begin{itemize}
    \item In Sec.~\ref{sec:groupactions}, we review the basics of the symmetry reduction of Lagrangians including PSC and the conditions that needs to be satisfied for its validity. Then, we present a list of all infinitesimal group actions from Hicks classifications showing which of them do satisfy these conditions and hence PSC. Generators of those that are compatible with PSC are then enumerated explicitly in the new coordinates and vector-field bases.
    \item In Sec.~\ref{sec:metrics}, we present the $\Gamma$-invariant metrics and $l$-chains calculated for the PSC-compatible infinitesimal group actions. This is everything that is needed for virtually any rigorous and successful symmetry reduction of Lagrangians of any metric theory of gravity.
    \item In Sec.~\ref{sec:insights}, we delve deeper into the classification of symmetries by providing our insights in various relations among the infinitesimal group actions at the level of the abstract algebras as well as for the vector fields themselves.
    \item In Sec.~\ref{sec:gaugecond}, we study residual diffeomorphism invariance in order to simplify the $\Gamma$-invariant metrics either at the level of the field equations or in the reduced Lagrangian without breaking PSC. The latter is based on the Noether identities which provide relations among the field equations and allow us (in some cases) fix the available freedom and remove redundancies in the symmetry reduction.
    \item In Sec.~\ref{sec:examples}, we discuss several examples of the symmetry reduction of gravitational Lagrangians to demonstrate the methodology described in previous sections.
\end{itemize}
The paper is concluded by Sec.~\ref{sec:concl}, which contains summary and discussions of our results. It is also supplemented by Apx.~\ref{apx:transfcoord} that contains transformations between the Hicks and the new coordinates and vector-field bases, and Apx.~\ref{apx:relinfgra} with details on relations among infinitesimal group action.


\section{PSC-compatible infinitesimal group actions} \label{sec:groupactions}
Consider a local purely gravitational theory (with no extra fields other than the metric) that is defined on a 4-dimensional (oriented) Lorentzian manifold $(M,\bs{g})$, for which we adopt the mostly positive signature convention $(-,+,+,+)$. The action is given by 
\begin{equation}
    S=\int_{M} \underline{\epsilon}(\bs{g}) \mathcal{L}[\bs{g}]\;,
\end{equation}
where $\underline{\epsilon}(\bs{g})$ is the Levi-Civita tensor (an antisymmetric 4-form), which defines the usual volume element $\sqrt{-g}\, d^4x$, and $\mathcal{L}[\bs{g}]$ stands for a scalar expression that is constructed from the metric $\bs{g}$, Riemann tensor $\bs{R}$, and its covariant derivatives $\bs{\nabla}\dots\bs{\nabla}\bs{R}$ of finite orders. Above, we use a convention in which the standard tensorial indices as well as indices of differential forms are suppressed. The former property is denoted by \textbf{boldface} while the latter by an \underline{underline}. Furthermore, we employ the default \texttt{xAct} conventions for the Riemann and Ricci tensors. For brevity, we will also introduce the \textit{Lagrangian (4-form)} $\underline{L}[\bs{g}]:= \underline{\epsilon}(\bs{g}) \mathcal{L}[\bs{g}]$. The overall dependence on the metric $\bs{g}$ and all its derivatives is denoted by square brackets $[\dots]$.

We also consider a $d$-dimensional \textit{infinitesimal group action} $\Gamma$ on $M$ of a Lie group $G$ that is characterized by the $d$-dimensional Lie algebra of vector fields $\bs{X}\in\Gamma$. It is important to note that  $\Gamma$ contains more information than the corresponding \textit{abstract Lie algebra} $\mathcal{A}$, which is completely independent of $M$. Indeed, there may exist several $\Gamma$ (or none at all) associated to the same $\mathcal{A}$, because a given group $G$ (locally described by $\mathcal{A}$) may act in several ways on the same manifold $M$. At a given point ${\mathrm{x}\in M}$, the \textit{isotropy subalgebra} $\Gamma_{\mathrm{x}}$ of $\Gamma$ is given by ${\Gamma_{\mathrm{x}}:=\{\bs{X}\in\Gamma: \bs{X}|_{\mathrm{x}}=0\}\subset\Gamma}$ and characterizes the subset of the infinitesimal group actions that leave the point $\mathrm{x}$ unchanged. (The isotropy subalgebras can be identified with subalgebras of the Lorentz algebra up to conjugation and classified according to~\cite{PateraWinternitzZassenhaus}.)

The infinitesimal group actions $\Gamma$ have been (re)classified from the viewpoint of Lorentzian Lie algebra-subalgebra pairs by Hicks in \cite{Hicks:thesis}, which contains a complete list of 92 distinct cases. Hicks classifies only the \textit{simple-$G$} group actions, for which the isotropy subgroups are all equivalent to a given subgroup $H\subset G$ via group conjugation. In particular, isometry groups with generically non-null orbits are simple-$G$ group actions.  Locally, simple-$G$ spacetimes can be written in coordinates $(\check{x}, \hat{x})$, where $\check{x}$ represents coordinates on the orbits, which are identified with the homogeneous space $G/H$, and $\hat{x}$ represents a complementary set of group invariants which label the orbits. The non-simple-$G$ spacetimes are known to evade PSC conditions, so we do not need to worry about them in this paper.

The \textit{$\Gamma$-invariant metric} $\hat{\bs{g}}$ is a metric of which $\bs{X}\in\Gamma$, are the \textit{isometry generators} also known as the \textit{Killing vectors}. Choosing a basis of the Killing vectors, ${\bs{X}_i\in\Gamma}$, ${i=1,\dots,d}$ the metric $\hat{\bs{g}}$ has to satisfy the Killing equations,
\begin{equation}\label{eq:Killingeq}
    \lie_{\bs{X}_i} \hat{\bs{g}}=0\;, \quad i=1,\dots,d\;.
\end{equation}
Ignoring the condition on the metric signature, such $\hat{\bs{g}}$ (for a given $\Gamma$) form an $s$-dimensional linear subspace of all $\binom{0}{(2)}$ tensor fields. (We use the round and square brackets in the tensor types to denote the symmetry or anti-symmetry of indices.) Hence, $\hat{\bs{g}}$ can be decomposed into a basis of $\Gamma$-invariant $\binom{0}{(2)}$ tensor fields $\bs{q}_i$, ${i=1,\dots,s}$,
\begin{equation}\label{eq:metricbase}
    \hat{\bs{g}}=\sum_{i=1}^{s}\phi_i\bs{q}_i\;,
\end{equation}
where the scalar fields $\phi_i$, ${i=1,\dots,s}$, are the corresponding symmetry-reduced metric components, which capture the entire freedom in the $\Gamma$-invariant metric. The Lorentzian signature adds an extra inequality constraint on $\phi_i$ ensuring that the determinant of $\hat{\bs{g}}$ is negative, ${\hat{g}<0}$.

Let us denote by $l$ the dimension of linearly independent Killing vectors at a given point $\mathrm{x}$, i.e., the dimension of $\Gamma/\Gamma_{\mathrm{x}}$. This also corresponds to the dimension of the orbit of $\mathrm{x}$ in $M$. Clearly, the dimension $p$ of $\Gamma_{\mathrm{x}}$ is related to $l$ by ${l=d-p}$; the isotropy subalgebra becomes trivial for ${p=0}$, i.e, ${d=l}$. The individual infinitesimal group actions in Hicks classification are distinguished by different isometry-isotropy pairs $(\Gamma, \Gamma_{\mathrm{x}})$ and denoted by the triplet $[d,l,c]$, which captures their dimensions and includes an extra label $c$ (further distinguishing individual pairs). Note that $\Gamma$ are also the Killing vectors of the induced metric on the orbit. Hence, its maximal possible symmetry then provides the inequality ${d\leq l(l+1)/2}$. On the other hand, the maximum dimension of the orbit is that of the manifold $M$, i.e., ${l=4}$, which corresponds to the spacetimes with transitive group actions known as the \textit{homogeneous spacetimes}. (If ${d\geq 7}$, then the spacetime defined by $\hat{\bs{g}}$ is necessarily a homogeneous spacetime.) The Hicks classification focuses on ${3\leq d\leq7}$, since the remaining cases are rather special: The case ${d=0}$ means no Killing vectors. The case of a single Killing vector has a trivial isotropy subalgebra, ${d=l=1}$. There are two possibilities for ${d=2}$ both of which have trivial isotropy subalgebra, ${d=l=2}$, the abelian and the non-abelian one. It can be shown \cite{Petrov}, there are no $\Gamma$-invariant metrics for ${d=8}$ or ${d=9}$. Finally, ${d=10}$ are the cases of maximal symmetry, which are the spacetimes of positive, negative, or vanishing constant curvature, i.e., the \textit{de Sitter} ($\mathrm{dS_4}$), \textit{anti-de Sitter} ($\mathrm{AdS_4}$), or \textit{Minkowski} ($\mathrm{M_4}$) \textit{spacetimes}, respectively.

Given an infinitesimal group action $\Gamma$ in a neighborhood of $\mathrm{x}\in M$, PSC asserts that the variation of the Lagrangian commutes with the symmetry reduction and this is true simultaneously for all possible theories \cite{Fels:2001rv}. Specifically, let $\underline{\bs{E}}(\underline{L})$ be the \textit{Euler-Lagrange expression} (a $\binom{(2)}{0}$-tensor-valued 4-form) that is obtained by the variation of $\underline{L}$,
\begin{equation}\label{eq:deltaL}
    \delta\underline{L}=\underline{\bs{E}}(\underline{L})\cdot\delta\bs{g}+\underline{\mathrm{d}}\underline{\eta}[\delta\bs{g}]\;,
\end{equation}
where $\cdot$ is contraction in the two symmetric tensor indices and $\underline{\eta}$ is a boundary 3-form \cite{Iyer:1994ys}. The field equations are then given by vanishing of the Euler-Lagrange expression, $\underline{\bs{E}}(\underline{L})[\bs{g}]=0$. PSC corresponds to the situation when the \textit{reduced field equations} ${\underline{\bs{E}}(\underline{L})[\hat{\bs{g}}]=0}$ are equivalent to the field equations ${\underline{{E}}_i(\hat{\underline{L}})[\phi_j]=0}$ of the \textit{reduced Lagrangian ($r$-form)} $\hat{\underline{L}}$ (defined below) for any gravitational Lagrangian $\underline{L}$, namely\footnote{This formulation of PSC from \cite{Fels:2001rv} is purely local (and better adapted to the gravitational setting) in contrast to the original global formulation of PSC by Palais \cite{Palais:1979rca}.}
\begin{equation}\label{eq:PSC}
    \forall \underline{L}: \underline{\bs{E}}(\underline{L})[\hat{\bs{g}}]=0 \Longleftrightarrow \underline{{E}}_i(\hat{\underline{L}})[\phi_j]=0\;.
\end{equation}
Here, the \textit{reduced Euler-Lagrange expression} ${\underline{\bs{E}}(\underline{L})[\hat{\bs{g}}]}$ is obtained by ordinary evaluation of the original Euler-Lagrange expression $\underline{\bs{E}}(\underline{L})$ on the $\Gamma$-invariant metric $\hat{\bs{g}}$ (e.g., insertion of spherically symmetric ansatz into left-hand side of Einstein's field equations, as it is typically done in the textbook derivations of the Schwarzschild spacetime). On the other hand, $\underline{{E}}_i(\hat{\underline{L}})$ are the Euler-Lagrange expressions ($r$-forms) that are obtained by taking the variation of the reduced Lagrangian $\hat{\underline{L}}$ with respect to the scalar fields $\phi_i$, i.e., \begin{equation}\label{eq:deltahatL}
    \delta\hat{\underline{L}}=\sum_{i=1}^s\underline{{E}}_i(\hat{\underline{L}})\delta\phi_i+\underline{\mathrm{d}}\underline{\hat\eta}[\delta\phi_j]\;,
\end{equation}
where $\underline{\hat\eta}$ is a boundary ${(r{-}1)}$-form. The quantity $\hat{\underline{L}}$ serves as a Lagrangian for the scalar fields $\phi_i$ on the set of group orbits, i.e., the \textit{reduced spacetime} ${\hat{M}=M/G}$; similarly, ${\underline{{E}}_i(\hat{\underline{L}})[\phi_i]=0}$ are the corresponding field equations satisfied by fields $\phi_i$.

The reduced Lagrangian $\hat{\underline{L}}$ can be rigorously defined as follows. Since the reduced spacetime $\hat{M}$ is of dimension ${r=4-l}$, the reduction of the 4-form $\underline{L}$ involves not just substitution of the $\Gamma$-invariant metric $\hat{\bs{g}}$, but also the reduction of the form degree from 4 to ${r}$. A common but unjustified approach in the physics literature is to integrate over $l$ coordinates of the orbit while ignoring the issues with diverging integrals. Rigorously, this can be accomplished by contraction with a specific $\Gamma$-invariant $\binom{[l]}{0}$ tensor~$\bs{\chi}$,
\begin{equation}
    \lie_{\bs{X}_i}\bs{\chi}=0\;, \quad i=1,\dots,d\;,
    \label{eq:ginvchain}
\end{equation}
called the \textit{$l$-chain}, that is of the form
\begin{equation}
    \bs{\chi}=\chi^{i_1\dots i_l} \bs{X}_{i_1}\cdots\bs{X}_{i_l}\;, \quad \chi^{i_1\dots i_l}=\chi^{[i_1\dots i_l]}\;.
\end{equation}
The reduced Lagrangian $\hat{\underline{L}}$ is then defined by
\begin{equation}\label{eq:reducedL}
    \hat{\underline{L}}[\phi_i]:=\underline{L}[\hat{\bs{g}}]\bullet\bs{\chi} = \hat{\underline{\epsilon}}(\phi_i) \mathcal{L}[\hat{\bs{g}}]\;,
\end{equation}
where $\bullet$ denotes contraction of all anti-symmetric indices of $\bs{\chi}$ with the last indices of $\underline{L}$. This maps $\underline{\epsilon}(\bs{g})$, the form describing the volume element on $M$, into $\hat{\underline{\epsilon}}(\phi_i):=\underline{\epsilon}(\hat{\bs{g}})\bullet\bs{\chi}$, which is the form describing the volume element on the reduced spacetime $\hat{M}$.

Since \eqref{eq:PSC} is demanded to hold for all possible theories, PSC is a property of the infinitesimal group action $\Gamma$ only (around a given point $\mathrm{x}$); it is completely independent of any specific $\underline{L}$. Naturally, the condition $\underline{\bs{E}}(\underline{L})[\hat{\bs{g}}]=0 \Longleftrightarrow \underline{{E}}_i(\hat{\underline{L}})[\phi_j]=0$ may also hold just for a subset of all theories (or for a single theory) and then the conditions become theory dependent. This is not called PSC and we will not study this situation here.

The necessary and sufficient condition for the validity of PSC is composed of two conditions that must be satisfied simultaneously, the ``Lie algebra condition'' (PSC1) and the ``(local) Palais condition'' (PSC2).\footnote{These two conditions are necessary and sufficient provided the Lagrangian admits a $\Gamma$-invariant boundary term.  This is always the case for generally covariant metric theories, see \cite{Iyer:1994ys}.} PSC1 guarantees that the reduction of the boundary term in \eqref{eq:deltaL} is a boundary term for the reduced Lagrangian, that is, it does not produce a volume term that would modify  the field equations. This condition is most easily formulated in the form of an extra condition on the $l$-chain, 
\begin{equation}\label{eq:lchainPSCcompcondition}
    \lie_{\bs{V}}\bs{\chi}=0\;, 
\end{equation}
for all $\Gamma$-invariant vector fields $\bs{V}$, i.e., ${[\bs{X}_i,\bs{V}]=0}$, ${i=1,\dots,d}$. This determines  $\bs{\chi}$ up to a multiplicative constant in all the cases considered in this paper. In \cite{Anderson1997} it was shown that the existence of a suitable $l$-chain is equivalent to the non-vanishing of the Lie algebra cohomology of $\Gamma$ relative to $\Gamma_{\mathrm{x}}$ at degree $l$, 
\begin{equation}\label{eq:PSCcohomcondition}
    \mathcal{H}^l(\Gamma, \Gamma_{\mathrm{x}}) \neq 0\;.
\end{equation}
This cohomology condition can be evaluated by a straightforward analysis of the structure constants of the symmetry algebra and its isotropy subalgebra. Geometrically, it is equivalent to the existence of a closed $\Gamma$-invariant $l$-form on the group orbits which is not the exterior derivative of a $\Gamma$-invariant ${(l-1)}$-form.

If PSC1 is satisfied then the field equations of the reduced Lagrangian always yield at least a subset of the reduced field equations. PSC2 arises by demanding this subset to contain all reduced field equations. This happens if and only if all the reduced field equations appear in the reduction of the first term on the right-hand side of \eqref{eq:deltaL}. To check this condition, let $S_{\mathrm{x}}$ and $S_{\mathrm{x}}^*$ denote the vector space of $\Gamma_{\mathrm{x}}$-invariant $\binom{0}{(2)}$ and $\binom{(2)}{0}$ tensors at $\mathrm{x}$, respectively. Denote by $V_{\mathrm{x}}$ the vector space of $\binom{(2)}{0}$ tensors which have a vanishing scalar contraction with all elements of $S_{\mathrm{x}}$. PSC2 can be expressed as the requirement that in the desired neighborhood of ${\mathrm{x}}$,
\begin{equation}\label{eq:PSCpalaiscondition}
    S_{\mathrm{x}}^* \cap V_{\mathrm{x}} = \{0\}\;,
\end{equation}
meaning that at each $\mathrm{x}$ there is no $\Gamma_{\mathrm{x}}$-invariant $\binom{(2)}{0}$ tensor that would contract to zero with all $\Gamma_{\mathrm{x}}$-invariant $\binom{0}{(2)}$ tensors. In  \cite{Fels:2001rv}, the intersection of the two vector spaces was computed for all 14 distinct subalgebras of the Lorentz Lie algebra given by Patera, Winternitz and Zassuenhaus \cite{PateraWinternitzZassenhaus}. It was found that \eqref{eq:PSCpalaiscondition} is satisfied if and only if the isotropy subalgebra is not equivalent to any of the null-rotation subalgebras of the Lorentz algebra. Clearly, PSC1 as well as PSC2 are conditions on the infinitesimal group action $\Gamma$ only. Interestingly, PSC1 is not even sensitive to the nature of the fields being considered here, while PSC2 could be modified if fields other than the metric were considered.

We checked PSC1 and PSC2 conditions, \eqref{eq:PSCcohomcondition} and \eqref{eq:PSCpalaiscondition}, for all 92 infinitesimal group actions classified by Hicks and summarized the results in Tab.~\ref{tab:psccond}. This part was done using tools from \texttt{DifferentialGeometry} package \cite{diffgeo} of \textsc{Maple} \cite{maple}. To check PSC1, let $\bs{Z}_i$ be a basis of Killing vectors that are ordered so that the last ${r=d-l}$ elements, ${\bs{Z}_{l+1}, \ldots, \bs{Z}_d}$, correspond to the isotropy subalgebra and we denote the structure constants by $C_{ij}^k$, ${[\bs{Z}_i,\bs{Z}_j]=C_{ij}^k\bs{Z}_k}$. Following \cite{Anderson1997}, PSC1 is  equivalent to two conditions that need to be satisfied simultaneously. The first condition is that the adjoint representation of the isotropy subalgebra on $\Gamma/\Gamma_{\textrm{x}}$ (the tangent space to the orbit) is  traceless,
\begin{equation}
    \sum^{l}_{i=1} C^i_{m i } = 0\;,
    \quad  {m = l+1, \ldots, d}.
\end{equation}
 The second condition states that if there are any vectors tangent to the orbit which are invariant under the adjoint representation of the isotropy, ${\bs{v}=\sum_{i=1}^l v^i\bs{Z}_i}$, then the adjoint representation of these invariant vectors on the tangent vectors to the orbit should also be traceless:
\begin{equation}
    \sum^{l}_{i=1} v^i C^k_{m i } = 0\;
     \implies 
    \sum^{l}_{i,j=1} v^i C^j_{i j } = 0\;,\quad {k= 1, \ldots, l}.
\end{equation}
 In order to check PSC2, we calculated the spaces $S_{\mathrm{x}}$, $S^*_{\mathrm{x}}$, and $V_{\mathrm{x}}$ for each element in Hicks classification and then checked (\ref{eq:PSCpalaiscondition}).  As expected, PSC2 was satisfied for the group actions whose isotropy subalgebras were not  equivalent to any of the null-rotation subalgebras of the Lorentz algebra. We remark that PSC2 is always satisfied if the isotropy subalgebra is trivial, i.e., for all [3,3,-] and [4,4,-].

\begin{table}[!ht]
\caption{Infinitesimal group actions $\Gamma$ according to the Hicks classification and showing which satisfy PSC1, PSC2, and PSC}\label{tab:psccond}
    \centering
    \begin{tabular}[t]{|l||c|c||c|}
    \hline
    {\scriptsize Hicks \#} & {\scriptsize PSC1} & {\scriptsize PSC2} & {\scriptsize PSC} \\ \hline \hline
        [3,2,1] & \cmark & \cmark & \cmark \\ \hline
        [3,2,2] & \cmark & \cmark & \cmark \\ \hline
        [3,2,3] & \cmark & \cmark & \cmark \\ \hline
        [3,2,4] & \cmark & \cmark & \cmark \\ \hline
        [3,2,5] & \cmark & \cmark & \cmark \\ \hline\hline
        [3,3,1] & \xmark & \cmark & \xmark \\ \hline
        [3,3,2] & \cmark & \cmark & \cmark \\ \hline
        [3,3,3] & \cmark & \cmark & \cmark \\ \hline
        [3,3,4] & \xmark & \cmark & \xmark \\ \hline
        [3,3,5] & \xmark & \cmark & \xmark \\ \hline
        [3,3,6] & \xmark & \cmark & \xmark \\ \hline
        [3,3,7] & \xmark & \cmark & \xmark \\ \hline
        [3,3,8] & \cmark & \cmark & \cmark \\ \hline
        [3,3,9] & \cmark & \cmark & \cmark \\ \hline\hline
        [4,3,1] & \cmark & \cmark & \cmark \\ \hline
        [4,3,2] & \cmark & \cmark & \cmark \\ \hline
        [4,3,3] & \cmark & \cmark & \cmark \\ \hline
        [4,3,4] & \cmark & \cmark & \cmark \\ \hline
        [4,3,5] & \cmark & \cmark & \cmark \\ \hline
        [4,3,6] & \cmark & \cmark & \cmark \\ \hline
        [4,3,7] & \xmark & \cmark & \xmark \\ \hline
        [4,3,8] & \cmark & \cmark & \cmark \\ \hline
        [4,3,9] & \cmark & \cmark & \cmark \\ \hline
    \end{tabular}
    \quad
    \begin{tabular}[t]{|l||c|c||c|}
    \hline
    {\scriptsize Hicks \#} & {\scriptsize PSC1} & {\scriptsize PSC2} & {\scriptsize PSC} \\ \hline \hline
        [4,3,10] & \cmark & \cmark & \cmark \\ \hline
        [4,3,11] & \cmark & \cmark & \cmark \\ \hline
        [4,3,12] & \xmark & \cmark & \xmark \\ \hline
        [4,3,13] & \cmark & \xmark & \xmark \\ \hline
        [4,3,14] & \cmark & \xmark & \xmark \\ \hline
        [4,3,15] & \cmark & \xmark & \xmark \\ \hline
        [4,3,16] & \cmark & \xmark & \xmark \\ \hline
        [4,3,17] & \cmark & \xmark & \xmark \\ \hline
        [4,3,18] & \cmark & \xmark & \xmark \\ \hline
        [4,3,19] & \cmark & \xmark & \xmark \\ \hline
        [4,3,20] & \cmark & \xmark & \xmark \\ \hline\hline
        [4,4,1] & \cmark & \cmark & \cmark \\ \hline
        [4,4,2] & \cmark & \cmark & \cmark \\ \hline
        [4,4,3] & \xmark & \cmark & \xmark \\ \hline
        [4,4,4] & \xmark & \cmark & \xmark \\ \hline
        [4,4,5] & \xmark & \cmark & \xmark \\ \hline
        [4,4,6] & \xmark & \cmark & \xmark \\ \hline
        [4,4,7] & \xmark & \cmark & \xmark \\ \hline
        [4,4,8] & \xmark & \cmark & \xmark \\ \hline
        [4,4,9] & \cmark & \cmark & \cmark \\ \hline
        [4,4,10] & \xmark & \cmark & \xmark \\ \hline
        [4,4,11] & \xmark & \cmark & \xmark \\ \hline
        [4,4,12] & \xmark & \cmark & \xmark \\ \hline
    \end{tabular}
        \quad
    \begin{tabular}[t]{|l||c|c||c|}
    \hline
    {\scriptsize Hicks \#} & {\scriptsize PSC1} & {\scriptsize PSC2} & {\scriptsize PSC} \\ \hline \hline
        [4,4,13] & \xmark & \cmark & \xmark \\ \hline
        [4,4,14] & \xmark & \cmark & \xmark \\ \hline
        [4,4,15] & \xmark & \cmark & \xmark \\ \hline
        [4,4,16] & \xmark & \cmark & \xmark \\ \hline
        [4,4,17] & \xmark & \cmark & \xmark \\ \hline
        [4,4,18] & \cmark & \cmark & \cmark \\ \hline
        [4,4,19] & \xmark & \cmark & \xmark \\ \hline
        [4,4,20] & \xmark & \cmark & \xmark \\ \hline
        [4,4,21] & \xmark & \cmark & \xmark \\ \hline
        [4,4,22] & \cmark & \cmark & \cmark \\ \hline
        [4,4,23] & \xmark & \cmark & \xmark \\ \hline\hline
        [5,4,-1] & \cmark & \xmark & \xmark \\ \hline
        [5,4,-2] & \xmark & \xmark & \xmark \\ \hline
        [5,4,-3] & \xmark & \xmark & \xmark \\ \hline
        [5,4,-4] & \xmark & \xmark & \xmark \\ \hline
        [5,4,-5] & \xmark & \xmark & \xmark \\ \hline
        [5,4,-6] & \xmark & \xmark & \xmark \\ \hline
        [5,4,1] & \cmark & \cmark & \cmark \\ \hline
        [5,4,2] & \cmark & \cmark & \cmark \\ \hline
        [5,4,3] & \cmark & \cmark & \cmark \\ \hline
        [5,4,4] & \xmark & \cmark & \xmark \\ \hline
        [5,4,5] & \xmark & \cmark & \xmark \\ \hline
        [5,4,6] & \cmark & \cmark & \cmark \\ \hline
    \end{tabular}
        \quad
    \begin{tabular}[t]{|l||c|c||c|}
    \hline
    {\scriptsize Hicks \#} & {\scriptsize PSC1} & {\scriptsize PSC2} & {\scriptsize PSC} \\ \hline \hline
        [5,4,7] & \cmark & \cmark & \cmark \\ \hline
        [5,4,8] & \xmark & \cmark & \xmark \\ \hline
        [5,4,9] & \xmark & \cmark & \xmark \\ \hline
        [5,4,10] & \xmark & \xmark & \xmark \\ \hline
        [5,4,11] & \cmark & \xmark & \xmark \\ \hline\hline
        [6,3,1] & \cmark & \cmark & \cmark \\ \hline
        [6,3,2] & \cmark & \cmark & \cmark \\ \hline
        [6,3,3] & \cmark & \cmark & \cmark \\ \hline
        [6,3,4] & \cmark & \cmark & \cmark \\ \hline
        [6,3,5] & \cmark & \cmark & \cmark \\ \hline
        [6,3,6] & \cmark & \cmark & \cmark \\ \hline\hline
        [6,4,-1] & \cmark & \xmark & \xmark \\ \hline
        [6,4,1] & \cmark & \cmark & \cmark \\ \hline
        [6,4,2] & \cmark & \cmark & \cmark \\ \hline
        [6,4,3] & \cmark & \cmark & \cmark \\ \hline
        [6,4,4] & \cmark & \cmark & \cmark \\ \hline
        [6,4,5] & \cmark & \cmark & \cmark \\ \hline
        [6,4,6] & \cmark & \xmark & \xmark \\ \hline\hline
        [7,4,1] & \cmark & \cmark & \cmark \\ \hline
        [7,4,2] & \cmark & \cmark & \cmark \\ \hline
        [7,4,3] & \cmark & \cmark & \cmark \\ \hline
        [7,4,4] & \cmark & \cmark & \cmark \\ \hline
        [7,4,5] & \cmark & \xmark & \xmark \\ \hline
    \end{tabular}
\end{table}

Let us also briefly discuss PSC in the cases of low and high dimensions $d$. PSC2 is always satisfied if the isotropy subalgebra is trivial, ${d=l}$, which includes all infinitesimal isometry group actions with ${d=1}$ and ${d=2}$. Also, the maximally symmetric cases with ${d=10}$ adhere to PSC2 because there is no $\Gamma$ invariant null vector field, the existence of which is a property of all null-rotation subalgebras of the Lorentz algebra. One can check that PSC1 is satisfied for the abelian infinitesimal group actions with ${d=1,2}$, and also for ${d=10}$, but not for the non-abelian one with ${d=2}$.

From this point onwards, we will only deal with the 44 cases of Hicks classification with ${3\leq d\leq 7}$ that do satisfy both conditions simultaneously, which we call the \textit{PSC-compatible} infinitesimal group actions and list them explicitly in Tab.~\ref{tab:PSCcompgroupactions}. It should be stressed that  these are not given in the original \textit{Hicks coordinates} $x_i$ nor the original \textit{Hicks vector-field bases} $\bs{X}_i$ from \cite{Hicks:thesis}. We transformed both to different ones as described in Apx.~\ref{apx:transfcoord}, see Tab.~\ref{tab:transf} and Tab.~\ref{tab:redefKV}. The new \textit{adapted coordinates} $y_i$ and the new \textit{adapted vector-field bases} $\bs{Y}_i$ are better suited to the well-known geometries and mutual relations among algebras as we will discuss in Sec.~\ref{sec:insights}; in some situations they also simplify the resulting field equations. Another difference is in the system used for the calculations. The Hicks classification was originally built on \texttt{DifferentialGeometry} \cite{diffgeo} package of \textsc{Maple} \cite{maple}, while we implemented the classification and all procedures required for the symmetry reduction of Lagrangians in the \texttt{xAct} \cite{xact} package of \textsc{Mathematica}~\cite{Mathematica}.

If PSC \eqref{eq:PSC} holds, then one can easily find a direct relation between the reduced Euler-Lagrange expression $\underline{\bs{E}}(\underline{L})[\hat{\bs{g}}]$ and the Euler-Lagrange expressions of the reduced Lagrangian $\underline{E}_i(\hat{\underline{L}})[\phi_j]$, 
\begin{equation}
    \underline{E}_i(\hat{\underline{L}})[\phi_j]=(\underline{\bs{E}}(\underline{L})[\hat{\bs{g}}]\bullet\bs{\chi})\cdot\bs{q}_i\;.
\end{equation}
Since $\bs{\chi}$ is determined up to an overall constant, the relation is not quite one-to-one. This problem disappears if we extract the Levi-Civita tensors by introducing ${\underline{\bs{E}}=:\underline{\epsilon}\bs{\mathcal{E}}}$,  ${\underline{{E}}_i=:\hat{\underline{\epsilon}}{\mathcal{E}}_i}$; then we obtain
\begin{equation}
    {\mathcal{E}}_i(\hat{\underline{L}})[\phi_j]={\bs{\mathcal{E}}}(\underline{L})[\hat{\bs{g}}]\cdot\bs{q}_i\;.
\end{equation}
The last relation can be also inverted,
\begin{equation}\label{eq:mathcalErel}
    {\bs{\mathcal{E}}}(\underline{L})[\hat{\bs{g}}]=\sum_{i=1}^{s}{\mathcal{E}}_i(\underline{\hat{L}})[\phi_j]\bs{p}_i\;,
\end{equation}
where we introduced the base of $\Gamma$-invariant $\binom{(2)}{0}$ tensor fields $\bs{p}_i$, ${i=1,\dots,s}$, that are chosen to be dual to $\bs{q}_i$ (introduced above),
\begin{equation}\label{eq:dual}
    \bs{p}_i \cdot \bs{q}_j = \delta_{ij}\;.
\end{equation}
Note that the existence of such dual $\Gamma$-invariant bases $\bs{q}_i$ and $\bs{p}_i$ is equivalent to PSC2, cf.~\eqref{eq:PSCpalaiscondition}.

The Lagrangian $\underline{L}$ may be split into the gravitational and matter part, ${\underline{L}=\underline{L}_{\textrm{g}}+\underline{L}_{\textrm{m}}}$, where we consider the matter fields to be non-dynamical and $\Gamma$-invariant. [The symmetry reduction with extra dynamical fields would modify \eqref{eq:deltaL} by variations with respect to the extra fields and could also alter PSC2. Nevertheless, the validity of PSC2 would remain unchanged if the extra fields were scalars or vectors \cite{Fels:2001rv}.] The Euler-Lagrange expression then gives rise to the field equations, ${\underline{\bs{E}}(\underline{L})=\underline{\bs{E}}(\underline{L}_{\textrm{g}})+\underline{\bs{E}}(\underline{L}_{\textrm{m}})=0}$. Employing the above notation, the reduced field equation read
\begin{equation}\label{eq:feqEE}
    -2{\bs{\mathcal{E}}}(\underline{L}_{\textrm{g}})[\hat{\bs{g}}]=\hat{\bs{T}}\;,
\end{equation}
where we introduced the usual \textit{reduced energy-momentum tensor} ${\hat{\bs{T}}:=2{\bs{\mathcal{E}}}(\underline{L}_{\textrm{m}})[\hat{\bs{g}}]}$. 


\begin{longtable}{|l||>{\raggedright\arraybackslash}p{16.5cm}|}
\caption{PSC-compatible infinitesimal group actions $\Gamma$}\label{tab:PSCcompgroupactions}\\\hline
{\scriptsize Hicks \#} & {\scriptsize Infinitesimal group action $\Gamma$} \\ \hline\hline
{[3,2,1]} & $\bs{\partial}_{y_3}$, $\bs{\partial}_{y_4}$, $- y_4\, \bs{\partial}_{y_3} \allowbreak + y_3\, \bs{\partial}_{y_4}$   \\ \hline
{[3,2,2]} & $\cos y_4\, \bs{\partial}_{y_3} \allowbreak -  \coth y_3 \sin y_4\, \bs{\partial}_{y_4}$, $- \sin y_4\, \bs{\partial}_{y_3} \allowbreak -  \cos y_4 \coth y_3\, \bs{\partial}_{y_4}$, $\bs{\partial}_{y_4}$   \\ \hline
{[3,2,3]} & $\cos y_4\, \bs{\partial}_{y_3} \allowbreak -  \cot y_3 \sin y_4\, \bs{\partial}_{y_4}$, $- \sin y_4\, \bs{\partial}_{y_3} \allowbreak -  \cos y_4 \cot y_3\, \bs{\partial}_{y_4}$, $\bs{\partial}_{y_4}$   \\ \hline
{[3,2,4]} & $\bs{\partial}_{y_1}$, $\bs{\partial}_{y_2}$, $y_2\, \bs{\partial}_{y_1} \allowbreak + y_1\, \bs{\partial}_{y_2}$   \\ \hline
{[3,2,5]} & $- \sin y_1 \tanh y_2\, \bs{\partial}_{y_1} \allowbreak + \cos y_1\, \bs{\partial}_{y_2}$, $- \cos y_1 \tanh y_2\, \bs{\partial}_{y_1} \allowbreak -  \sin y_1\, \bs{\partial}_{y_2}$, $\bs{\partial}_{y_1}$   \\ \hline\hline
{[3,3,2]} & $\bs{\partial}_{y_1}$, $\bs{\partial}_{y_2}$, $\bs{\partial}_{y_4}$   \\ \hline
{[3,3,3]} & $\bs{\partial}_{y_1}$, $\bs{\partial}_{y_4}$, $y_4\, \bs{\partial}_{y_1} \allowbreak +\, \bs{\partial}_{y_3}$   \\ \hline
{[3,3,8]} & $- \csch y_3 \sin y_4\, \bs{\partial}_{y_1} \allowbreak + \cos y_4\, \bs{\partial}_{y_3} \allowbreak -  \coth y_3 \sin y_4\, \bs{\partial}_{y_4}$, $- \cos y_4 \csch y_3\, \bs{\partial}_{y_1} \allowbreak -  \sin y_4\, \bs{\partial}_{y_3} \allowbreak -  \cos y_4 \coth y_3\, \bs{\partial}_{y_4}$, $\bs{\partial}_{y_4}$   \\ \hline
{[3,3,9]} & $- \csc y_3 \sin y_4\, \bs{\partial}_{y_1} \allowbreak + \cos y_4\, \bs{\partial}_{y_3} \allowbreak -  \cot y_3 \sin y_4\, \bs{\partial}_{y_4}$, $- \cos y_4 \csc y_3\, \bs{\partial}_{y_1} \allowbreak -  \sin y_4\, \bs{\partial}_{y_3} \allowbreak -  \cos y_4 \cot y_3\, \bs{\partial}_{y_4}$, $\bs{\partial}_{y_4}$   \\ \hline\hline
{[4,3,1]} & $\cos y_4\, \bs{\partial}_{y_3} \allowbreak -  \coth y_3 \sin y_4\, \bs{\partial}_{y_4}$, $- \sin y_4\, \bs{\partial}_{y_3} \allowbreak -  \cos y_4 \coth y_3\, \bs{\partial}_{y_4}$, $\bs{\partial}_{y_4}$, $\bs{\partial}_{y_1}$   \\ \hline
{[4,3,2]} & $- \csch y_3 \sin y_4\, \bs{\partial}_{y_1} \allowbreak + \cos y_4\, \bs{\partial}_{y_3} \allowbreak -  \coth y_3 \sin y_4\, \bs{\partial}_{y_4}$, $- \cos y_4 \csch y_3\, \bs{\partial}_{y_1} \allowbreak -  \sin y_4\, \bs{\partial}_{y_3} \allowbreak -  \cos y_4 \coth y_3\, \bs{\partial}_{y_4}$, $\bs{\partial}_{y_4}$, $\bs{\partial}_{y_1}$   \\ \hline
{[4,3,3]} & $\cos y_4\, \bs{\partial}_{y_3} \allowbreak -  \cot y_3 \sin y_4\, \bs{\partial}_{y_4}$, $- \sin y_4\, \bs{\partial}_{y_3} \allowbreak -  \cos y_4 \cot y_3\, \bs{\partial}_{y_4}$, $\bs{\partial}_{y_4}$, $\bs{\partial}_{y_1}$   \\ \hline
{[4,3,4]} & $- \csc y_3 \sin y_4\, \bs{\partial}_{y_1} \allowbreak + \cos y_4\, \bs{\partial}_{y_3} \allowbreak -  \cot y_3 \sin y_4\, \bs{\partial}_{y_4}$, $- \cos y_4 \csc y_3\, \bs{\partial}_{y_1} \allowbreak -  \sin y_4\, \bs{\partial}_{y_3} \allowbreak -  \cos y_4 \cot y_3\, \bs{\partial}_{y_4}$, $\bs{\partial}_{y_4}$, $\bs{\partial}_{y_1}$   \\ \hline
{[4,3,5]} & $\bs{\partial}_{y_1}$, $\bs{\partial}_{y_4}$, $y_4\, \bs{\partial}_{y_1} \allowbreak +\, \bs{\partial}_{y_3}$, $\tfrac{1}{2} (- y_3^2 \allowbreak + y_4^2)\, \bs{\partial}_{y_1} \allowbreak + y_4\, \bs{\partial}_{y_3} \allowbreak -  y_3\, \bs{\partial}_{y_4}$   \\ \hline
{[4,3,6]} & $\bs{\partial}_{y_3}$, $\bs{\partial}_{y_4}$, $- y_4\, \bs{\partial}_{y_3} \allowbreak + y_3\, \bs{\partial}_{y_4}$, $\bs{\partial}_{y_1}$   \\ \hline
{[4,3,8]} & $- \sin y_1 \tanh y_2\, \bs{\partial}_{y_1} \allowbreak + \cos y_1\, \bs{\partial}_{y_2}$, $- \cos y_1 \tanh y_2\, \bs{\partial}_{y_1} \allowbreak -  \sin y_1\, \bs{\partial}_{y_2}$, $\bs{\partial}_{y_1}$, $\bs{\partial}_{y_4}$   \\ \hline
{[4,3,9]} & $- \sin y_1 \tanh y_2\, \bs{\partial}_{y_1} \allowbreak + \cos y_1\, \bs{\partial}_{y_2} \allowbreak -  \sech y_2 \sin y_1\, \bs{\partial}_{y_3}$, $- \cos y_1 \tanh y_2\, \bs{\partial}_{y_1} \allowbreak -  \sin y_1\, \bs{\partial}_{y_2} \allowbreak -  \cos y_1 \sech y_2\, \bs{\partial}_{y_3}$, $\bs{\partial}_{y_1}$, $\bs{\partial}_{y_3}$   \\ \hline
{[4,3,10]} & $\bs{\partial}_{y_1}$, $\bs{\partial}_{y_4}$, $y_4\, \bs{\partial}_{y_1} \allowbreak +\, \bs{\partial}_{y_3}$, $y_3\, \bs{\partial}_{y_3} \allowbreak -  y_4\, \bs{\partial}_{y_4}$   \\ \hline
{[4,3,11]} & $\bs{\partial}_{y_1}$, $\bs{\partial}_{y_2}$, $\bs{\partial}_{y_4}$, $y_2\, \bs{\partial}_{y_1} \allowbreak + y_1\, \bs{\partial}_{y_2}$   \\ \hline\hline
{[4,4,1]} & $- \csc y_3 \sin y_4\, \bs{\partial}_{y_1} \allowbreak + \cos y_4\, \bs{\partial}_{y_3} \allowbreak -  \cot y_3 \sin y_4\, \bs{\partial}_{y_4}$, $- \cos y_4 \csc y_3\, \bs{\partial}_{y_1} \allowbreak -  \sin y_4\, \bs{\partial}_{y_3} \allowbreak -  \cos y_4 \cot y_3\, \bs{\partial}_{y_4}$, $\bs{\partial}_{y_4}$, $\bs{\partial}_{y_2}$   \\ \hline
{[4,4,2]} & $- \csch y_3 \sin y_4\, \bs{\partial}_{y_1} \allowbreak + \cos y_4\, \bs{\partial}_{y_3} \allowbreak -  \coth y_3 \sin y_4\, \bs{\partial}_{y_4}$, $- \cos y_4 \csch y_3\, \bs{\partial}_{y_1} \allowbreak -  \sin y_4\, \bs{\partial}_{y_3} \allowbreak -  \cos y_4 \coth y_3\, \bs{\partial}_{y_4}$, $\bs{\partial}_{y_4}$, $\bs{\partial}_{y_2}$   \\ \hline
{[4,4,9]} & $\bs{\partial}_{y_1}$, $\bs{\partial}_{y_2}$, $\bs{\partial}_{y_4}$, $- y_4\, \bs{\partial}_{y_1} \allowbreak +\, \bs{\partial}_{y_3} \allowbreak + y_2\, \bs{\partial}_{y_4}$   \\ \hline
{[4,4,18]} & $\bs{\partial}_{y_1}$, $\bs{\partial}_{y_4}$, $y_4\, \bs{\partial}_{y_1} \allowbreak +\, \bs{\partial}_{y_3}$, $\bs{\partial}_{y_2} \allowbreak -  y_3\, \bs{\partial}_{y_3} \allowbreak + y_4\, \bs{\partial}_{y_4}$   \\ \hline
{[4,4,22]} & $\bs{\partial}_{y_1}$, $\bs{\partial}_{y_4}$, $y_4\, \bs{\partial}_{y_1} \allowbreak +\, \bs{\partial}_{y_3}$, $\tfrac{1}{2} (- y_3^2 \allowbreak + y_4^2)\, \bs{\partial}_{y_1} \allowbreak +\, \bs{\partial}_{y_2} \allowbreak + y_4\, \bs{\partial}_{y_3} \allowbreak -  y_3\, \bs{\partial}_{y_4}$   \\ \hline\hline
{[5,4,1]} & $- \csch y_3 \sin y_4\, \bs{\partial}_{y_1} \allowbreak + \cos y_4\, \bs{\partial}_{y_3} \allowbreak -  \coth y_3 \sin y_4\, \bs{\partial}_{y_4}$, $- \cos y_4 \csch y_3\, \bs{\partial}_{y_1} \allowbreak -  \sin y_4\, \bs{\partial}_{y_3} \allowbreak -  \cos y_4 \coth y_3\, \bs{\partial}_{y_4}$, $\bs{\partial}_{y_4}$, $\bs{\partial}_{y_1}$, $\bs{\partial}_{y_2}$   \\ \hline
{[5,4,2]} & $- \csc y_3 \sin y_4\, \bs{\partial}_{y_1} \allowbreak + \cos y_4\, \bs{\partial}_{y_3} \allowbreak -  \cot y_3 \sin y_4\, \bs{\partial}_{y_4}$, $- \cos y_4 \csc y_3\, \bs{\partial}_{y_1} \allowbreak -  \sin y_4\, \bs{\partial}_{y_3} \allowbreak -  \cos y_4 \cot y_3\, \bs{\partial}_{y_4}$, $\bs{\partial}_{y_4}$, $\bs{\partial}_{y_1}$, $\bs{\partial}_{y_2}$   \\ \hline
{[5,4,3]} & $\bs{\partial}_{y_1}$, $\bs{\partial}_{y_4}$, $y_4\, \bs{\partial}_{y_1} \allowbreak +\, \bs{\partial}_{y_3}$, $\tfrac{1}{2} (- y_3^2 \allowbreak + y_4^2)\, \bs{\partial}_{y_1} \allowbreak + y_4\, \bs{\partial}_{y_3} \allowbreak -  y_3\, \bs{\partial}_{y_4}$, $\bs{\partial}_{y_2}$   \\ \hline
{[5,4,6]} & $\bs{\partial}_{y_1}$, $\bs{\partial}_{y_4}$, $y_4\, \bs{\partial}_{y_1} \allowbreak +\, \bs{\partial}_{y_3}$, $y_3\, \bs{\partial}_{y_3} \allowbreak -  y_4\, \bs{\partial}_{y_4}$, $\bs{\partial}_{y_2}$   \\ \hline
{[5,4,7]} & $- \sin y_1 \tanh y_2\, \bs{\partial}_{y_1} \allowbreak + \cos y_1\, \bs{\partial}_{y_2} \allowbreak -  \sech y_2 \sin y_1\, \bs{\partial}_{y_3}$, $- \cos y_1 \tanh y_2\, \bs{\partial}_{y_1} \allowbreak -  \sin y_1\, \bs{\partial}_{y_2} \allowbreak -  \cos y_1 \sech y_2\, \bs{\partial}_{y_3}$, $\bs{\partial}_{y_1}$, $\bs{\partial}_{y_3}$, $\bs{\partial}_{y_4}$   \\ \hline\hline
{[6,3,1]} & $\cos y_4\, \bs{\partial}_{y_3} \allowbreak -  \cot y_3 \sin y_4\, \bs{\partial}_{y_4}$, $- \sin y_4\, \bs{\partial}_{y_3} \allowbreak -  \cos y_4 \cot y_3\, \bs{\partial}_{y_4}$, $\bs{\partial}_{y_4}$, $\cos y_4 \sin y_3\, \bs{\partial}_{y_2} \allowbreak + \cos y_3 \cos y_4 \cot y_2\, \bs{\partial}_{y_3} \allowbreak -  \cot y_2 \csc y_3 \sin y_4\, \bs{\partial}_{y_4}$, $\sin y_3 \sin y_4\, \bs{\partial}_{y_2} \allowbreak + \cos y_3 \cot y_2 \sin y_4\, \bs{\partial}_{y_3} \allowbreak + \cos y_4 \cot y_2 \csc y_3\, \bs{\partial}_{y_4}$, $\cos y_3\, \bs{\partial}_{y_2} \allowbreak -  \cot y_2 \sin y_3\, \bs{\partial}_{y_3}$   \\ \hline
{[6,3,2]} & $\bs{\partial}_{y_3}$, $\bs{\partial}_{y_4}$, $- y_4\, \bs{\partial}_{y_3} \allowbreak + y_3\, \bs{\partial}_{y_4}$, $y_4\, \bs{\partial}_{y_2} \allowbreak -  y_2\, \bs{\partial}_{y_4}$, $y_3\, \bs{\partial}_{y_2} \allowbreak -  y_2\, \bs{\partial}_{y_3}$, $-\, \bs{\partial}_{y_2}$   \\ \hline
{[6,3,3]} & $\cos y_4\, \bs{\partial}_{y_3} \allowbreak -  \cot y_3 \sin y_4\, \bs{\partial}_{y_4}$, $- \sin y_4\, \bs{\partial}_{y_3} \allowbreak -  \cos y_4 \cot y_3\, \bs{\partial}_{y_4}$, $\bs{\partial}_{y_4}$, $\cos y_4 \sin y_3\, \bs{\partial}_{y_2} \allowbreak + \cos y_3 \cos y_4 \coth y_2\, \bs{\partial}_{y_3} \allowbreak -  \coth y_2 \csc y_3 \sin y_4\, \bs{\partial}_{y_4}$, $\sin y_3 \sin y_4\, \bs{\partial}_{y_2} \allowbreak + \cos y_3 \coth y_2 \sin y_4\, \bs{\partial}_{y_3} \allowbreak + \cos y_4 \coth y_2 \csc y_3\, \bs{\partial}_{y_4}$, $\cos y_3\, \bs{\partial}_{y_2} \allowbreak -  \coth y_2 \sin y_3\, \bs{\partial}_{y_3}$   \\ \hline
{[6,3,4]} & $- \sin(y_1 \allowbreak -  y_3) \tanh y_2\, \bs{\partial}_{y_1} \allowbreak + \cos(y_1 \allowbreak -  y_3)\, \bs{\partial}_{y_2} \allowbreak + \coth y_2 \sin(y_1 \allowbreak -  y_3)\, \bs{\partial}_{y_3}$, $\cos(y_1 \allowbreak -  y_3) \tanh y_2\, \bs{\partial}_{y_1} \allowbreak + \sin(y_1 \allowbreak -  y_3)\, \bs{\partial}_{y_2} \allowbreak -  \cos(y_1 \allowbreak -  y_3) \coth y_2\, \bs{\partial}_{y_3}$, $-\, \bs{\partial}_{y_1} \allowbreak +\, \bs{\partial}_{y_3}$, $\sin(y_1 \allowbreak + y_3) \tanh y_2\, \bs{\partial}_{y_1} \allowbreak -  \cos(y_1 \allowbreak + y_3)\, \bs{\partial}_{y_2} \allowbreak + \coth y_2 \sin(y_1 \allowbreak + y_3)\, \bs{\partial}_{y_3}$, $\cos(y_1 \allowbreak + y_3) \tanh y_2\, \bs{\partial}_{y_1} \allowbreak + \sin(y_1 \allowbreak + y_3)\, \bs{\partial}_{y_2} \allowbreak + \cos(y_1 \allowbreak + y_3) \coth y_2\, \bs{\partial}_{y_3}$, $\bs{\partial}_{y_1} \allowbreak +\, \bs{\partial}_{y_3}$   \\ \hline
{[6,3,5]} & $\bs{\partial}_{y_1}$, $\bs{\partial}_{y_2}$, $\bs{\partial}_{y_4}$, $y_2\, \bs{\partial}_{y_1} \allowbreak + y_1\, \bs{\partial}_{y_2}$, $y_4\, \bs{\partial}_{y_1} \allowbreak + y_1\, \bs{\partial}_{y_4}$, $- y_4\, \bs{\partial}_{y_2} \allowbreak + y_2\, \bs{\partial}_{y_4}$   \\ \hline
{[6,3,6]} & $\cos y_3\, \bs{\partial}_{y_2} \allowbreak -  \cot y_2 \sin y_3\, \bs{\partial}_{y_3}$, $- \sin y_3\, \bs{\partial}_{y_2} \allowbreak -  \cos y_3 \cot y_2\, \bs{\partial}_{y_3}$, $\bs{\partial}_{y_3}$, $\cos y_3 \sin y_2\, \bs{\partial}_{y_1} \allowbreak + \cos y_2 \cos y_3 \tanh y_1\, \bs{\partial}_{y_2} \allowbreak -  \tanh y_1 \csc y_2 \sin y_3\, \bs{\partial}_{y_3}$, $\sin y_2 \sin y_3\, \bs{\partial}_{y_1} \allowbreak + \cos y_2 \tanh y_1 \sin y_3\, \bs{\partial}_{y_2} \allowbreak + \cos y_3 \tanh y_1 \csc y_2\, \bs{\partial}_{y_3}$, $\cos y_2\, \bs{\partial}_{y_1} \allowbreak -  \tanh y_1 \sin y_2\, \bs{\partial}_{y_2}$   \\ \hline\hline
{[6,4,1]} & $\cos y_4\, \bs{\partial}_{y_3} \allowbreak -  \cot y_3 \sin y_4\, \bs{\partial}_{y_4}$, $- \sin y_4\, \bs{\partial}_{y_3} \allowbreak -  \cos y_4 \cot y_3\, \bs{\partial}_{y_4}$, $\bs{\partial}_{y_4}$, $- \sin y_1 \tanh y_2\, \bs{\partial}_{y_1} \allowbreak + \cos y_1\, \bs{\partial}_{y_2}$, $- \cos y_1 \tanh y_2\, \bs{\partial}_{y_1} \allowbreak -  \sin y_1\, \bs{\partial}_{y_2}$, $\bs{\partial}_{y_1}$   \\ \hline
{[6,4,2]} & $\cos y_4\, \bs{\partial}_{y_3} \allowbreak -  \coth y_3 \sin y_4\, \bs{\partial}_{y_4}$, $- \sin y_4\, \bs{\partial}_{y_3} \allowbreak -  \cos y_4 \coth y_3\, \bs{\partial}_{y_4}$, $\bs{\partial}_{y_4}$, $- \sin y_1 \tanh y_2\, \bs{\partial}_{y_1} \allowbreak + \cos y_1\, \bs{\partial}_{y_2}$, $- \cos y_1 \tanh y_2\, \bs{\partial}_{y_1} \allowbreak -  \sin y_1\, \bs{\partial}_{y_2}$, $\bs{\partial}_{y_1}$   \\ \hline
{[6,4,3]} & $- \sin y_1 \tanh y_2\, \bs{\partial}_{y_1} \allowbreak + \cos y_1\, \bs{\partial}_{y_2}$, $- \cos y_1 \tanh y_2\, \bs{\partial}_{y_1} \allowbreak -  \sin y_1\, \bs{\partial}_{y_2}$, $\bs{\partial}_{y_1}$, $\bs{\partial}_{y_3}$, $\bs{\partial}_{y_4}$, $- y_4\, \bs{\partial}_{y_3} \allowbreak + y_3\, \bs{\partial}_{y_4}$   \\ \hline
{[6,4,4]} & $\cos y_4\, \bs{\partial}_{y_3} \allowbreak -  \cot y_3 \sin y_4\, \bs{\partial}_{y_4}$, $- \sin y_4\, \bs{\partial}_{y_3} \allowbreak -  \cos y_4 \cot y_3\, \bs{\partial}_{y_4}$, $\bs{\partial}_{y_4}$, $\bs{\partial}_{y_1}$, $\bs{\partial}_{y_2}$, $y_2\, \bs{\partial}_{y_1} \allowbreak + y_1\, \bs{\partial}_{y_2}$   \\ \hline
{[6,4,5]} & $\cos y_4\, \bs{\partial}_{y_3} \allowbreak -  \coth y_3 \sin y_4\, \bs{\partial}_{y_4}$, $- \sin y_4\, \bs{\partial}_{y_3} \allowbreak -  \cos y_4 \coth y_3\, \bs{\partial}_{y_4}$, $\bs{\partial}_{y_4}$, $y_2\, \bs{\partial}_{y_1} \allowbreak + y_1\, \bs{\partial}_{y_2}$, $\bs{\partial}_{y_1}$, $\bs{\partial}_{y_2}$   \\ \hline\hline
{[7,4,1]} & $\cos y_4\, \bs{\partial}_{y_3} \allowbreak -  \cot y_3 \sin y_4\, \bs{\partial}_{y_4}$, $- \sin y_4\, \bs{\partial}_{y_3} \allowbreak -  \cos y_4 \cot y_3\, \bs{\partial}_{y_4}$, $\bs{\partial}_{y_4}$, $\cos y_4 \sin y_3\, \bs{\partial}_{y_2} \allowbreak + \cos y_3 \cos y_4 \cot y_2\, \bs{\partial}_{y_3} \allowbreak -  \cot y_2 \csc y_3 \sin y_4\, \bs{\partial}_{y_4}$, $\sin y_3 \sin y_4\, \bs{\partial}_{y_2} \allowbreak + \cos y_3 \cot y_2 \sin y_4\, \bs{\partial}_{y_3} \allowbreak + \cos y_4 \cot y_2 \csc y_3\, \bs{\partial}_{y_4}$, $\cos y_3\, \bs{\partial}_{y_2} \allowbreak -  \cot y_2 \sin y_3\, \bs{\partial}_{y_3}$, $\bs{\partial}_{y_1}$   \\ \hline
{[7,4,2]} & $\cos y_4\, \bs{\partial}_{y_3} \allowbreak -  \cot y_3 \sin y_4\, \bs{\partial}_{y_4}$, $- \sin y_4\, \bs{\partial}_{y_3} \allowbreak -  \cos y_4 \cot y_3\, \bs{\partial}_{y_4}$, $\bs{\partial}_{y_4}$, $\cos y_4 \sin y_3\, \bs{\partial}_{y_2} \allowbreak + \cos y_3 \cos y_4 \coth y_2\, \bs{\partial}_{y_3} \allowbreak -  \coth y_2 \csc y_3 \sin y_4\, \bs{\partial}_{y_4}$, $\sin y_3 \sin y_4\, \bs{\partial}_{y_2} \allowbreak + \cos y_3 \coth y_2 \sin y_4\, \bs{\partial}_{y_3} \allowbreak + \cos y_4 \coth y_2 \csc y_3\, \bs{\partial}_{y_4}$, $\cos y_3\, \bs{\partial}_{y_2} \allowbreak -  \coth y_2 \sin y_3\, \bs{\partial}_{y_3}$, $\bs{\partial}_{y_1}$   \\ \hline
{[7,4,3]} & $\cos y_3\, \bs{\partial}_{y_2} \allowbreak -  \cot y_2 \sin y_3\, \bs{\partial}_{y_3}$, $- \sin y_3\, \bs{\partial}_{y_2} \allowbreak -  \cos y_3 \cot y_2\, \bs{\partial}_{y_3}$, $\bs{\partial}_{y_3}$, $\cos y_3 \sin y_2\, \bs{\partial}_{y_1} \allowbreak + \cos y_2 \cos y_3 \tanh y_1\, \bs{\partial}_{y_2} \allowbreak -  \tanh y_1 \csc y_2 \sin y_3\, \bs{\partial}_{y_3}$, $\sin y_2 \sin y_3\, \bs{\partial}_{y_1} \allowbreak + \cos y_2 \tanh y_1 \sin y_3\, \bs{\partial}_{y_2} \allowbreak + \cos y_3 \tanh y_1 \csc y_2\, \bs{\partial}_{y_3}$, $\cos y_2\, \bs{\partial}_{y_1} \allowbreak -  \tanh y_1 \sin y_2\, \bs{\partial}_{y_2}$, $\bs{\partial}_{y_4}$ \\ \hline
{[7,4,4]} & $- \sin(y_1 \allowbreak -  y_3) \tanh y_2\, \bs{\partial}_{y_1} \allowbreak + \cos(y_1 \allowbreak -  y_3)\, \bs{\partial}_{y_2} \allowbreak + \coth y_2 \sin(y_1 \allowbreak -  y_3)\, \bs{\partial}_{y_3}$, $\cos(y_1 \allowbreak -  y_3) \tanh y_2\, \bs{\partial}_{y_1} \allowbreak + \sin(y_1 \allowbreak -  y_3)\, \bs{\partial}_{y_2} \allowbreak -  \cos(y_1 \allowbreak -  y_3) \coth y_2\, \bs{\partial}_{y_3}$, $-\, \bs{\partial}_{y_1} \allowbreak +\, \bs{\partial}_{y_3}$, $\sin(y_1 \allowbreak + y_3) \tanh y_2\, \bs{\partial}_{y_1} \allowbreak -  \cos(y_1 \allowbreak + y_3)\, \bs{\partial}_{y_2} \allowbreak + \coth y_2 \sin(y_1 \allowbreak + y_3)\, \bs{\partial}_{y_3}$, $\cos(y_1 \allowbreak + y_3) \tanh y_2\, \bs{\partial}_{y_1} \allowbreak + \sin(y_1 \allowbreak + y_3)\, \bs{\partial}_{y_2} \allowbreak + \cos(y_1 \allowbreak + y_3) \coth y_2\, \bs{\partial}_{y_3}$, $\bs{\partial}_{y_1} \allowbreak +\, \bs{\partial}_{y_3}$, $\bs{\partial}_{y_4}$  \\ \hline
\end{longtable}


\section{\texorpdfstring{$\Gamma$-invariant metrics and $l$-chains}{Gamma-invariant metrics and l-chains}}\label{sec:metrics}

Given the set of PSC-compatible infinitesimal group actions, we can now search for the $\Gamma$-invariant metrics $\hat{\bs{g}}$ by solving the Killing equations \eqref{eq:Killingeq}. It should be emphasized that this is an opposite approach to what one is typically used to as we do not derive the Killing vectors for a given metric. Instead, we start with a base of vector fields $\bs{X}_i\in\Gamma$ on a manifold $M$ but no geometry is given {\it a priori}. Then we search for the metrics $\hat{\bs{g}}$ on $M$ of which $\bs{X}_i$ are the isometry generators. A calculation of this type was done by Petrov \cite{Petrov} for some infinitesimal group actions and also by Hicks \cite{Hicks:thesis}, both of which contain typos and mistakes of various kinds. We (re-)derived and corrected the latter for the PSC-compatible subset of infinitesimal group actions and expressed them in the adapted coordinates $y_i$. The results are summarized in Tab.~\ref{tab:invmetrforPSCcomp}. These are all possible four-dimensional spacetimes with symmetries that allow for symmetry reduction of arbitrary Lagrangians and lead to fully equivalent field equations.

Let us briefly comment on how the calculations of $\Gamma$-invariant metrics were performed. The Killing equations \eqref{eq:Killingeq} lead to a system of first-order partial differential equations for the unknown functions $\varphi_i$, i.e., the metric components in adapted coordinates $y_i$, ${\varphi_i = \{ \hat{g}_{11},\ldots,\hat{g}_{14},\hat{g}_{22},\ldots,\hat{g}_{24},\hat{g}_{33},\ldots,\hat{g}_{34},\hat{g}_{44}\}}$, which generally depend on all the coordinates $y_i$. Such a system is highly coupled and seemingly overdetermined. Therefore, we processed it in the following three steps to ensure the \textsc{Mathematica} differential equation solver \texttt{DSolve} is able to handle it. Step 1: We construct the matrix $\bs{\mathsf{A}}$ for a system of linear homogeneous equations $\bs{\mathsf{Ax}}=0$, where the unknowns $\mathsf{x}_p$ for ${p = 1,\ldots,i_\text{max}}$ correspond directly to the components $\varphi_i$, and for ${p=i_\text{max} + 1,\ldots, 5 i_\text{max}}$ to all their possible derivatives $\partial_i \varphi_j$. Note that the elements of $\bs{\mathsf{A}}$ are not necessarily constants in general. The matrix $\bs{\mathsf{A}}$ is then reduced using the Gaussian elimination method. We extract the equations involving only $\varphi_i$ without derivatives (i.e., $\mathsf{x}_p$ for $p = 1,\ldots,i_\text{max}$), solve for $\varphi_i$, and substitute the result back into the remaining system. Step 2: We construct the matrix $\bs{\mathsf{A}}$ again and reduce it as in step one. Now, all equations involving only one $\partial_i \varphi_j$ are extracted and this simple subsystem is solved using \texttt{DSolve}. The result is inserted back into the remaining system and the matrix $\bs{\mathsf{A}}$ is constructed and reduced again. Step 3: We go through the system sequentially, equation by equation, until \texttt{DSolve} is able to find a solution to any of the single equations. The result is substituted back and the remaining system is processed in the same way. If, during this iteration process, \texttt{DSolve} is unable to solve any equation in the remaining system, we restart step three with rearranged columns of matrix $\bs{\mathsf{A}}$ before its reduction, resulting in a different but equivalent system. This algorithm allowed us to solve the Killing equations \eqref{eq:Killingeq} for all the cases. Similar methods were used also for calculations of $\Gamma$-invariant vector fields, $l$-chains, and residual diffeomorphism generators below.

By looking at the Hicks labels $[d,l,c]$ in Tab.~\ref{tab:invmetrforPSCcomp}, it becomes obvious that the dimensions of isometry algebra $d$ and of the orbit $l$ (encoding the dimension $p$ of the isotropy subalgebra through ${p=d-l}$) crucially affect the effectiveness of the symmetry reduction. The value of $l$ is directly related to the number of variables upon which these functions depend, the dimension $r$ of the reduced spacetime $\hat{M}$ (recall that ${r=4-l}$). It dramatically changes the type of the equations we solve. For the homogeneous spacetimes, ${l=4}$, we have ${r=0}$, meaning that the field equations are purely \textit{algebraic equations} for the constants $\phi_i$. The metrics with ${l=3}$ lead to ${r=1}$, which is of a high interest as it leads to \textit{ordinary differential equations} and contain the most interesting metric ansatzes with single-variable functions $\phi_i$. The metrics with ${l=2}$, i.e., ${r=2}$, give rise to \textit{partial differential equations}, for which the ansatzes are very generic as they contain two-variable functions $\phi_i$. The value of $p$, on the other hand, may further influence the number $s$ of undetermined functions/constants $\phi_i$, in the $\Gamma$-invariant metric \eqref{eq:metricbase}, i.e., the number of independent $\bs{q}_i$. Larger $p$ generally leads to smaller or equal $s$. Naturally, smaller $s$ means fewer field equations since they are obtained by variation of $\hat{\underline{L}}$ with respect to $\phi_i$. The categories [3,3,-] and [4,4,-] have ${s=10}$, [3,2,-], [4,3,-], and [5,4,-] have ${s=4}$, while [6,3,-], [6,4,-], and [7,4,-] have ${s=2}$. (These numbers are given purely by invariance under the isotropy generators. Hence, they can be deduced by looking at the number of solutions of algebraic matrix similarity equations for adjoint isotropy representations.) As we will discuss in Sec.\ref{sec:gaugecond}, the number $s$ can be often effectively reduced by fixing some $\phi_i$ through the residual diffeomorphism invariance, but not always at the level of the Lagrangian (i.e., before we take the variation). Considering all these aspects, it is not surprising that the attractive setups [4,3,-] and [6,3,-] contain some of the most popular metric ansatzes such as the stationary spherically symmetric metric and spatially homogeneous isotropic cosmology, respectively.

There are some remarks to be made here. The Hicks classification classifies spacetimes by enumerating the possible Lorentzian Lie algebra-subalgebra pairs $(\Gamma,\Gamma_{\mathrm{x}})$ that can occur for a Lorentzian manifold admitting $\Gamma$ as Killing vectors. Hicks is, in effect, classifying spacetimes by listing the homogeneous spaces/spacetimes which can arise as orbits of the isometry groups. The classification is a local one: given a simple-$G$ spacetime, each point ${\mathrm{x}\in M}$ will have a neighborhood in which the algebra of its Killing vectors and isotropy subalgebra at $\mathrm{x}$, i.e., the pair $(\Gamma,\Gamma_{\mathrm{x}})$, is one of those enumerated by Hicks. It may happen that $\Gamma$ for two distinct cases are given by the same expressions, but they are valid at different ${\mathrm{x}\in M}$ that are stabilized by different $\Gamma_{\mathrm{x}}$. [The same is true for the $\Gamma$-invariant metric $\hat{\bs{g}}$, which are determined by $\Gamma$ through \eqref{eq:Killingeq}.] This is exactly the case of [6,3,3] and [6,3,6], or [7,4,2] and [7,4,3], in the original Hicks coordinates $x_i$. In our adapted coordinates $y_i$, however, these expressions are genuinely different as they are parameterized by coordinates that only admit the respective isotropy.

As mentioned in the previous sections, a rigorous reduction of a Lagrangian $\underline{L}$ requires not just substitution of the $\Gamma$-invariant metrics but also mapping this 4-form to an $r$-form by means of the $l$-chain $\bs{\chi}$, see \eqref{eq:reducedL}. Since we focus on PSC-compatible infinitesimal group actions, the $l$-chains are fixed uniquely (up to the constant scaling) by the conditions  \eqref{eq:ginvchain}, \eqref{eq:lchainPSCcompcondition}. [The reduction of PSC-non-compatible metrics is still possible via $l$-chains. Nevertheless, if PSC1 is violated, i.e., \eqref{eq:lchainPSCcompcondition} or \eqref{eq:PSCcohomcondition} is not met, the $l$-chains are not unique.] We solved these conditions and listed the $l$-chains for PSC-compatible infinitesimal group actions in Tab.~\ref{tab:lchainsPSCcomp}. As expressing the condition \eqref{eq:lchainPSCcompcondition} explicitly requires the $\Gamma$-invariant vector fields $\bs{V}$, ${[\bs{X}_i,\bs{V}]=0}$, ${i=1,\dots,d}$ which are of their own interest; we list them in Tab.~\ref{tab:invvectfieldsPSCcomp}. The number $t$ of undetermined functions/constants $f_i$, 
\begin{equation}
    \bs{V}=\sum_{i=1}^{t}f_i\bs{V}_i\;,
\end{equation}
i.e., the number of independent $\bs{V}_i$, again depends on $d$ and $l$. Specifically, the categories [3,3,-] and [4,4,-] have ${t=4}$, [3,2,-], [4,3,-], and [5,4,-] have $t=2$, [6,3,-] and [7,4,-] have ${t=1}$, while [6,4,-] have ${t=0}$ meaning that it possesses only a trivial $\Gamma$-invariant vector field, ${\bs{V}=0}$, which provides no extra constraints on $\bs{\chi}$ through \eqref{eq:lchainPSCcompcondition}. The $\Gamma$-invariant vector fields will become useful also in the context of residual diffeomorphism generators in Sec.~\ref{sec:gaugecond}.

\begin{longtable}{|l||>{\raggedright\arraybackslash}p{16.5cm}|}
\caption{$\Gamma$-invariant metrics}\label{tab:invmetrforPSCcomp}\\\hline
{\scriptsize Hicks \#} & {\scriptsize $\Gamma$-invariant metric $\hat{\bs{g}}$}\\\hline\hline
{[3,2,1]} & $- \phi_1(y_1, y_2)\, \bs{\mathrm{d}}{y_1}^2 \allowbreak + \phi_2(y_1, y_2) (\bs{\mathrm{d}}{y_1}\vee\bs{\mathrm{d}}{y_2}) \allowbreak + \phi_3(y_1, y_2)\, \bs{\mathrm{d}}{y_2}^2 \allowbreak + \phi_4(y_1, y_2) (\bs{\mathrm{d}}{y_3}^2 \allowbreak +\, \bs{\mathrm{d}}{y_4}^2)$ \\ \hline
{[3,2,2]} & $- \phi_1(y_1, y_2)\, \bs{\mathrm{d}}{y_1}^2 \allowbreak + \phi_2(y_1, y_2) (\bs{\mathrm{d}}{y_1}\vee\bs{\mathrm{d}}{y_2}) \allowbreak + \phi_3(y_1, y_2)\, \bs{\mathrm{d}}{y_2}^2 \allowbreak + \phi_4(y_1, y_2) (\bs{\mathrm{d}}{y_3}^2 \allowbreak + \sinh^2y_3\, \bs{\mathrm{d}}{y_4}^2)$ \\ \hline
{[3,2,3]} & $- \phi_1(y_1, y_2)\, \bs{\mathrm{d}}{y_1}^2 \allowbreak + \phi_2(y_1, y_2) (\bs{\mathrm{d}}{y_1}\vee\bs{\mathrm{d}}{y_2}) \allowbreak + \phi_3(y_1, y_2)\, \bs{\mathrm{d}}{y_2}^2 \allowbreak + \phi_4(y_1, y_2) (\bs{\mathrm{d}}{y_3}^2 \allowbreak + \sin^2y_3\, \bs{\mathrm{d}}{y_4}^2)$ \\ \hline
{[3,2,4]} & $\phi_1(y_3, y_4) (-\, \bs{\mathrm{d}}{y_1}^2 \allowbreak +\, \bs{\mathrm{d}}{y_2}^2) \allowbreak + \phi_2(y_3, y_4)\, \bs{\mathrm{d}}{y_3}^2 \allowbreak + \phi_3(y_3, y_4) (\bs{\mathrm{d}}{y_3}\vee\bs{\mathrm{d}}{y_4}) \allowbreak + \phi_4(y_3, y_4)\, \bs{\mathrm{d}}{y_4}^2$ \\ \hline
{[3,2,5]} & $\phi_1(y_3, y_4) (-\cosh^2y_2\, \bs{\mathrm{d}}{y_1}^2 \allowbreak + \, \bs{\mathrm{d}}{y_2}^2) \allowbreak + \phi_2(y_3, y_4)\, \bs{\mathrm{d}}{y_3}^2 \allowbreak + \phi_3(y_3, y_4) (\bs{\mathrm{d}}{y_3}\vee\bs{\mathrm{d}}{y_4}) \allowbreak + \phi_4(y_3, y_4)\, \bs{\mathrm{d}}{y_4}^2$ \\ \hline\hline
{[3,3,2]} & $- \phi_1(y_3)\, \bs{\mathrm{d}}{y_1}^2 \allowbreak + \phi_2(y_3)\, \bs{\mathrm{d}}{y_3}^2 \allowbreak + \phi_3(y_3) (\bs{\mathrm{d}}{y_3}\vee\bs{\mathrm{d}}{y_4}) \allowbreak + \phi_4(y_3)\, \bs{\mathrm{d}}{y_4}^2 \allowbreak + \phi_5(y_3)\, \bs{\mathrm{d}}{y_2}^2 \allowbreak + \phi_6(y_3) (\bs{\mathrm{d}}{y_1}\vee\bs{\mathrm{d}}{y_2}) \allowbreak + \phi_7(y_3) (\bs{\mathrm{d}}{y_1}\vee\bs{\mathrm{d}}{y_3}) \allowbreak + \phi_8(y_3) (\bs{\mathrm{d}}{y_1}\vee\bs{\mathrm{d}}{y_4}) \allowbreak + \phi_9(y_3) (\bs{\mathrm{d}}{y_2}\vee\bs{\mathrm{d}}{y_3}) \allowbreak + \phi_{10}(y_3) (\bs{\mathrm{d}}{y_2}\vee\bs{\mathrm{d}}{y_4})$ \\ \hline
{[3,3,3]} & $-\phi_1(y_2) (\bs{\mathrm{d}}{y_1} \allowbreak - y_3\, \bs{\mathrm{d}}{y_4})^2 \allowbreak + \phi_2(y_2) \bigl(\bs{\mathrm{d}}{y_2}\vee(\bs{\mathrm{d}}{y_1} \allowbreak - y_3\, \bs{\mathrm{d}}{y_4})\bigr) \allowbreak + \phi_3(y_2)\, \bs{\mathrm{d}}{y_2}^2 \allowbreak + \phi_4(y_2)\, \bs{\mathrm{d}}{y_3}^2 \allowbreak + \phi_5(y_2) \bigl(\bs{\mathrm{d}}{y_3}\vee(\bs{\mathrm{d}}{y_1} \allowbreak - y_3\, \bs{\mathrm{d}}{y_4})\bigr) \allowbreak + \phi_6(y_2) \bigl(\bs{\mathrm{d}}{y_4}\vee(\bs{\mathrm{d}}{y_1} \allowbreak - y_3\, \bs{\mathrm{d}}{y_4})\bigr) \allowbreak + \phi_7(y_2) (\bs{\mathrm{d}}{y_2}\vee\bs{\mathrm{d}}{y_3}) \allowbreak + \phi_8(y_2) (\bs{\mathrm{d}}{y_2}\vee\bs{\mathrm{d}}{y_4}) \allowbreak + \phi_9(y_2) (\bs{\mathrm{d}}{y_3}\vee\bs{\mathrm{d}}{y_4}) \allowbreak + \phi_{10}(y_2)\, \bs{\mathrm{d}}{y_4}^2$ \\ \hline
{[3,3,8]} & $- \phi_1(y_2) (\bs\omega^1)^2 + \phi_2(y_2) (\bs{\mathrm{d}}{y_2}\vee\bs\omega^1) + \phi_3(y_2) \,\bs{\mathrm{d}}{y_2}^2 + \phi_4(y_2) (\bs\omega^2)^2 \allowbreak + \phi_5(y_2) (\bs{\mathrm{d}}{y_2}\vee\bs\omega^3) + \phi_6(y_2) (\bs\omega^1\vee\bs\omega^3) \allowbreak + \phi_7(y_2) (\bs\omega^2 \vee \bs\omega^3) \allowbreak + \phi_8(y_2) (\bs\omega^1 \vee \bs\omega^2) + \phi_9(y_2) (\bs{\mathrm{d}}{y_2}\vee\bs\omega^2) \allowbreak + \phi_{10}(y_2) (\bs\omega^3)^2$, \newline where $\bs{\omega}^1 = \bs{\mathrm{d}}{y_1} - \cosh y_3 \, \bs{\mathrm{d}}{y_4}$, $\bs\omega^2 = \sin y_1 \,\bs{\mathrm{d}}{y_3} + \cos y_1 \sinh y_3 \,\bs{\mathrm{d}}{y_4}$, $\bs\omega^3 = \cos y_1 \,\bs{\mathrm{d}}{y_3} -  \sin y_1 \sinh y_3 \,\bs{\mathrm{d}}{y_4}$ \\ \hline
{[3,3,9]} & $- \phi_1(y_2) (\bs\omega^1)^2 + \phi_2(y_2) (\bs{\mathrm{d}}{y_2}\vee\bs\omega^1) + \phi_3(y_2) \,\bs{\mathrm{d}}{y_2}^2 + \phi_4(y_2) (\bs\omega^2)^2 \allowbreak + \phi_5(y_2) (\bs{\mathrm{d}}{y_2}\vee\bs\omega^3) + \phi_6(y_2) (\bs\omega^1\vee\bs\omega^3) \allowbreak + \phi_7(y_2) (\bs\omega^2 \vee \bs\omega^3) \allowbreak + \phi_8(y_2) (\bs\omega^1 \vee \bs\omega^2) + \phi_9(y_2) (\bs{\mathrm{d}}{y_2}\vee\bs\omega^2) \allowbreak + \phi_{10}(y_2) (\bs\omega^3)^2$, \newline where $\bs{\omega}^1 = \bs{\mathrm{d}}{y_1} - \cos y_3 \, \bs{\mathrm{d}}{y_4}$, $\bs\omega^2 = \sin y_1 \,\bs{\mathrm{d}}{y_3} + \cos y_1 \sin y_3 \,\bs{\mathrm{d}}{y_4}$, $\bs\omega^3 = \cos y_1 \,\bs{\mathrm{d}}{y_3} -  \sin y_1 \sin y_3 \,\bs{\mathrm{d}}{y_4}$ \\ \hline \hline
{[4,3,1]} & $- \phi_1(y_2)\, \bs{\mathrm{d}}{y_1}^2 \allowbreak + \phi_2(y_2) (\bs{\mathrm{d}}{y_1}\vee\bs{\mathrm{d}}{y_2}) \allowbreak + \phi_3(y_2)\, \bs{\mathrm{d}}{y_2}^2 \allowbreak + \phi_4(y_2) (\bs{\mathrm{d}}{y_3}^2 \allowbreak + \sinh^2y_3\, \bs{\mathrm{d}}{y_4}^2)$ \\ \hline
{[4,3,2]} & $-\phi_1(y_2) (\bs{\mathrm{d}}{y_1} \allowbreak - \cosh y_3\, \bs{\mathrm{d}}{y_4})^2 \allowbreak + \phi_2(y_2) \bigl(\bs{\mathrm{d}}{y_2}\vee(\bs{\mathrm{d}}{y_1} \allowbreak - \cosh y_3\, \bs{\mathrm{d}}{y_4})\bigr) \allowbreak + \phi_3(y_2)\, \bs{\mathrm{d}}{y_2}^2 \allowbreak + \phi_4(y_2) (\bs{\mathrm{d}}{y_3}^2 \allowbreak + \sinh^2y_3\, \bs{\mathrm{d}}{y_4}^2)$ \\ \hline
{[4,3,3]} & $- \phi_1(y_2)\, \bs{\mathrm{d}}{y_1}^2 \allowbreak + \phi_2(y_2) (\bs{\mathrm{d}}{y_1}\vee\bs{\mathrm{d}}{y_2}) \allowbreak + \phi_3(y_2)\, \bs{\mathrm{d}}{y_2}^2 \allowbreak + \phi_4(y_2) (\bs{\mathrm{d}}{y_3}^2 \allowbreak + \sin^2y_3\, \bs{\mathrm{d}}{y_4}^2)$ \\ \hline
{[4,3,4]} & $-\phi_1(y_2) (\bs{\mathrm{d}}{y_1} \allowbreak - \cos y_3\, \bs{\mathrm{d}}{y_4})^2 \allowbreak + \phi_2(y_2) \bigl(\bs{\mathrm{d}}{y_2}\vee(\bs{\mathrm{d}}{y_1} \allowbreak - \cos y_3\, \bs{\mathrm{d}}{y_4})\bigr) \allowbreak + \phi_3(y_2)\, \bs{\mathrm{d}}{y_2}^2 \allowbreak + \phi_4(y_2) (\bs{\mathrm{d}}{y_3}^2 \allowbreak + \sin^2y_3\, \bs{\mathrm{d}}{y_4}^2)$ \\ \hline
{[4,3,5]} & $-\phi_1(y_2) (\bs{\mathrm{d}}{y_1} \allowbreak - y_3\, \bs{\mathrm{d}}{y_4})^2 \allowbreak + \phi_2(y_2) \bigl(\bs{\mathrm{d}}{y_2}\vee(\bs{\mathrm{d}}{y_1} \allowbreak - y_3\, \bs{\mathrm{d}}{y_4})\bigr) \allowbreak + \phi_3(y_2)\, \bs{\mathrm{d}}{y_2}^2 \allowbreak + \phi_4(y_2) (\bs{\mathrm{d}}{y_3}^2 \allowbreak +\, \bs{\mathrm{d}}{y_4}^2)$ \\ \hline
{[4,3,6]} & $- \phi_1(y_2)\, \bs{\mathrm{d}}{y_1}^2 \allowbreak + \phi_2(y_2) (\bs{\mathrm{d}}{y_1}\vee\bs{\mathrm{d}}{y_2}) \allowbreak + \phi_3(y_2)\, \bs{\mathrm{d}}{y_2}^2 \allowbreak + \phi_4(y_2) (\bs{\mathrm{d}}{y_3}^2 \allowbreak +\, \bs{\mathrm{d}}{y_4}^2)$ \\ \hline
{[4,3,8]} & $\phi_1(y_3) (-\cosh^2y_2\, \bs{\mathrm{d}}{y_1}^2 \allowbreak + \, \bs{\mathrm{d}}{y_2}^2) \allowbreak + \phi_2(y_3)\, \bs{\mathrm{d}}{y_3}^2 \allowbreak + \phi_3(y_3) (\bs{\mathrm{d}}{y_3}\vee\bs{\mathrm{d}}{y_4}) \allowbreak + \phi_4(y_3)\, \bs{\mathrm{d}}{y_4}^2$ \\ \hline
{[4,3,9]} & $\phi_1(y_4) (-\cosh^2y_2\, \bs{\mathrm{d}}{y_1}^2 \allowbreak + \, \bs{\mathrm{d}}{y_2}^2) \allowbreak + \phi_2(y_4) (\sinh y_2\, \bs{\mathrm{d}}{y_1} \allowbreak +\, \bs{\mathrm{d}}{y_3})^2 \allowbreak + \phi_3(y_4) \bigl(\bs{\mathrm{d}}{y_4}\vee(\sinh y_2\, \bs{\mathrm{d}}{y_1} \allowbreak +\, \bs{\mathrm{d}}{y_3})\bigr) \allowbreak + \phi_4(y_4)\, \bs{\mathrm{d}}{y_4}^2$ \\ \hline
{[4,3,10]} & $-\phi_1(y_2) (\bs{\mathrm{d}}{y_1} \allowbreak - y_3\, \bs{\mathrm{d}}{y_4})^2 \allowbreak + \phi_2(y_2) \bigl(\bs{\mathrm{d}}{y_2}\vee(\bs{\mathrm{d}}{y_1} \allowbreak - y_3\, \bs{\mathrm{d}}{y_4})\bigr) \allowbreak + \phi_3(y_2)\, \bs{\mathrm{d}}{y_2}^2 \allowbreak + \phi_4(y_2) (\bs{\mathrm{d}}{y_3}\vee\bs{\mathrm{d}}{y_4})$ \\ \hline
{[4,3,11]} & $\phi_1(y_3) (-\, \bs{\mathrm{d}}{y_1}^2 \allowbreak +\, \bs{\mathrm{d}}{y_2}^2) \allowbreak + \phi_2(y_3)\, \bs{\mathrm{d}}{y_3}^2 \allowbreak + \phi_3(y_3) (\bs{\mathrm{d}}{y_3}\vee\bs{\mathrm{d}}{y_4}) \allowbreak + \phi_4(y_3)\, \bs{\mathrm{d}}{y_4}^2$ \\ \hline \hline
{[4,4,1]} & $- \phi_1 (\bs\omega^1)^2 + \phi_2 (\bs{\mathrm{d}}{y_2}\vee\bs\omega^1) + \phi_3 \,\bs{\mathrm{d}}{y_2}^2 + \phi_4 (\bs\omega^2)^2 \allowbreak + \phi_5 (\bs{\mathrm{d}}{y_2}\vee\bs\omega^3) + \phi_6 (\bs\omega^1\vee\bs\omega^3) \allowbreak + \phi_7 (\bs\omega^2 \vee \bs\omega^3) \allowbreak + \phi_8 (\bs\omega^1 \vee \bs\omega^2) + \phi_9 (\bs{\mathrm{d}}{y_2}\vee\bs\omega^2) \allowbreak + \phi_{10} (\bs\omega^3)^2$, \newline where $\bs{\omega}^1 = \bs{\mathrm{d}}{y_1} - \cos y_3 \, \bs{\mathrm{d}}{y_4}$, $\bs\omega^2 = \sin y_1 \,\bs{\mathrm{d}}{y_3} + \cos y_1 \sin y_3 \,\bs{\mathrm{d}}{y_4}$, $\bs\omega^3 = \cos y_1 \,\bs{\mathrm{d}}{y_3} -  \sin y_1 \sin y_3 \,\bs{\mathrm{d}}{y_4}$ \\ \hline
{[4,4,2]} & $- \phi_1 (\bs\omega^1)^2 + \phi_2 (\bs{\mathrm{d}}{y_2}\vee\bs\omega^1) + \phi_3 \,\bs{\mathrm{d}}{y_2}^2 + \phi_4 (\bs\omega^2)^2 \allowbreak + \phi_5 (\bs{\mathrm{d}}{y_2}\vee\bs\omega^3) + \phi_6 (\bs\omega^1\vee\bs\omega^3) \allowbreak + \phi_7 (\bs\omega^2 \vee \bs\omega^3) \allowbreak + \phi_8 (\bs\omega^1 \vee \bs\omega^2) + \phi_9 (\bs{\mathrm{d}}{y_2}\vee\bs\omega^2) \allowbreak + \phi_{10} (\bs\omega^3)^2$, \newline where $\bs{\omega}^1 = \bs{\mathrm{d}}{y_1} - \cosh y_3 \, \bs{\mathrm{d}}{y_4}$, $\bs\omega^2 = \sin y_1 \,\bs{\mathrm{d}}{y_3} + \cos y_1 \sinh y_3 \,\bs{\mathrm{d}}{y_4}$, $\bs\omega^3 = \cos y_1 \,\bs{\mathrm{d}}{y_3} -  \sin y_1 \sinh y_3 \,\bs{\mathrm{d}}{y_4}$ \\ \hline
{[4,4,9]} & $- \phi_1 (\bs{\mathrm{d}}{y_1} - \tfrac{1}{2} y_3^2 \, \bs{\mathrm{d}}{y_2} + y_3 \, \bs{\mathrm{d}}{y_4})^2 + \phi_2 \, \bs{\mathrm{d}}{y_3}^2 + \phi_3 \bigl(\bs{\mathrm{d}}{y_3}\vee(- y_3 \, \bs{\mathrm{d}}{y_2} + \bs{\mathrm{d}}{y_4})\bigr) + \phi_4 (- y_3 \, \bs{\mathrm{d}}{y_2} + \bs{\mathrm{d}}{y_4})^2 + \phi_5 \, \bs{\mathrm{d}}{y_2}^2 \allowbreak + \phi_6 \bigl(\bs{\mathrm{d}}{y_2}\vee(\bs{\mathrm{d}}{y_1} -  \tfrac{1}{2} y_3^2 \, \bs{\mathrm{d}}{y_2} + y_3 \, \bs{\mathrm{d}}{y_4})\bigr) + \phi_7 \bigl(\bs{\mathrm{d}}{y_3}\vee(\bs{\mathrm{d}}{y_1} - \tfrac{1}{2} y_3^2 \, \bs{\mathrm{d}}{y_2} + y_3 \, \bs{\mathrm{d}}{y_4})\bigr) \allowbreak + \phi_8 \bigl((- y_3 \, \bs{\mathrm{d}}{y_2} + \bs{\mathrm{d}}{y_4})\vee(\bs{\mathrm{d}}{y_1} - \tfrac{1}{2} y_3^2 \, \bs{\mathrm{d}}{y_2} + y_3 \, \bs{\mathrm{d}}{y_4})\bigr) + \phi_9 (\bs{\mathrm{d}}{y_2}\vee\bs{\mathrm{d}}{y_3}) + \phi_{10} \bigl(\bs{\mathrm{d}}{y_2}\vee(- y_3 \, \bs{\mathrm{d}}{y_2} + \bs{\mathrm{d}}{y_4})\bigr)$ \\ \hline
{[4,4,18]} & $-\phi_1 (\bs{\mathrm{d}}{y_1} \allowbreak - y_3\, \bs{\mathrm{d}}{y_4})^2 \allowbreak + \phi_2 \bigl(\bs{\mathrm{d}}{y_2}\vee(\bs{\mathrm{d}}{y_1} \allowbreak - y_3\, \bs{\mathrm{d}}{y_4})\bigr) \allowbreak + \phi_3\, \bs{\mathrm{d}}{y_2}^2 \allowbreak + e^{2 y_2} \phi_4\, \bs{\mathrm{d}}{y_3}^2 \allowbreak + e^{y_2} \phi_5 \bigl(\bs{\mathrm{d}}{y_3}\vee(\bs{\mathrm{d}}{y_1} \allowbreak - y_3\, \bs{\mathrm{d}}{y_4})\bigr) \allowbreak + e^{- y_2} \phi_6 \bigl(\bs{\mathrm{d}}{y_4}\vee(\bs{\mathrm{d}}{y_1} \allowbreak - y_3\, \bs{\mathrm{d}}{y_4})\bigr) \allowbreak + e^{y_2} \phi_7 (\bs{\mathrm{d}}{y_2}\vee\bs{\mathrm{d}}{y_3}) \allowbreak + e^{- y_2} \phi_8 (\bs{\mathrm{d}}{y_2}\vee\bs{\mathrm{d}}{y_4}) \allowbreak + \phi_9 (\bs{\mathrm{d}}{y_3}\vee\bs{\mathrm{d}}{y_4}) \allowbreak + e^{-2 y_2} \phi_{10}\, \bs{\mathrm{d}}{y_4}^2$ \\ \hline
{[4,4,22]} & $- \phi_1 (\bs\omega_1)^2 + \phi_2 (\bs{\mathrm{d}}{y_2}\vee\bs\omega^1) + \phi_3 \, \bs{\mathrm{d}}{y_2}^2 + \phi_4 (\bs\omega^2)^2 \allowbreak + \phi_5 (\bs\omega^1 \vee \bs\omega^2) + \phi_6 (\bs\omega^1 \vee \bs\omega^3) \allowbreak + \phi_7 (\bs{\mathrm{d}}{y_2}\vee\bs\omega^2) + \phi_8 (\bs{\mathrm{d}}{y_2}\vee\bs\omega^3) \allowbreak + \phi_9 (\bs\omega^2 \vee \bs\omega^3) \allowbreak + \phi_{10} (\bs\omega^3)^2$, \newline where $\bs{\omega}^1 = \bs{\mathrm{d}}{y_1} - y_3 \, \bs{\mathrm{d}}{y_4}$, $\bs\omega^2 = \cos y_2 \,\bs{\mathrm{d}}{y_3} - \sin y_2 \,\bs{\mathrm{d}}{y_4}$, $\bs\omega^3 = \sin y_2 \,\bs{\mathrm{d}}{y_3} + \cos y_2 \,\bs{\mathrm{d}}{y_4}$ \\ \hline \hline
{[5,4,1]} & $-\phi_1 (\bs{\mathrm{d}}{y_1} \allowbreak - \cosh y_3\, \bs{\mathrm{d}}{y_4})^2 \allowbreak + \phi_2 \bigl(\bs{\mathrm{d}}{y_2}\vee(\bs{\mathrm{d}}{y_1} \allowbreak - \cosh y_3\, \bs{\mathrm{d}}{y_4})\bigr) \allowbreak + \phi_3\, \bs{\mathrm{d}}{y_2}^2 \allowbreak + \phi_4 (\bs{\mathrm{d}}{y_3}^2 \allowbreak + \sinh^2y_3\, \bs{\mathrm{d}}{y_4}^2)$ \\ \hline
{[5,4,2]} & $-\phi_1 (\bs{\mathrm{d}}{y_1} \allowbreak - \cos y_3\, \bs{\mathrm{d}}{y_4})^2 \allowbreak + \phi_2 \bigl(\bs{\mathrm{d}}{y_2}\vee(\bs{\mathrm{d}}{y_1} \allowbreak - \cos y_3\, \bs{\mathrm{d}}{y_4})\bigr) \allowbreak + \phi_3\, \bs{\mathrm{d}}{y_2}^2 \allowbreak + \phi_4 (\bs{\mathrm{d}}{y_3}^2 \allowbreak + \sin^2y_3\, \bs{\mathrm{d}}{y_4}^2)$ \\ \hline
{[5,4,3]} & $-\phi_1 (\bs{\mathrm{d}}{y_1} \allowbreak - y_3\, \bs{\mathrm{d}}{y_4})^2 \allowbreak + \phi_2 \bigl(\bs{\mathrm{d}}{y_2}\vee(\bs{\mathrm{d}}{y_1} \allowbreak - y_3\, \bs{\mathrm{d}}{y_4})\bigr) \allowbreak + \phi_3\, \bs{\mathrm{d}}{y_2}^2 \allowbreak + \phi_4 (\bs{\mathrm{d}}{y_3}^2 \allowbreak +\, \bs{\mathrm{d}}{y_4}^2)$ \\ \hline
{[5,4,6]} & $-\phi_1 (\bs{\mathrm{d}}{y_1} \allowbreak - y_3\, \bs{\mathrm{d}}{y_4})^2 \allowbreak + \phi_2 \bigl(\bs{\mathrm{d}}{y_2}\vee(\bs{\mathrm{d}}{y_1} \allowbreak - y_3\, \bs{\mathrm{d}}{y_4})\bigr) \allowbreak + \phi_3\, \bs{\mathrm{d}}{y_2}^2 \allowbreak + \phi_4 (\bs{\mathrm{d}}{y_3}\vee\bs{\mathrm{d}}{y_4})$ \\ \hline
{[5,4,7]} & $\phi_1 (-\cosh^2y_2\, \bs{\mathrm{d}}{y_1}^2 \allowbreak + \, \bs{\mathrm{d}}{y_2}^2) \allowbreak + \phi_2 (\sinh y_2\, \bs{\mathrm{d}}{y_1} \allowbreak +\, \bs{\mathrm{d}}{y_3})^2 \allowbreak + \phi_3 \bigl(\bs{\mathrm{d}}{y_4}\vee(\sinh y_2\, \bs{\mathrm{d}}{y_1} \allowbreak +\, \bs{\mathrm{d}}{y_3})\bigr) \allowbreak + \phi_4\, \bs{\mathrm{d}}{y_4}^2$ \\ \hline\hline
{[6,3,1]} & $- \phi_1(y_1)\, \bs{\mathrm{d}}{y_1}^2 \allowbreak + \phi_2(y_1) \bigl(\bs{\mathrm{d}}{y_2}^2 \allowbreak + \sin^2y_2 (\bs{\mathrm{d}}{y_3}^2 \allowbreak + \sin^2y_3\, \bs{\mathrm{d}}{y_4}^2)\bigr)$ \\ \hline
{[6,3,2]} & $- \phi_1(y_1)\, \bs{\mathrm{d}}{y_1}^2 \allowbreak + \phi_2(y_1) (\bs{\mathrm{d}}{y_2}^2 \allowbreak +\, \bs{\mathrm{d}}{y_3}^2 \allowbreak +\, \bs{\mathrm{d}}{y_4}^2)$ \\ \hline
{[6,3,3]} & $- \phi_1(y_1)\, \bs{\mathrm{d}}{y_1}^2 \allowbreak + \phi_2(y_1) \bigl(\bs{\mathrm{d}}{y_2}^2 \allowbreak + \sinh^2y_2 (\bs{\mathrm{d}}{y_3}^2 \allowbreak + \sin^2y_3\, \bs{\mathrm{d}}{y_4}^2)\bigr)$ \\ \hline
{[6,3,4]} & $\phi_1(y_4) (- \cosh^2y_2\, \bs{\mathrm{d}}{y_1}^2 \allowbreak +\, \bs{\mathrm{d}}{y_2}^2 \allowbreak + \sinh^2y_2\, \bs{\mathrm{d}}{y_3}^2) \allowbreak + \phi_2(y_4)\, \bs{\mathrm{d}}{y_4}^2$ \\ \hline
{[6,3,5]} & $\phi_1(y_3) (-\, \bs{\mathrm{d}}{y_1}^2 \allowbreak +\, \bs{\mathrm{d}}{y_2}^2 \allowbreak +\, \bs{\mathrm{d}}{y_4}^2) \allowbreak + \phi_2(y_3)\, \bs{\mathrm{d}}{y_3}^2$ \\ \hline
{[6,3,6]} & $\phi_1(y_4) \bigl(- \bs{\mathrm{d}}{y_1}^2 \allowbreak + \cosh^2y_1\, ( \bs{\mathrm{d}}{y_2}^2 \allowbreak + \sin^2y_2\, \bs{\mathrm{d}}{y_3}^2)\bigr) \allowbreak + \phi_2(y_4)\, \bs{\mathrm{d}}{y_4}^2$ \\ \hline\hline
{[6,4,1]} & $\phi_1 (-\cosh^2y_2\, \bs{\mathrm{d}}{y_1}^2 \allowbreak + \, \bs{\mathrm{d}}{y_2}^2) \allowbreak + \phi_2 (\bs{\mathrm{d}}{y_3}^2 \allowbreak + \sin^2y_3\, \bs{\mathrm{d}}{y_4}^2)$ \\ \hline
{[6,4,2]} & $\phi_1 (-\cosh^2y_2\, \bs{\mathrm{d}}{y_1}^2 \allowbreak + \, \bs{\mathrm{d}}{y_2}^2) \allowbreak + \phi_2 (\bs{\mathrm{d}}{y_3}^2 \allowbreak + \sinh^2y_3\, \bs{\mathrm{d}}{y_4}^2)$ \\ \hline
{[6,4,3]} & $\phi_1 (-\cosh^2y_2\, \bs{\mathrm{d}}{y_1}^2 \allowbreak + \, \bs{\mathrm{d}}{y_2}^2) \allowbreak + \phi_2 (\bs{\mathrm{d}}{y_3}^2 \allowbreak +\, \bs{\mathrm{d}}{y_4}^2)$ \\ \hline
{[6,4,4]} & $\phi_1 (-\, \bs{\mathrm{d}}{y_1}^2 \allowbreak +\, \bs{\mathrm{d}}{y_2}^2) \allowbreak + \phi_2 (\bs{\mathrm{d}}{y_3}^2 \allowbreak + \sin^2y_3\, \bs{\mathrm{d}}{y_4}^2)$ \\ \hline
{[6,4,5]} & $\phi_1 (-\, \bs{\mathrm{d}}{y_1}^2 \allowbreak +\, \bs{\mathrm{d}}{y_2}^2) \allowbreak + \phi_2 (\bs{\mathrm{d}}{y_3}^2 \allowbreak + \sinh^2y_3\, \bs{\mathrm{d}}{y_4}^2)$ \\ \hline\hline
{[7,4,1]} & $- \phi_1\, \bs{\mathrm{d}}{y_1}^2 \allowbreak + \phi_2 \bigl(\bs{\mathrm{d}}{y_2}^2 \allowbreak + \sin^2y_2 (\bs{\mathrm{d}}{y_3}^2 \allowbreak + \sin^2y_3\, \bs{\mathrm{d}}{y_4}^2)\bigr)$ \\ \hline
{[7,4,2]} & $- \phi_1\, \bs{\mathrm{d}}{y_1}^2 \allowbreak + \phi_2 \bigl(\bs{\mathrm{d}}{y_2}^2 \allowbreak + \sinh^2y_2 (\bs{\mathrm{d}}{y_3}^2 \allowbreak + \sin^2y_3\, \bs{\mathrm{d}}{y_4}^2)\bigr)$ \\ \hline
{[7,4,3]} & $\phi_1 \bigl(-\bs{\mathrm{d}}{y_1}^2 \allowbreak + \cosh^2y_1 (\bs{\mathrm{d}}{y_2}^2 \allowbreak + \sin^2y_2\, \bs{\mathrm{d}}{y_3}^2)\bigr) \allowbreak + \phi_2\, \bs{\mathrm{d}}{y_4}^2$ \\ \hline
{[7,4,4]} & $\phi_1 (- \cosh^2y_2\, \bs{\mathrm{d}}{y_1}^2 \allowbreak +\, \bs{\mathrm{d}}{y_2}^2 \allowbreak + \sinh^2y_2\, \bs{\mathrm{d}}{y_3}^2) \allowbreak + \phi_2\, \bs{\mathrm{d}}{y_4}^2$ \\ \hline
\end{longtable}

\begin{longtable}{|l||>{\raggedright\arraybackslash}p{10cm}|}
\caption{$\Gamma$-invariant vector fields}\label{tab:invvectfieldsPSCcomp}
\\ \hline
{\scriptsize Hicks \#} & {\scriptsize $\Gamma$-invariant vector fields $\bs{V}$} \\\hline\hline
{[3,2,\{1,2,3\}]}& $f_1(y_1, y_2)\, \bs{\partial}_{y_1} \allowbreak + f_2(y_1, y_2)\, \bs{\partial}_{y_2}$ \\ \hline
{[3,2,\{4,5\}]} & $f_1(y_3, y_4)\, \bs{\partial}_{y_3} \allowbreak + f_2(y_3, y_4)\, \bs{\partial}_{y_4}$ \\ \hline\hline
{[3,3,2]} & $f_1(y_3)\, \bs{\partial}_{y_1} \allowbreak + f_2(y_3)\, \bs{\partial}_{y_2} \allowbreak + f_3(y_3)\, \bs{\partial}_{y_3} \allowbreak + f_4(y_3)\, \bs{\partial}_{y_4}$ \\ \hline
{[3,3,3]} & $f_1(y_2)\, \bs{\partial}_{y_1} \allowbreak + f_2(y_2)\, \bs{\partial}_{y_2} \allowbreak + f_3(y_2)\, \bs{\partial}_{y_3} \allowbreak + f_4(y_2) (y_3\, \bs{\partial}_{y_1} \allowbreak +\, \bs{\partial}_{y_4})$ \\ \hline
{[3,3,8]} & $f_1(y_2)\, \bs{\partial}_{y_1} \allowbreak + f_2(y_2)\, \bs{\partial}_{y_2} \allowbreak + f_3(y_2) \bigl(\sin y_1\, \bs{\partial}_{y_3} \allowbreak + \cos y_1 \csch y_3 (\cosh y_3\, \bs{\partial}_{y_1} \allowbreak +\, \bs{\partial}_{y_4})\bigr) \allowbreak + f_4(y_2) \bigl(\cos y_1\, \bs{\partial}_{y_3} \allowbreak -  \csch y_3 \sin y_1 (\cosh y_3\, \bs{\partial}_{y_1} \allowbreak +\, \bs{\partial}_{y_4})\bigr)$ \\ \hline
{[3,3,9]} & $f_1(y_2)\, \bs{\partial}_{y_1} \allowbreak + f_2(y_2)\, \bs{\partial}_{y_2} \allowbreak + f_3(y_2) \bigl(\sin y_1\, \bs{\partial}_{y_3} \allowbreak + \cos y_1 \csc y_3 (\cos y_3\, \bs{\partial}_{y_1} \allowbreak +\, \bs{\partial}_{y_4})\bigr) \allowbreak + f_4(y_2) \bigl(\cos y_1\, \bs{\partial}_{y_3} \allowbreak -  \csc y_3 \sin y_1 (\cos y_3\, \bs{\partial}_{y_1} \allowbreak +\, \bs{\partial}_{y_4})\bigr)$ \\ \hline\hline
{[4,3,\{1,2,3,4,5,6,10\}]} & $f_1(y_2)\, \bs{\partial}_{y_1} \allowbreak + f_2(y_2)\, \bs{\partial}_{y_2}$ \\ \hline
{[4,3,\{8,11\}]} & $f_1(y_3)\, \bs{\partial}_{y_3} \allowbreak + f_2(y_3)\, \bs{\partial}_{y_4}$ \\ \hline
{[4,3,9]} & $f_1(y_4)\, \bs{\partial}_{y_3} \allowbreak + f_2(y_4)\, \bs{\partial}_{y_4}$ \\ \hline\hline
{[4,4,1]} & $f_1\, \bs{\partial}_{y_1} \allowbreak + f_2\, \bs{\partial}_{y_2} \allowbreak + f_3 \bigl(\sin y_1\, \bs{\partial}_{y_3} \allowbreak + \cos y_1 \csc y_3 (\cos y_3\, \bs{\partial}_{y_1} \allowbreak +\, \bs{\partial}_{y_4})\bigr) \allowbreak + f_4 \bigl(\cos y_1\, \bs{\partial}_{y_3} \allowbreak -  \csc y_3 \sin y_1 (\cos y_3\, \bs{\partial}_{y_1} \allowbreak +\, \bs{\partial}_{y_4})\bigr)$ \\ \hline
{[4,4,2]} & $f_1\, \bs{\partial}_{y_1} \allowbreak + f_2\, \bs{\partial}_{y_2} \allowbreak + f_3 \bigl(\sin y_1\, \bs{\partial}_{y_3} \allowbreak + \cos y_1 \csch y_3 (\cosh y_3\, \bs{\partial}_{y_1} \allowbreak +\, \bs{\partial}_{y_4})\bigr) \allowbreak + f_4 \bigl(\cos y_1\, \bs{\partial}_{y_3} \allowbreak -  \csch y_3 \sin y_1 (\cosh y_3\, \bs{\partial}_{y_1} \allowbreak +\, \bs{\partial}_{y_4})\bigr)$ \\ \hline
{[4,4,9]} & $f_1\, \bs{\partial}_{y_1} \allowbreak + f_2\, \bs{\partial}_{y_3} \allowbreak + f_3 (y_3\, \bs{\partial}_{y_1} \allowbreak - \, \bs{\partial}_{y_4}) \allowbreak + f_4 \bigl(-2\, \bs{\partial}_{y_2} \allowbreak + y_3 (y_3\, \bs{\partial}_{y_1} \allowbreak - 2\, \bs{\partial}_{y_4})\bigr)$ \\ \hline
{[4,4,18]} & $f_1\, \bs{\partial}_{y_1} \allowbreak + f_2\, \bs{\partial}_{y_2} \allowbreak + e^{- y_2} f_3\, \bs{\partial}_{y_3} \allowbreak + e^{y_2} f_4 (y_3\, \bs{\partial}_{y_1} \allowbreak +\, \bs{\partial}_{y_4})$ \\ \hline
{[4,4,22]} & $f_1\, \bs{\partial}_{y_1} \allowbreak + f_2\, \bs{\partial}_{y_2} \allowbreak + f_3 \bigl(\sin y_2\, \bs{\partial}_{y_3} \allowbreak + \cos y_2 (y_3\, \bs{\partial}_{y_1} \allowbreak +\, \bs{\partial}_{y_4})\bigr) \allowbreak + f_4 \bigl(\cos y_2\, \bs{\partial}_{y_3} \allowbreak -  \sin y_2 (y_3\, \bs{\partial}_{y_1} \allowbreak +\, \bs{\partial}_{y_4})\bigr)$ \\ \hline\hline
{[5,4,\{1,2,3,6\}]} & $f_1\, \bs{\partial}_{y_1} \allowbreak + f_2\, \bs{\partial}_{y_2}$ \\ \hline
{[5,4,7]} & $f_1\, \bs{\partial}_{y_3} \allowbreak + f_2\, \bs{\partial}_{y_4}$ \\ \hline\hline
{[6,3,\{1,2,3\}]} & $f_1(y_1)\, \bs{\partial}_{y_1}$ \\ \hline
{[6,3,\{4,6\}]} & $f_1(y_4)\, \bs{\partial}_{y_4}$ \\ \hline
{[6,3,5]} & $f_1(y_3)\, \bs{\partial}_{y_3}$ \\ \hline\hline
{[6,4,\{1,2,3,4,5\}]} & $\emptyset$ \\ \hline\hline
{[7,4,\{1,2\}]} & $f_1\, \bs{\partial}_{y_1}$ \\ \hline
{[7,4,\{3,4\}]} & $f_1\, \bs{\partial}_{y_4}$ \\ \hline
\end{longtable}


\begin{table}[!ht]
\caption{$l$-chains}\label{tab:lchainsPSCcomp}
    \centering
    \begin{tabular}[t]{|l||l|}
    \hline
{\scriptsize Hicks \#} & {\scriptsize $l$-chain $\bs{\chi}$} \\\hline\hline
{[3,2,1]} & $\bs{\partial}_{y_3} \wedge \bs{\partial}_{y_4}$ \\ \hline
{[3,2,2]} & $\csch y_3\, \bs{\partial}_{y_3} \wedge \bs{\partial}_{y_4}$ \\ \hline
{[3,2,3]} & $\csc y_3\, \bs{\partial}_{y_3} \wedge \bs{\partial}_{y_4}$ \\ \hline
{[3,2,4]} & $\bs{\partial}_{y_1} \wedge \bs{\partial}_{y_2}$ \\ \hline
{[3,2,5]} & $\sech y_2\, \bs{\partial}_{y_1} \wedge \bs{\partial}_{y_2}$ \\ \hline\hline
{[3,3,2]} & $\bs{\partial}_{y_1} \wedge\, \bs{\partial}_{y_2} \wedge \bs{\partial}_{y_4}$ \\ \hline
{[3,3,3]} & $\bs{\partial}_{y_1} \wedge\, \bs{\partial}_{y_3} \wedge \bs{\partial}_{y_4}$ \\ \hline
{[3,3,8]} & $\csch y_3\, \bs{\partial}_{y_1} \wedge \bs{\partial}_{y_3} \wedge \bs{\partial}_{y_4}$ \\ \hline
{[3,3,9]} & $\csc y_3\, \bs{\partial}_{y_1} \wedge \bs{\partial}_{y_3} \wedge \bs{\partial}_{y_4}$ \\ \hline\hline
{[4,3,\{1,2\}]} & $\csch y_3\, \bs{\partial}_{y_1} \wedge \bs{\partial}_{y_3} \wedge \bs{\partial}_{y_4}$ \\ \hline
{[4,3,\{3,4\}]} & $\csc y_3\, \bs{\partial}_{y_1} \wedge \bs{\partial}_{y_3} \wedge \bs{\partial}_{y_4}$ \\ \hline
{[4,3,\{5,6,10\}]} & $\bs{\partial}_{y_1} \wedge\, \bs{\partial}_{y_3} \wedge \bs{\partial}_{y_4}$ \\ \hline
{[4,3,8]} & $\sech y_2\, \bs{\partial}_{y_1} \wedge \bs{\partial}_{y_2} \wedge \bs{\partial}_{y_4}$ \\ \hline
{[4,3,9]} & $\sech y_2\, \bs{\partial}_{y_1} \wedge \bs{\partial}_{y_2} \wedge \bs{\partial}_{y_3}$ \\ \hline
{[4,3,11]} & $\bs{\partial}_{y_1} \wedge\, \bs{\partial}_{y_2} \wedge \bs{\partial}_{y_4}$ \\ \hline\hline
{[4,4,1]} & $\csc y_3\, \bs{\partial}_{y_1} \wedge \bs{\partial}_{y_2} \wedge \bs{\partial}_{y_3} \wedge \bs{\partial}_{y_4}$ \\ \hline
{[4,4,2]} & $\csch y_3\, \bs{\partial}_{y_1} \wedge \bs{\partial}_{y_2} \wedge \bs{\partial}_{y_3} \wedge \bs{\partial}_{y_4}$ \\ \hline
{[4,4,\{9,18,22\}]} & $\bs{\partial}_{y_1} \wedge\, \bs{\partial}_{y_2} \wedge \bs{\partial}_{y_3} \wedge \bs{\partial}_{y_4}$ \\ \hline
\end{tabular}
    \quad
    \begin{tabular}[t]{|l||l|}
    \hline
{\scriptsize Hicks \#} & {\scriptsize $l$-chain $\bs{\chi}$} \\\hline\hline
{[5,4,\{1,7\}]} & $\csch y_3\, \bs{\partial}_{y_1} \wedge \bs{\partial}_{y_2} \wedge \bs{\partial}_{y_3} \wedge \bs{\partial}_{y_4}$ \\ \hline
{[5,4,2]} & $\csc y_3\, \bs{\partial}_{y_1} \wedge \bs{\partial}_{y_2} \wedge \bs{\partial}_{y_3} \wedge \bs{\partial}_{y_4}$ \\ \hline
{[5,4,\{3,6\}]} & $\bs{\partial}_{y_1} \wedge\, \bs{\partial}_{y_2} \wedge \bs{\partial}_{y_3} \wedge \bs{\partial}_{y_4}$ \\ \hline\hline
{[6,3,1]} & $\csc^2y_2 \csc y_3\, \bs{\partial}_{y_2} \wedge \bs{\partial}_{y_3} \wedge \bs{\partial}_{y_4}$ \\ \hline
{[6,3,2]} & $\bs{\partial}_{y_2} \wedge\, \bs{\partial}_{y_3} \wedge \bs{\partial}_{y_4}$ \\ \hline
{[6,3,3]} & $\csc y_3 \csch^2y_2\, \bs{\partial}_{y_2} \wedge \bs{\partial}_{y_3} \wedge \bs{\partial}_{y_4}$ \\ \hline
{[6,3,4]} & $\csch y_2 \sech y_2\, \bs{\partial}_{y_1} \wedge \bs{\partial}_{y_2} \wedge \bs{\partial}_{y_3}$ \\ \hline
{[6,3,5]} & $\bs{\partial}_{y_1} \wedge\, \bs{\partial}_{y_2} \wedge \bs{\partial}_{y_4}$ \\ \hline
{[6,3,6]} & $\csc y_2 \sech^2y_1\, \bs{\partial}_{y_1} \wedge \bs{\partial}_{y_2} \wedge \bs{\partial}_{y_3}$ \\ \hline\hline
{[6,4,1]} & $\csc y_3 \sech y_2\, \bs{\partial}_{y_1} \wedge \bs{\partial}_{y_2} \wedge \bs{\partial}_{y_3} \wedge \bs{\partial}_{y_4}$ \\ \hline
{[6,4,2]} & $\csch y_3 \sech y_2\, \bs{\partial}_{y_1} \wedge \bs{\partial}_{y_2} \wedge \bs{\partial}_{y_3} \wedge \bs{\partial}_{y_4}$ \\ \hline
{[6,4,3]} & $\sech y_2\, \bs{\partial}_{y_1} \wedge \bs{\partial}_{y_2} \wedge \bs{\partial}_{y_3} \wedge \bs{\partial}_{y_4}$ \\ \hline
{[6,4,4]} & $\csc y_3\, \bs{\partial}_{y_1} \wedge \bs{\partial}_{y_2} \wedge \bs{\partial}_{y_3} \wedge \bs{\partial}_{y_4}$ \\ \hline
{[6,4,5]} & $\csch y_3\, \bs{\partial}_{y_1} \wedge \bs{\partial}_{y_2} \wedge \bs{\partial}_{y_3} \wedge \bs{\partial}_{y_4}$ \\ \hline\hline
{[7,4,1]} & $\csc^2y_2 \csc y_3\, \bs{\partial}_{y_1} \wedge \bs{\partial}_{y_2} \wedge \bs{\partial}_{y_3} \wedge \bs{\partial}_{y_4}$ \\ \hline
{[7,4,2]} & $\csc y_3 \csch^2y_2\, \bs{\partial}_{y_1} \wedge \bs{\partial}_{y_2} \wedge \bs{\partial}_{y_3} \wedge \bs{\partial}_{y_4}$ \\ \hline
{[7,4,3]} & $\csc y_2 \sech^2y_1\, \bs{\partial}_{y_1} \wedge \bs{\partial}_{y_2} \wedge \bs{\partial}_{y_3} \wedge \bs{\partial}_{y_4}$ \\ \hline
{[7,4,4]} & $\csch y_2 \sech y_2\, \bs{\partial}_{y_1} \wedge \bs{\partial}_{y_2} \wedge \bs{\partial}_{y_3} \wedge \bs{\partial}_{y_4}$ \\ \hline
    \end{tabular}
\end{table}


\section{Symmetry classification insights}\label{sec:insights}

As the tables above may look somewhat intimidating, we provide a geometrical and physical context for the above (PSC-compatible) symmetries. Specifically, we will discuss relations among infinitesimal group actions and their manifestation in famous geometries. This should shed more light at the overall structure of symmetries and stimulate various applications within and mainly beyond the general relativity.

Since higher-dimensional Lie algebras often admit many lower-dimensional subalgebras, there exist various relations among the infinitesimal group actions too. As mentioned above, there is an important difference between the abstract Lie algebras $\mathcal{A}$, which are given purely by the multiplication tables with no reference to the manifold $M$, and the infinitesimal group actions $\Gamma$, i.e., the Lie algebras of vector fields on $M$. Specifically, two inequivalent $\Gamma$ can be very different realizations of the same $\mathcal{A}$, as the same group $G$ may act differently on $M$; it may even possess the orbits of distinct dimensions.

\begin{figure}
    \centering
    \includegraphics[width=\textwidth]{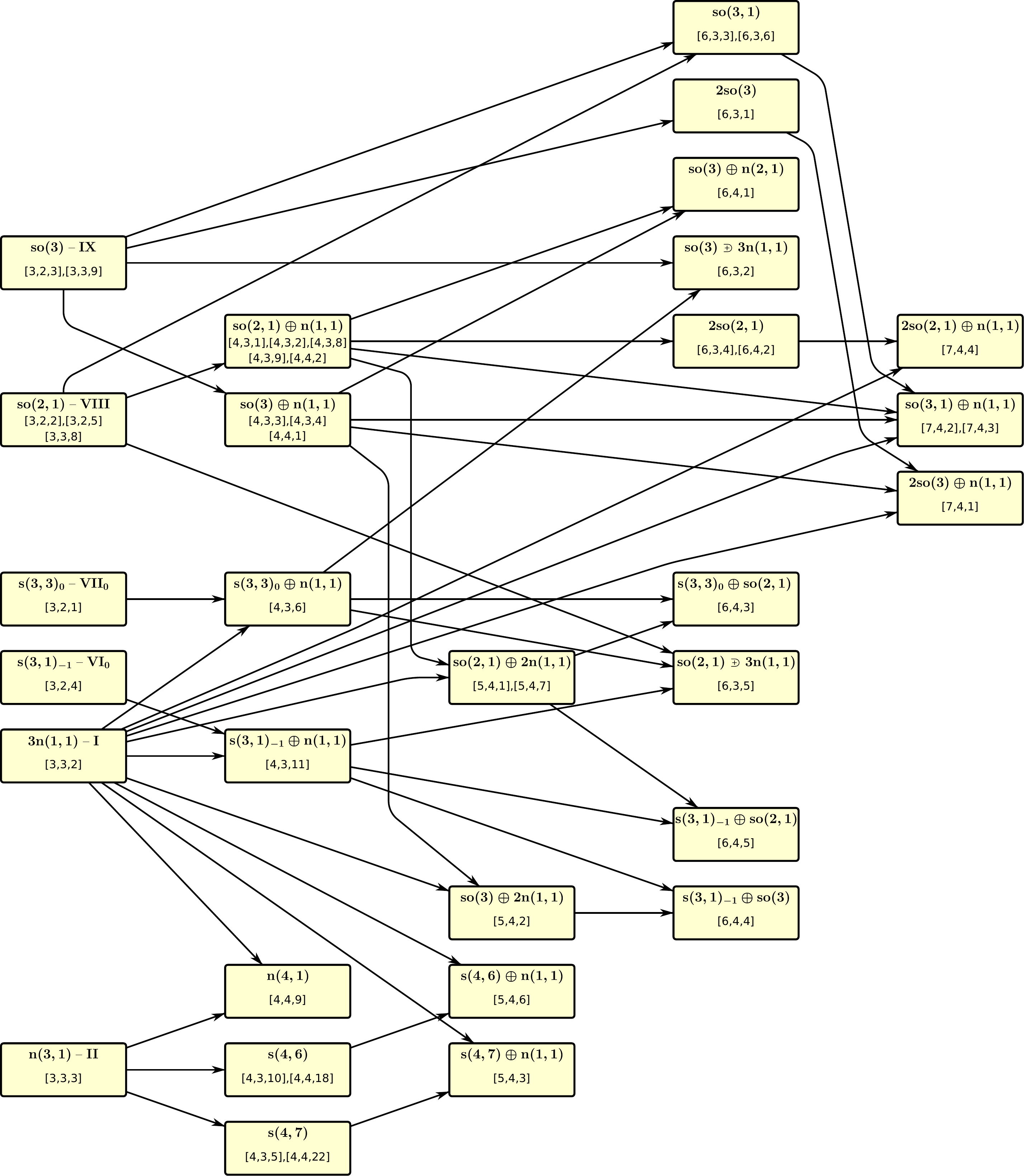}
    \caption{Relations among abstract Lie algebras $\mathcal{A}$. The relations are represented by a graph with nodes for $\mathcal{A}$s and directed edges (arrows) pointing from their subalgebras. The algebras are arranged horizontally with their increasing dimension $d$ when going from left to right. For simplicity, multiple occurrences are denoted just by one arrow and the arrows corresponding to subalgebras that are already contained within larger subalgebras are omitted. Each node contains: i) the name of $\mathcal{A}$ according to the \textit{\v{S}nobl-Winternitz classification} \cite{Snobl2014-te} and the \textit{Bianchi classification} for ${d=3}$ ii) the list of (PSC-compatible) $\Gamma$s with such $\mathcal{A}$. Note that `n' stands for \textit{nilpotent}, specifically, `n(1,1)' is the \textit{abelian} algebra, `s' stands for \textit{solvable}; the value of the parameter $a$ for the cases `s(3,1)' and `s(3,3)' is marked by the subscript; $\oplus$ and $\niplus$ denote the \textit{direct} and \textit{semi-direct} sums, respectively.}
    \label{fig:abstractspider}
\end{figure}

Taking into account just the multiplication tables, we constructed a diagram relating the abstract algebras $\mathcal{A}$ (appearing within PSC-compatible $\Gamma$) by their subalgebras, see Fig.~\ref{fig:abstractspider}. It was obtained by identifying all subalgebras constructed from all subsets of the basis vectors. Naturally, although this simple method captures many subalgebras, it is possible that some relations among $\mathcal{A}$s may not have been discovered.

\begin{figure}
    \centering
    \includegraphics[width=0.92\textwidth]{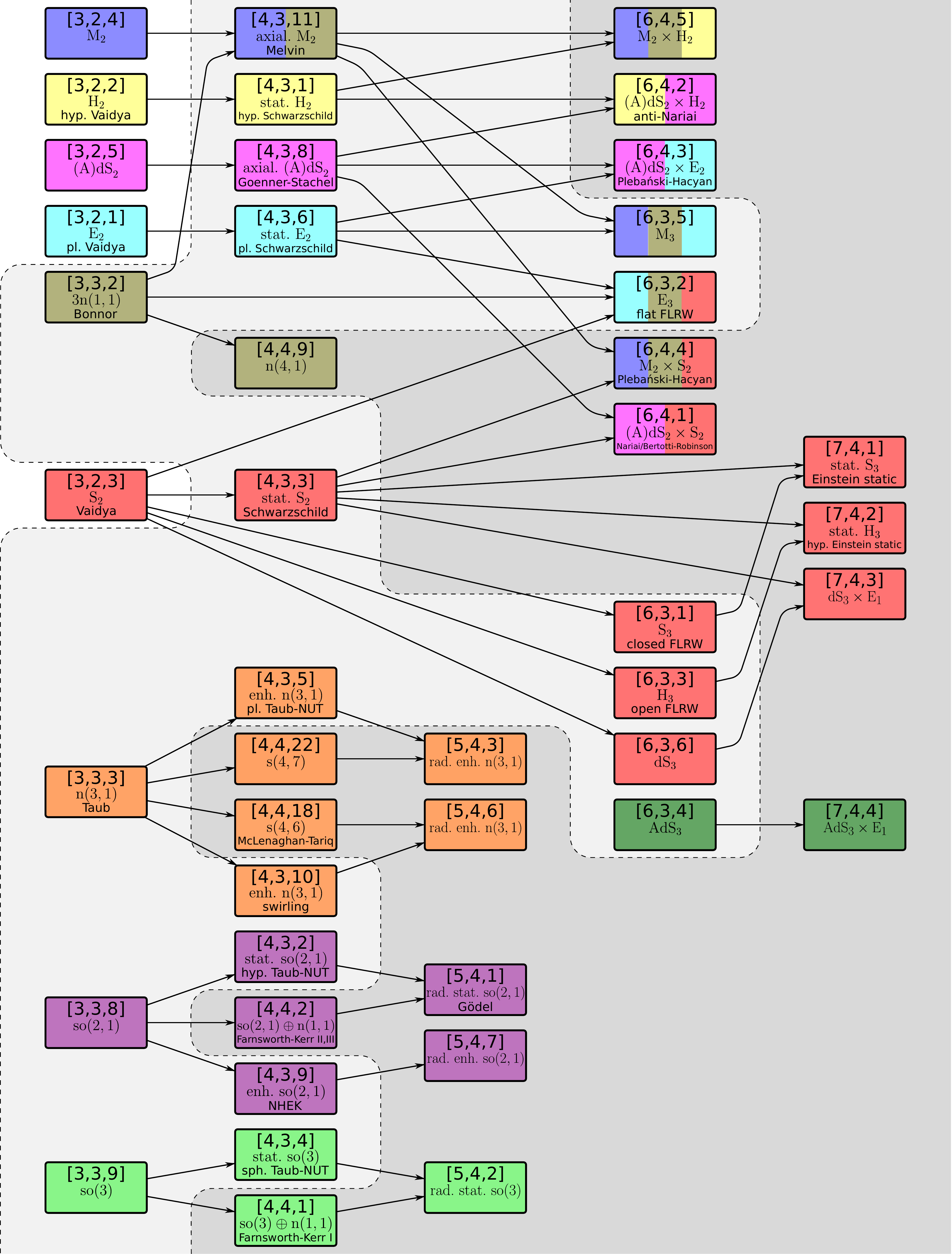}
    \caption{Relations among infinitesimal group actions $\Gamma$ (for details, see Tab~\ref{tab:arrows}). The relations are represented by a graph with color-coded nodes for $\Gamma$s and directed edges (arrows) pointing from their `subactions'. The actions are arranged horizontally with increasing dimension of the algebras $d$ when going from left to right. The background colors (and the dashed lines) highlight different dimensions of the orbits $l$. Each node contains: i) the Hicks label $[d,l,c]$ of $\Gamma$ ii) a brief description (either concerning the orbit, or the abstract algebra $\mathcal{A}$ if ${p=0}$, or relation to another $\Gamma$) iii) a known example of a spacetime with such symmetry. The following acronyms are used: \abbrev{sph}erical, \abbrev{pl}anar, \abbrev{hyp}erbolic, \abbrev{axial}, \abbrev{rad}ial, \abbrev{stat}ionary, and \abbrev{enh}anced as well as \abbrev{F}riedmann-\abbrev{L}ema\^{i}tre-\abbrev{R}obertson-\abbrev{W}alker, \abbrev{N}ewman-\abbrev{U}nti-\abbrev{T}amburino, and \abbrev{N}ear \abbrev{H}orizon \abbrev{E}xtreme \abbrev{K}err.}
    \label{fig:actionspider}
\end{figure}

A relation between abstract algebras $\mathcal{A}$ does not necessarily imply a relation between the infinitesimal group actions $\Gamma$. To address this issue, we constructed another diagram in Fig.~\ref{fig:actionspider}, which takes into account the equivalence of $\Gamma$ (rather than just $\mathcal{A}$). Here, an arrow is only present if there exists a diffeomorphism between the sets of vector fields, which we found explicitly in all the cases. Actually, the adapted coordinates $y_i$ were adjusted so that the cases reduce to each other in the same coordinates (i.e., they were unified across related infinitesimal group actions) whenever possible and unless it led to more complicated or unnatural expressions. As before, although the structure we revealed is quite rich, this diagram may not be complete. Our search for relations between $\Gamma$s was based on various observations (transformations of subsets of vector fields and their coordinates) rather than a systematic analysis, which is very hard because finding diffeomorphisms (or proving their non-existence) is not easily algorithmizable. The extra Killing vectors and the choices of $\phi_i$ specializing the metrics (to those with higher degree of symmetry) for each relation (arrow) are listed in Tab.~\ref{tab:arrows} in Apx.~\ref{apx:relinfgra}.

In Fig.~\ref{fig:actionspider}, we also provided brief descriptions that capture the main essence of the symmetries in the most common situations (often motivated by relations to other infinitesimal group actions). For example, the well-known case of stationary spherically symmetric metric [4,3,3] is abbreviated as `stationary $\mathrm{S_2}$' while the flat FLRW metric [6,3,3] as `$\mathrm{E_3}$'. (Naturally, these `names' are somewhat subjective, but they may help the reader navigating the diagram.) It may be instructive to compare [4,3,3] with [4,3,4], which we call `stat. $\mathrm{so(3)}$', and [4,4,2], which we refer to simply as `$\mathrm{so(3)\oplus n(1,1)}$'; all three possess the same abstract algebra $\mathrm{so(3)\oplus n(1,1)}$ but describe its different action differently on $M$. Specifically, [4,4,2] is the action on the entire $M$ while [4,3,3] and [4,3,4] have 3-dimensional orbits only. The infinitesimal group actions [4,3,3] and [4,3,4] can be thought of as stationary [3,2,3] and [3,3,9]. (Again, although [3,2,3] and [3,3,9] are both actions of so(3). The latter has 3-dimensional orbits, so we refer to it simply as `so(3)', while the former is referred to directly by its orbits `$\mathrm{S_2}$'.)

Another type of information that may help with the orientation in the graph of infinitesimal group actions above (apart from the mutual relations), is to associate as many $\Gamma$-invariant metrics as one can with the well-known geometries, e.g., the exact solutions of general relativity. This is especially useful when looking for their counterparts in modified theories of gravity. Such a task often starts with a general metric ansatz that shares the same isometries $\Gamma$ with the well-known metric, but contains unknown functions that are determined by solving the modified field equations.  Whenever possible, we also adjusted the adapted coordinates $y_i$ so that the solutions were easily identifiable in the most common and/or most natural coordinates.

Roughly half of the solutions appearing in \cite{Stephani2009-dt} have been classified according to the Hicks classification in \cite{Hicks:thesis,Hwang:thesis} by identifying Lorentzian Lie algebra-subalgebra pairs of their Killing vectors. (Note that these studies included also some metrics that are technically not `solutions' in a sense that their energy-momentum tensor is not analyzed, i.e., Einstein's equations are not imposed.) Nevertheless, they were not explicitly compared to the $\Gamma$-invariant metrics $\hat{\bs{g}}$, as the Hicks coordinates $x_i$ are often quite different from more traditional ones. In Fig.~\ref{fig:actionspider}, we mentioned the most iconic geometries within the infinitesimal group actions for which the solutions (with recognized energy-momentum tensor) exist and are known. Let us now comment on how some of them can be identified explicitly within the $\Gamma$-invariant metrics from Tab.~\ref{tab:invmetrforPSCcomp}.

\begin{description}[style=nextline,leftmargin=0.5cm]

\item[\descstyle{[3,2,-]}]
Well-known pure-radiation solutions of the Einstein equations that exhibit only the 2-spherical, 2-planar, or 2-hyperbolic symmetries [3,2,\{3,1,2\}] (i.e., with 2-dimensional spacelike orbits of constant curvature $\mathrm{S_2}$, $\mathrm{E_2}$, or $\mathrm{H_2}$) are the \textit{Vaidya spacetime} [(15.20) \cite{Stephani2009-dt}, (9.32) \cite{Griffiths:2009dfa}] and its analogues with vanishing or negative spacelike curvatures [(15.17) with Tab.~15.1 \cite{Stephani2009-dt}, (9.35) \cite{Griffiths:2009dfa}],
\begin{equation}\label{eq:Vaidya}
    \bs{g}=-\big(\epsilon-\tfrac{2 m(u)}{r}\big) \bs{\mathrm{d}} u^2-\bs{\mathrm{d}} u \vee \bs{\mathrm{d}} r+r^2\bs{g}_2\;.
\end{equation}
Here, the function $m$ is determined by the null matter, $\epsilon=+1,0,-1$, and $\bs{g}_2$ are metrics of 2-spaces of constant unit curvature,
\begin{equation}
    \bs{g}_2=\begin{cases}
        \bs{\mathrm{d}}\vartheta^2+\sin^2\vartheta \bs{\mathrm{d}}\varphi^2\;, &\epsilon=+1\;,\\
        \bs{\mathrm{d}}\vartheta^2+\bs{\mathrm{d}}\varphi^2\;, &\epsilon=0\;,\\
        \bs{\mathrm{d}}\vartheta^2+\sinh^2\vartheta \bs{\mathrm{d}}\varphi^2 &\epsilon=-1\;.\\
    \end{cases}
\end{equation}
The spacetimes \eqref{eq:Vaidya} can be obtained from the $\Gamma$-invariant metrics $\hat{\bs{g}}$ by denoting the adapted coordinates $y_i$ by $(u,r,\vartheta,\varphi)$ and setting ${\phi_1=\epsilon-{2 m(u)}/{r}}$, ${\phi_2=-1}$, ${\phi_4=r^2}$, ${\phi_3=0}$. We are not aware of any well-known metrics that would only admit (1+1)-Poincar\'e or (2+1)-Lorentzian symmetries [3,2,\{4,5\}] (i.e., with 2-dimensional timelike orbits $\mathrm{M_2}$ or $\mathrm{(A)dS_2}$). As we will see below, there exist various solutions possessing these symmetries when enhanced by extra symmetries.

\item[\descstyle{[3,3,-]}]
An example of a solution of the Einstein-Maxwell equations with the stationary cylindrical symmetry [3,3,2] (i.e., 3n(1,1) acting on 3-dimensional orbits), is the \textit{Bonnor spacetime} [(22.11) \cite{Stephani2009-dt}],
\begin{equation}
    \bs{g}=\rho^{2m^2}(a\rho^m+b\rho^{-m})^2\, \big(-\bs{\mathrm{d}}{t}^2 + \bs{\mathrm{d}}{\rho}^2\big)  + (a\rho^m+b\rho^{-m})^{-2}\bs{\mathrm{d}}{z}^2 +\rho^2 (a\rho^m+b\rho^{-m})^2\, \bs{\mathrm{d}}{\varphi}^2 
\end{equation}
where $a$, $b$, and $m$ are constants. This spacetime is obtainable from $\hat{\bs{g}}$ by denoting the adapted coordinates by $(t,z,\rho,\varphi)$ and setting ${\phi_1/\rho^{2m^2}=\phi_2/\rho^{2m^2}=\phi_4/\rho^2=1/\phi_5=(a\rho^m+b\rho^{-m})^2}$, ${\phi_3=\phi_6=\phi_7=\phi_8=\phi_9=\phi_{10}=0}$. One vacuum solution with only n(3,1) acting on 3-dimensional orbits [3,3,3] is the \textit{Taub spacetime} [(13.55) \cite{Stephani2009-dt}]
\begin{equation}\label{eq:Taub}
    \bs{g}=\pm\tfrac{2\sqrt{ab}}{\cosh (2\sqrt{ab}\tau)}(\bs{\mathrm{d}}{x} \allowbreak - z\, \bs{\mathrm{d}}{y})^2   +\tfrac{\cosh (2\sqrt{ab}\tau)}{2\sqrt{ab}}\left[\mp e^{2(a+b)\tau}\bs{\mathrm{d}}{\tau}^2 \allowbreak + e^{2a\tau}\, \bs{\mathrm{d}}{y}^2+ e^{2b\tau}\, \bs{\mathrm{d}}{z}^2 \right]\;,
\end{equation}
where $a$ and $b$ are constants. (If ${a=b}$, an additional Killing vector appears.) Denoting the adapted coordinates by $(x,\tau,z,y)$, this spacetime corresponds to ${\mp\phi_1(y_2)=\phi_{10}/e^{2a\tau}=\phi_{4}/e^{2b\tau}=\mp\phi_3/e^{2(a+b)\tau}={\cosh (2\sqrt{ab}\tau)}/{(2\sqrt{ab})}}$, ${\phi_2=\phi_5=\phi_6=\phi_7=\phi_8=\phi_9=0}$. We are not aware of a well-known metric with only so(2,1) acting on 3-dimensional orbits [3,3,8] or only so(3) acting on 3-dimensional orbits [3,3,9]. There are various examples with extra symmetries though; see below.

\item[\descstyle{[4,3,-]}]
The stationary 2-spherical, 2-planar, or 2-hyperbolic symmetries [4,3,\{3,6,1\}] contain very well-known vacuum solutions of the Einstein field equations, the \textit{Schwarzschild spacetime} [(15.19) \cite{Stephani2009-dt}, (8.1) \cite{Griffiths:2009dfa}] and its counterparts with vanishing or negative constant spatial curvature [(15.17) with Tab.~15.1  \cite{Stephani2009-dt}, (9.1) \cite{Griffiths:2009dfa}] (also known as the AI, AIII, or AII metrics, respectively),
\begin{equation}\label{eq:schw}
    \bs{g}=-\big(\epsilon-\tfrac{2m}{r}\big)\, \bs{\mathrm{d}}{t}^2 \allowbreak +(\epsilon-\tfrac{2m}{r})^{-1}\, \bs{\mathrm{d}}{r}^2+ r^2 \bs{g}_2\;,
\end{equation}
where $m$ is a constant and $\epsilon=+1,0,-1$.
Naming the adapted coordinates $(t,r,\vartheta,\varphi)$, it is given by the choice $\phi_4=r^2$, $\phi_1=1/\phi_3=\epsilon-{2m}/{r}$, $\phi_2=0$. 
Somewhat curious analogy of the above spacetimes but with (1+1)-Poincar\'e and (2+1)-Lorentzian symmetries enhanced by axial symmetry [4,3,\{11,8\}] are the \textit{Goenner-Stachel spacetimes} [(15.3) with (15.12) \cite{Stephani2009-dt}],
\begin{equation}\label{eq:GS}
    \bs{g}=\rho^2 \tilde{\bs{g}}_2 \allowbreak + \big(\varepsilon-\tfrac{2m}{\rho}\big)^{-1}\, \bs{\mathrm{d}}{\rho}^2 +  \big(\varepsilon-\tfrac{2m}{\rho}\big)\, \bs{\mathrm{d}}{\varphi}^2\;,
\end{equation}
where $m$ is a constant, $\varepsilon=0,\pm1$, and $\tilde{\bs{g}}_2$ are the metrics of 2-spacetimes of constant unit curvature,
\begin{equation}
    \tilde{\bs{g}}_2=\begin{cases}
        -\bs{\mathrm{d}}\tau^2+\bs{\mathrm{d}}\sigma^2\;,  &\varepsilon=0\;,\\
        \mp\bs{\mathrm{d}}\sigma^2\pm\cosh^2\sigma\,\bs{\mathrm{d}}\tau^2\;,  &\varepsilon=\pm 1\;.
    \end{cases}
\end{equation}
Since \eqref{eq:GS} differs from \eqref{eq:schw} only by signs, it is a vacuum solution. Denoting the adapted coordinates $(\tau,\sigma,\rho,\varphi)$, the spacetimes \eqref{eq:GS} correspond to $\hat{\bs{g}}$ with ${\phi_1=\sgn(1-2\varepsilon)\rho^2}$, ${\phi_3=0}$, ${\phi_4=1/\phi_2=\varepsilon-2m/\rho}$.\footnote{To compare with \cite{Stephani2009-dt}, we note that the metric of i) $\mathrm{M_2}$, ii) $\mathrm{dS_2}$, and iii) $\mathrm{AdS_2}$, may be cast to the form i) $\tilde{\bs{g}}_2=-z^2\,\bs{\mathrm{d}}t^2+\bs{\mathrm{d}}z^2$, ii) $\tilde{\bs{g}}_2=-\sin^2 z\,\bs{\mathrm{d}}t^2+\bs{\mathrm{d}}z^2$, and iii) $\tilde{\bs{g}}_2=-\sinh^2 z\,\bs{\mathrm{d}}t^2+\bs{\mathrm{d}}z^2$ by applying the coordinate transformations i) ${\sigma = z\cosh t}$, ${\tau = z\sinh t}$, ii) ${\sigma = \sinh ^{-1}(\sinh t \sin z)}$, ${\tau = \sec ^{-1}\left(\cosh t \sec z \sqrt{\tanh ^2 t \sin ^2 z+\sech^2 t}\right)}$, and iii) ${\sigma = \log \left(\sqrt{\cosh ^2 t \sinh ^2 z+1}+\cosh t \sinh z\right)}$, ${\tau = \tan ^{-1}(\csch t \coth z)}$.} A more familiar example with the axial (1+1)-Poincar\'e symmetry [4,3,11] (which may be alternatively viewed as stationary cylindrical symmetry enhanced by boost symmetry) is the \textit{Melvin spacetime} [(22.13) \cite{Stephani2009-dt}, (7.21) \cite{Griffiths:2009dfa}],
\begin{equation}
    \bs{g}=(1+b^2\rho^2)^2 (- \, \bs{\mathrm{d}}{t}^2 +\bs{\mathrm{d}}{z}^2+\bs{\mathrm{d}}{\rho}^2 )+\tfrac{\rho^2}{(1+b^2\rho^2)^2}\, \bs{\mathrm{d}}{\varphi}^2\;,
\end{equation}
where $b$ is a constant. This solution of Einstein-Maxwell equations is, in the adapted coordinates $(t,z,\rho,\varphi)$, given by the choice ${\phi_1=\phi_2=\rho^2/\phi_4=(1+b^2\rho^2)^2}$, ${\phi_3=0}$. Stationary so(3) and so(2,1) symmetries and the enhanced n(3,1) symmetry [4,3,\{4,5,2\}] include other well-known vacuum solutions, the \textit{Taub-NUT spacetime} [(13.49) \cite{Stephani2009-dt}, (12.1) \cite{Griffiths:2009dfa}] and its planar and hyperbolic counterparts [(13.48) \cite{Stephani2009-dt}, (12.13) \cite{Griffiths:2009dfa}],
\begin{equation}\label{eq:TaubNUT}
    \bs{g}=-\tfrac{\epsilon (r^2-l^2)-2mr}{r^2+l^2}\bigl(\bs{\mathrm{d}}{t}-2l\Upsilon\,\bs{\mathrm{d}}{\varphi}\bigr)^2 + \tfrac{r^2+l^2}{\epsilon (r^2-l^2)-2mr}\, \bs{\mathrm{d}}{r}^2+(r^2+l^2) \bs{g}_2\;,
\end{equation}
where $m$ and $l$ are constants, ${\epsilon=\pm1,0}$, and $\Upsilon$ is a shorthand for
\begin{equation}
    \Upsilon=\begin{cases}
        \cos\vartheta\;, &\epsilon=1\;,\\
        \vartheta\;, &\epsilon=0\;,\\
        \cosh\vartheta\;, &\epsilon=-1\;.
    \end{cases}
\end{equation}
The spacetimes \eqref{eq:TaubNUT}, with the adapted coordinates denoted by $(t/(2l),r,\vartheta,\varphi)$, correspond to the following components of the $\Gamma$-invariant metrics: ${\phi_1/(2l)^2=1/\phi_3={(\epsilon (r^2-l^2)-2mr)}/{(r^2+l^2)}}$, ${\phi_4=r^2+l^2}$, ${\phi_2=0}$. A very popular vacuum solution, the \textit{NHEK spacetime} [(31.61) \cite{Stephani2009-dt}, (2.6) \cite{Bardeen:1999px}, (1) \cite{Kunduri:2007vf}],
\begin{equation}\label{eq:NHEK}
    \bs{g}= (a^2+x^2) \left(-\cosh^2\sigma\,\bs{\mathrm{d}}{\tau}^2+\bs{\mathrm{d}}{\sigma}^2\right)+  \tfrac{a^2-x^2}{a^2+x^2} \bigl(\bs{\mathrm{d}}{z}+2a\sinh\sigma\,\bs{\mathrm{d}}{\tau}\bigr)^2  + \tfrac{a^2+x^2}{a^2-x^2}\, \bs{\mathrm{d}}{x}^2\;,
\end{equation}
with $a$ being a constant, is included in the enhanced so(2,1) symmetry [4,3,9]. Indeed, in the adapted coordinates given by $(\tau,\sigma,z/(2a),x)$ one can obtain this metric by setting ${\phi_1=a^2+x^2}$, ${\phi_2/(2a)^2=1/\phi_4=(a^2-x^2)/(a^2+x^2)}$, ${\phi_3=0}$.\footnote{In order to compare our metric with the one from \cite{Kunduri:2007vf}, we should introduce the coordinates $v=\frac{2 a^2 (\sin \tau-\sech\sigma)}{\cos \tau-\tanh \sigma}$, $r=\cosh \sigma (\cos \tau-\tanh \sigma)$, $\phi=-\frac{z}{2 a}-2 \tanh ^{-1}\left(\csc \tau-\tanh \frac{\sigma }{2} \cot \tau\right)-\log (\cosh \sigma \cos \tau-\sinh \sigma)+2 \tanh ^{-1}\left(\tan \frac{\tau }{2}\right)$, $x=a \cos\theta$, which brings the metric to the form ${\bs{g}=\frac{1+\cos ^2 \theta}{2}\Big[-\frac{r^2}{r_0^2} \bs{\mathrm{d}} v^2+\bs{\mathrm{d}} v \vee \bs{\mathrm{d}} r+r_0^2 \bs{\mathrm{d}} \theta^2\Big]+\frac{2 r_0^2 \sin ^2 \theta}{1+\cos ^2 \theta}\Big(\bs{\mathrm{d}} \phi+\frac{r}{r_0^2} \bs{\mathrm{d}} v\Big)^2}$, where we denoted ${r_0^2=2a^2}$.} Another vacuum solution, which recently received renewed attention, is the \textit{swirling spacetime} [(24.22) \cite{Stephani2009-dt}, (16.27) \cite{Griffiths:2009dfa}, (3.3) \cite{Astorino:2022aam}],
\begin{equation}\label{eq:swirl}
    \bs{g}=\big(1+j^2\rho^4\big) (\bs{\mathrm{d}}{u}\vee \bs{\mathrm{d}}{v} +\bs{\mathrm{d}}{\rho}^2)+\tfrac{\rho^2}{1+j^2\rho^4}  (\bs{\mathrm{d}}{\sigma} + 4ju\, \bs{\mathrm{d}}{v})^2\;,
\end{equation}
where $j$ is a constant. This geometry can be identified within the enhanced n(3,1) symmetry [4,3,10]. Denoting the adapted coordinates $(-\sigma/(4j),\rho,u,v)$, \eqref{eq:swirl} is given by ${\phi_3=\phi_4=(1+j^2\rho^4)=-(4j)^2\rho^2/\phi_1}$, ${\phi_2=0}$.\footnote{The form $\bs{g}=\big(1+j^2\rho^4\big)(-\bs{\mathrm{d}} t^{{2}} +\bs{\mathrm{d}} z^{{2}}+\bs{\mathrm{d}} \rho^{{2}} )+\tfrac{\rho^2}{1+j^2\rho^4}(\bs{\mathrm{d}} \phi+4j z\,\bs{\mathrm{d}} t )^{{2}}$ from \cite{Astorino:2022aam} can be achieved by the coordinate transformation ${u = \tfrac{1}{\sqrt{2}}(z-t)}$, ${v= \tfrac{1}{\sqrt{2}}(z+t)}$, ${\sigma = \phi+j t (t+2 z)-j z^2}$.} Let us also remark that this metric is also obtainable from the planar Taub-NUT spacetime \eqref{eq:TaubNUT} by a double Wick rotation \cite{Astorino:2022aam}; this may hint at possible relation between [4,3,5] and [4,3,10] when analytically continued to the complex manifolds.

\item[\descstyle{[4,4,-]}]
There exist perfect-fluid solutions of the Einstein equations that admit either $\mathrm{so(3)\oplus n(1,1)}$ or $\mathrm{so(2,1)\oplus n(1,1)}$ symmetries [4,4,\{1,2\}] known as the \textit{Farnsworth-Kerr spacetimes} [(12.27)--(12.29) \cite{Stephani2009-dt}]. The former symmetry is exhibited by the solution of class I,
\begin{equation}
\begin{aligned}
    \bs{g} &= a^2 \big[(1-k) (\bs{\mathrm{d}}{\xi} - \cos \varsigma \,\bs{\mathrm{d}}{\zeta})^2  + (1+k) (\sin \xi \,\bs{\mathrm{d}}{\varsigma} + \cos \xi \sin \varsigma \,\bs{\mathrm{d}}{\zeta})^2  \\
    &\feq- \bs{\mathrm{d}}{t}^2+  \sqrt{1 - 2k^2} \bigl(\bs{\mathrm{d}}{t}\vee(\cos \xi \,\bs{\mathrm{d}}{\varsigma} -  \sin \varsigma \sin \xi\, \bs{\mathrm{d}}{\zeta})\bigr) + (1 + 2 k^2) (\cos \xi \,\bs{\mathrm{d}}{\varsigma} -  \sin \varsigma \sin \xi \,\bs{\mathrm{d}}{\zeta})^2\big]\;,
\end{aligned}
\end{equation}
which is given by the choice ${\phi_1 = -a^2 (1-k)}$, ${\phi_3 = - a^2}$, ${\phi_4 = a^2(1+k)}$, ${\phi_5 = a^2 \sqrt{1 - 2k^2}}$, ${\phi_{10} = a^2 (1 + 2 k^2)}$, ${\phi_2=\phi_6=\phi_7 = \phi_8 = \phi_9 = 0}$ in the adapted coordinates $(\xi,t,\varsigma,\zeta)$.\footnote{In order to compare with \cite{Stephani2009-dt}, one has to perform the transformation ${\xi = \cot^{-1}(\csc y \tan z) -  x}$, ${\sin \varsigma = -\sin z \sqrt{1 + \cot^2z \sin^2y}}$, ${\csc \zeta = \csc y \sin z \sqrt{1 + \cot^2z \sin^2y}}$,
} The latter symmetry is possessed by the solution of class II, 
\begin{equation}
\begin{aligned}
    \bs{g} &= a^2\big[(1-k) (\bs{\mathrm{d}}{\xi} - \cosh \varsigma \,\bs{\mathrm{d}}{\zeta})^2  + (1+k) (\sin \xi \,\bs{\mathrm{d}}{\varsigma} + \cos \xi \sinh \varsigma \,\bs{\mathrm{d}}{\zeta})^2 \\
    &\feq + \bs{\mathrm{d}}{u}^2-\bigl(\bs{\mathrm{d}}{u}\vee(\cos \xi\, \bs{\mathrm{d}}{\varsigma} -  \sin \xi \sinh \varsigma \,\bs{\mathrm{d}}{\zeta})\bigr) -(1+2 k^2) (\cos \xi \, \bs{\mathrm{d}}{\varsigma} - \sin \xi \sinh \varsigma \, \bs{\mathrm{d}}{\zeta})^2\big]\;,
\end{aligned}
\end{equation}
as well as the solution of class III,
\begin{equation}
\begin{aligned}
    \bs{g} &= a^2\big[(1-s) (\bs{\mathrm{d}}{\xi} - \cosh \varsigma \,\bs{\mathrm{d}}{\zeta})^2  +  (1+s) (\sin \xi \,\bs{\mathrm{d}}{\varsigma} + \cos \xi \sinh \varsigma \,\bs{\mathrm{d}}{\zeta})^2 
    \\
    &\feq+\bs{\mathrm{d}}{u}^2-2 (\cos \xi \, \bs{\mathrm{d}}{\varsigma} - \sin \xi \sinh \varsigma \, \bs{\mathrm{d}}{\zeta})^2\big]\;.
\end{aligned}
\end{equation}
Denoting the adapted coordinates by $(\xi,u,\varsigma,\zeta)$, the class II corresponds to the choice ${\phi_1 = -a^2(1-k)}$, ${\phi_3 = a^2}$, ${\phi_4 = a^2(1+k)}$, ${\phi_5 = - a^2 \sqrt{1-2k^2}}$, ${\phi_{10} = - a^2(1+2 k^2)}$, ${\phi_2=\phi_6 = \phi_7 = \phi_8 = \phi_9 = 0}$, while the class III to the choice ${\phi_1 = -a^2(1-s)}$, ${\phi_3 = a^2}$, ${\phi_4 = a^2(1+s)}$, ${\phi_{10} = -2a^2}$, ${\phi_2 =\phi_5 = \phi_6 = \phi_7 = \phi_8 = \phi_9 = 0}$.\footnote{In order to compare with \cite{Stephani2009-dt}, one has to perform the transformation ${\xi = \cot^{-1}(\csch y \tanh z) - x}$, ${\sinh \varsigma = -\sinh z \sqrt{1 + \sinh^2y \coth^2z}}$, ${\csc \zeta = \csch y \sinh z \sqrt{1 + \sinh^2y \coth^2z}}$.} Here, $a$, $k$, and $s$ are arbitrary constants. The s(4,6) symmetry [4,4,18] is admitted by the solution of the Einstein-Maxwell equations which we refer to as the \textit{McLenaghan-Tariq spacetime} [(12.21) \cite{Stephani2009-dt}],
\begin{equation}
    \bs{g}=-(\bs{\mathrm{d}}{t}  - 2y\, \bs{\mathrm{d}}{\varphi})^2  + \tfrac{a^2}{x^2}\,\big(\bs{\mathrm{d}}{x}^2 +   \bs{\mathrm{d}}{y}^2\big)   + x^2\, \bs{\mathrm{d}}{\varphi}^2\;,
\end{equation}
where $a$ is a constant. The above spacetime correspond to the following choice: ${\phi_1=4}$, ${\phi_3=\phi_4=a^2}$, ${\phi_{10}=1}$, ${\phi_2=\phi_5=\phi_6=\phi_7=\phi_8=\phi_9=0}$ of the $\Gamma$-invariant metric $\hat{\bs{g}}$ when written in the adapted coordinates $(t/2,-\log x,y,\varphi)$. There seem to be no known spacetimes corresponding to the symmetries [4,4,\{9,22\}].

\item[\descstyle{[5,4,-]}]
The radially-invariant stationary so(2,1) symmetry [5,4,1] is manifested in the \textit{G\"odel spacetime} [(12.26) \cite{Stephani2009-dt}, (22.11) \cite{Griffiths:2009dfa}], a well-known perfect-fluid solution of the Einstein field equations given by the metric
\begin{equation}\label{eq:Godel}
    \bs{g}=-\tfrac{1}{2\omega^2} (\bs{\mathrm{d}}{\tau} - \cosh \vartheta\, \bs{\mathrm{d}}{\varphi})^2 + \bs{\mathrm{d}}{z}^2 + \tfrac{1}{4\omega^2} (\bs{\mathrm{d}}{\vartheta}^2 \allowbreak + \sinh^2\vartheta\, \bs{\mathrm{d}}{\varphi}^2)\;,
\end{equation}
with $\omega$ being a constant. If the adapted coordinates are denoted by $(\tau,z,\vartheta,\varphi)$, the spacetime is given by ${\phi_1 = {1}/{(2\omega^2)}}$, ${\phi_3 = 1}$, ${\phi_4 = {1}/{(4\omega^2)}}$,  ${\phi_2 = 0}$.\footnote{To recover the metric from \cite{Griffiths:2009dfa}, one should perform the coordinate transformation ${\tau = - \sqrt{2} \omega t - 2 \tan^{-1} \tfrac{1 - e^{-2 \omega x}}{\sqrt{2}\omega y} - \tan^{-1} \frac{2 \sqrt{2} \omega y}{1 - e^{-4\omega x} - 2 \omega^2 y^2}}$, ${\cosh \vartheta = \tfrac{1}{2} e^{2\omega x} (1 + e^{-4\omega x} + 2 \omega^2 y^2)}$, ${\tan \varphi = \frac{2\sqrt{2} \omega  y}{1 - e^{-4\omega x} - 2 \omega^2 y^2}}$.} To the best of our knowledge, there are no well-known examples of spacetimes admitting symmetries [5,4,\{2,3,6,7\}].

\item[\descstyle{[6,3,-]}]
Families of all closed, flat, and open \textit{FLRW spacetimes} [(12.9), (14.2) \cite{Stephani2009-dt}, (6.4) \cite{Griffiths:2009dfa}], the well-known perfect-fluid solutions of the Einstein field equations, are equivalent to the $\Gamma$-invariant metrics $\hat{\bs{g}}$ with 3-spherical, 3-planar, and 3-hyperbolic symmetries [6,3,\{1,2,3\}] (i.e., with 3-dimensional spacelike orbits of constant curvature $\mathrm{S_3}$, $\mathrm{E_3}$, or $\mathrm{H_3}$),
\begin{equation}\label{eq:FLRW}
    \bs{g}=-\bs{\mathrm{d}}t^2+a^2(t)\bs{g}_3\;.
\end{equation}
One can see this by denoting $(t,\chi,\vartheta,\varphi)$, setting ${\phi_1=1}$ by re-parametrization of $t$, and renaming ${\phi_2=a^2}$ with $a$ being a function of $t$. Here, $\bs{g}_3$ is the metric of a 3-space of constant unit curvature,
\begin{equation}
    \bs{g}_3=\begin{cases}
        \bs{\mathrm{d}}\chi^2+ \sin^2\chi\big(\bs{\mathrm{d}}\vartheta^2+\sin^2\vartheta \bs{\mathrm{d}}\varphi^2\big)\;, &\epsilon=1\;,\\
        \bs{\mathrm{d}}\chi^2+\bs{\mathrm{d}}\vartheta^2+\bs{\mathrm{d}}\varphi^2\;, &\epsilon=0\;,\\
        \bs{\mathrm{d}}\chi^2+ \sinh^2\chi\big(\bs{\mathrm{d}}\vartheta^2+\sin^2\vartheta \bs{\mathrm{d}}\varphi^2\big)\;, &\epsilon=-1\;.
    \end{cases}
\end{equation}
Analogous spacetimes to these but with the (2+1)-Poincar\'e, (3+1)-Lorentzian, or sum-of-two (2+1)-Lorentzian symmetries [6,3,\{5,6,4\}] (i.e., with 3-dimensional timelike orbits $\mathrm{M_3}$, $\mathrm{dS_3}$, or $\mathrm{AdS_3}$) do not seem to occur in the well-known spacetimes.

\item[\descstyle{[6,4,-]}]
The cases [6,4,\{1,2,3,4,5\}] correspond to the (1+1)-Poincar\'e or (2+1)-Lorentzian symmetries combined with the 2-spherical, 2-planar, or 2-hyperbolic symmetries. The respective $\Gamma$-invariant metrics $\hat{\bs{g}}$ are simply the direct product spacetimes of 2-dimensional constant curvature geometries, i.e., products of $\mathrm{M_2}$ or $\mathrm{(A)dS_2}$ with $\mathrm{S_2}$, $\mathrm{E_2}$, or $\mathrm{H_2}$,
\begin{equation}
    \bs{g}=a^2\tilde{\bs{g}}_2+b^2\bs{g}_2\;,
\end{equation}
where $a$ and $b$ are two constants. Depending on the specific combinations of signs of curvatures of the 2-dimensional geometries, the metrics bear names: the \textit{Bertotti-Robinson}, \textit{(anti-)Nariai}, or \textit{Pleba\'nski-Hacyan spacetimes} [(12.8) \cite{Stephani2009-dt}, (7.10) \cite{Griffiths:2009dfa}]. To match the notation, we just write the $\Gamma$-invariant metrics $\hat{\bs{g}}$ in the adapted coordinates $(\tau,\sigma,\vartheta,\varphi)$ and denote ${\phi_1=\pm a^2}$, ${\phi_2=b^2}$. Notice that the combination ${\varepsilon=\epsilon=0}$ is missing here as it would correspond to $\mathrm{M_4}$, which is a maximally symmetric spacetime (i.e., it admits 10 Killing vectors).

\item[\descstyle{[7,4,-]}]
The stationary 3-spherical and 3-hyperbolic symmetries [7,4,\{1,2\}] correspond to the \textit{Einstein static spacetime} [(12.9) with (12.23) \cite{Stephani2009-dt}, (3.8) \cite{Griffiths:2009dfa}] and its hyperbolic version [(12.9) with (12.23) \cite{Stephani2009-dt}],
\begin{equation}
    \bs{g}=-\bs{\mathrm{d}}t^2+a^2\bs{g}_3\;.
\end{equation}
Again, the 3-planar version, ${\epsilon=0}$, is absent as it would reduce to $\mathrm{M_4}$. Choosing the adapted coordinates $(t,\chi,\vartheta,\varphi)$, the above metric is equivalent to setting ${\phi_1=1}$, which can be adjusted just by constant re-scaling of $t$, and renaming the constant ${\phi_2=a^2}$. We are not aware of any well-known spacetimes with the stationary (3+1)-Lorentzian or sum-of-two (2+1)-Lorentzian symmetries [7,4,\{3,4\}].
\end{description}


\section{Imposing gauge conditions}\label{sec:gaugecond}

The $\Gamma$-invariant metric $\hat{\bs{g}}$ may be cast into a simpler form by a subgroup of the diffeomorphism group, called the \textit{residual diffeomorphism group}, that preserves the form of $\hat{\bs{g}}$. The \textit{residual diffeomorphisms} $\Phi$ are diffeomorphisms on $M$ that pullback the $\Gamma$-invariant metric $\hat{\bs{g}}$ to $\Phi^*\hat{\bs{g}}$ so that it remains $\Gamma$-invariant \cite{Anderson:1999cn}. This group is obtainable by constructing the \textit{residual diffeomorphism generators} $\bs{W}$, which are the vector fields on $M$ defined by
\begin{equation}\label{eq:resgeneqn}
    [\bs{W},\bs{X}_i]=\sum_{j=1}^{d} a_{ij} \bs{X}_j\;,\quad i=1,\dots,d\;,
\end{equation}
for some constants $a_{ij}$. It is easy to see that $\lie_{\bs{W}}\hat{\bs{g}}$ is $\Gamma$-invariant, ${\lie_{\bs{X}}\lie_{\bs{W}}\hat{\bs{g}}=(\lie_{\bs{X}}\lie_{\bs{W}}-\lie_{\bs{W}}\lie_{\bs{X}})\hat{\bs{g}}=\lie_{[\bs{W},\bs{X}]}\hat{\bs{g}}=0}$, which implies that the flow $\Phi_{\tau}$ of the vector field $\bs{W}$ transforms the components $\phi_i$ of $\hat{\bs{g}}$ to the components~${\psi_i(\tau)}$ of~$\Phi_{\tau}^*\hat{\bs{g}}$, i.e.,
\begin{equation}\label{eq:transf}
    \Phi_{\tau}^*\left(\sum_{i=1}^s \phi_i \bs{q}_i\right)=\sum_{i=1}^s \psi_i(\tau) \bs{q}_i\;,
\end{equation}
where $\psi_i(\tau)$ are some $\Gamma$-invariant scalar fields on $M$. Hence, a clever application of the residual diffeomorphism generators may allow one to fix some of the components $\phi_i$ to specific functions/constants and thus effectively reduce the number $s$ we introduced above.

Although the physics is not affected by action of diffeomorphisms in generally covariant theories (since the Lagrangian is diffeomorphism equivariant), the variational principle can be sensitive to it! In other words, there is no issue in choosing a gauge at the level of the field equations (after the variation), but fixing some $\phi_i$ at the level of the Lagrangian (before the variation) may result in a loss of possibly indispensable equations corresponding to the variations with respect to these $\phi_i$. As we will discuss below, imposing gauge conditions in the reduced Lagrangian without breaking PSC is still possible in some cases as a consequence of Noether identities \cite{Anderson:1999cn}.

The residual diffeomorphims group may be generated by several types of residual diffeomorphism generators. The Killing vector fields $\bs{X}$ from Tab.~\ref{tab:PSCcompgroupactions} belong to somewhat trivial part of it in a sense that they preserve the metric $\hat{\bs{g}}$ exactly as it is, i.e., without changing $\phi_i$. More interesting residual diffeomorphism generators are the $\Gamma$-invariant vector fields $\bs{V}$, which we list in Tab.~\ref{tab:invvectfieldsPSCcomp}; they satisfy \eqref{eq:resgeneqn} with the vanishing right-hand side, ${a_{ij}=0}$. Apart from that, however, there may exist further non-$\Gamma$-invariant residual diffeomorphism generators $\bs{P}$ [with non-zero right-hand side of \eqref{eq:resgeneqn}], which we collect in Tab.~\ref{tab:residualgenerators}. Therefore, the non-trivial residual diffeomorphism generators $\bs{W}$ are either~$\bs{V}$ or $\bs{P}$. 

Notice that some of the vector fields $\bs{V}$ in Tab.~\ref{tab:invvectfieldsPSCcomp} contain a freedom in functions $f_i$ (rather than just constants). These form an \textit{infinite-dimensional} part of the residual diffeomorphism group and give rise to the \textit{Noether identities} for the Euler-Lagrange expressions $\underline{E}_i$, which we list in Tab.~\ref{tab:Noetherid}. (Here, the partial derivatives can be understood as only acting on the components $E_i$ of the $r$-forms ${\underline{E}_i={E}_i\,\underline{\mathrm{d}}\hat{y}_{i_1}\wedge\dots\wedge\underline{\mathrm{d}}\hat{y}_{i_r}}$, where ${\underline{\mathrm{d}}\hat{y}_{i_1}\wedge\dots\wedge\underline{\mathrm{d}}\hat{y}_{i_r}}$ is the coordinate volume form on $\hat{M}$.)  The Noether identities hold for every $\phi_i$ regardless whether they correspond to a solution or not. 

The Noether identities for the symmetry-reduced theory can be obtained by either of two methods: i) Recall that the field equations arising from diffeomorphism-equivariant Lagrangians $\underline{L}$ are necessarily divergence-free as a consequence of Noether's second theorem; one can symmetry-reduce this divergence-free condition. ii) Alternatively, one can use the equivariance of the reduced Lagrangian $\hat{\underline{L}}$ with respect to the residual diffeomorphisms and apply Noether's second theorem directly to $\hat{\underline{L}}$. PSC guarantees that both methods yield the same Noether identities.\footnote{i) By contracting the vanishing divergence of the Euler-Lagrange expression ${\bs{\nabla}\cdot\underline{\bs{E}}=0}$ with the infinite-dimensional residual diffeomorphism generator ${\bs{W}=f\bs{W}_0}$ (where $f$ is an arbitrary function) and lowering the index $\bs{W}_0^\flat:=\bs{g}\cdot\bs{W}_0$, one obtains $0=\bs{W}_0^\flat\cdot(\bs{\nabla}\cdot\underline{\bs{E}})=\bs{\partial}\cdot(\bs{\underline{E}}\cdot\bs{W}_0^\flat)-(\bs{\nabla}\bs{W}_0^\flat)\cdot\bs{\underline{E}}$. Here, we treated 4-forms as densities, used the Leibniz rule, and rewrote the divergence of the vector-density $\bs{\underline{E}}\cdot\bs{W}_0^\flat$ using the coordinate derivative. The Noether identities correspond to the symmetry-reduction of this equation [similar to the reduction of $\underline{L}$ in \eqref{eq:reducedL}], i.e., $0=\sum_{i=1}^s\left[\bs{\partial}\cdot(\underline{E}_i\bs{p}_i\cdot\bs{W}_0^{\hat{\flat}})-\underline{E}_i\bs{p}_i\cdot(\hat{\bs{\nabla}}\bs{W}_0^{\hat{\flat}})\right]$. 
ii) Consider the infinitesimal version of the transformation \eqref{eq:transf}, ${\lie_{\bs{W}}\hat{\bs{g}}=\sum_{i=1}^s \bs{q}_i \delta \phi_i}$, where we denoted the variation of the fields $\phi_i$ by ${\delta \phi_i := \psi'_i(0)}$. Using \eqref{eq:dual}, we can write ${\delta \phi_i=\bs{p}_i\cdot\lie_{\bs{W}}\hat{\bs{g}}}$ and substitute this into \eqref{eq:deltahatL}, $\delta_{\bs{W}}\hat{\underline{L}}=\sum_{i=1}^s \underline{{E}}_i \,\bs{p}_i\cdot \lie_{\bs{W}}\hat{\bs{g}}+\underline{\mathrm{d}}(\dots)$, where $\underline{\mathrm{d}}(\dots)$ denotes the boundary terms. Employing $\lie_{\bs{W}}\hat{\bs{g}}=\textrm{Sym}(\bs{W}_0^{\hat{\flat}}\bs{\mathrm{d}}f+f\hat{\bs{\nabla}}\bs{W}_0^{\hat{\flat}})$ and $\underline{{E}}_i \,\bs{p}_i\cdot(\bs{W}_0^{\hat{\flat}}\bs{\mathrm{d}}f)=\bs{\partial}\cdot(\underline{E}_i\bs{p}_i\cdot\bs{W}^{\hat{\flat}})-f\bs{\partial}\cdot(\underline{E}_i\bs{p}_i\cdot\bs{W}_0^{\hat{\flat}})$ with 4-forms being again treated as densities, we get $\delta_{\bs{W}}\hat{\underline{L}}=f\sum_{i=1}^s\left[-\bs{\partial}\cdot(\underline{E}_i\bs{p}_i\cdot\bs{W}_0^{\hat{\flat}})+\underline{E}_i\bs{p}_i\cdot(\hat{\bs{\nabla}}\bs{W}_0^{\hat{\flat}})\right]+\underline{\mathrm{d}}(\dots)$. Finally, since the variation $\delta\phi_i$ is a symmetry of the reduced Lagrangian, $\delta_{\bs{W}}\hat{\underline{L}}=\underline{\mathrm{d}}(\dots)$, and $f$ is an arbitrary compactly supported function, we may drop the boundary term $\underline{\mathrm{d}}(\dots)$ within the integral over $\hat{M}$ and regain the above formula for the Noether identities.} In any case, the Noether identities for $\underline{E}_{i}$ imply that the field equations ${\underline{E}_{i}=0}$ are not all independent. Under favorable circumstances (depending upon the symmetry group) the Noether identities may be used to eliminate one or more field equations as being redundant. One can then use the residual diffeomorphism group to gauge fix --- in the reduced Lagrangian $\hat{\underline{L}}$ --- the symmetry-reduced metric components $\phi_i$ corresponding to the redundant equations without losing any reduced field equations.

Those $\bs{V}$ or $\bs{P}$ with freedom in constants (not functions) form the \textit{finite-dimensional} part of the residual diffeomorphism group. These finite-dimensional transformations can be employed in gauge fixing in two ways: i) They can either lead to some \textit{scaling transformations} that further simplify the previous Noether identities (see Sec.~\ref{scc:433} and Sec.~\ref{scc:436}). ii) They may possibly give rise to the \textit{algebraic identities} among the field equations, ${\underline{E}_{i}=0}$, which could be used in fixing constants in the homogeneous spacetimes. Unfortunately, Noether's second theorem is not applicable here and it is unclear to us how to proceed in deriving the algebraic identities in general.

Finally, it should be stressed that any gauge fixing regardless whether it is performed in the reduced Lagrangian or the field equations may sometimes result in a loss of solutions for which such a gauge choice is impossible (or it may no longer be permissible prior to variation if the Noether identities degenerate). This can only be rectified by considering multiple (allowed) gauge choices that cover all possibilities.


\begin{longtable}{|l||>{\raggedright\arraybackslash}p{12cm}|}
\caption{Residual diffeomorphism generators. Here, we exclude vector fields that are either Killing $\bs{X}$ or $\Gamma$-invariant $\bs{V}$ (i.e., ${a_{ij}=0}$), which are already contained in Tab.~\ref{tab:PSCcompgroupactions} and Tab.~\ref{tab:invvectfieldsPSCcomp}, and only list the non-$\Gamma$-invariant vector fields $\bs{P}$.}\label{tab:residualgenerators}
\\ \hline
{\scriptsize Hicks \#} & {\scriptsize Non-$\Gamma$-invariant residual diffeomorphism generators $\bs{P}$} \\\hline\hline
{[3,2,1]} & $y_3 \, \bs{\partial}_{y_3} \allowbreak + y_4 \, \bs{\partial}_{y_4}$ \\ \hline
{[3,2,\{2,3,5\}]} & $\emptyset$ \\ \hline
{[3,2,4]} & $y_1 \, \bs{\partial}_{y_1} \allowbreak + y_2 \, \bs{\partial}_{y_2}$ \\ \hline\hline
{[3,3,2]} &  $y_1 \, \bs{\partial}_{y_1}, \allowbreak y_2 \, \bs{\partial}_{y_1}, \allowbreak y_4 \, \bs{\partial}_{y_1}, \allowbreak y_1 \, \bs{\partial}_{y_2}, \allowbreak y_2 \, \bs{\partial}_{y_2}, \allowbreak y_4 \, \bs{\partial}_{y_2}, \allowbreak y_1 \, \bs{\partial}_{y_4}, \allowbreak y_2 \, \bs{\partial}_{y_4}, \allowbreak y_4 \, \bs{\partial}_{y_4}$ \\ \hline
{[3,3,3]} & $y_3 \, \bs{\partial}_{y_1}, \allowbreak y_4 \, \bs{\partial}_{y_1}, \allowbreak y_1 \, \bs{\partial}_{y_1} + y_3 \, \bs{\partial}_{y_3}, \allowbreak y_3 \, \bs{\partial}_{y_3} - y_4 \, \bs{\partial}_{y_4}, \allowbreak y_4 \, \bs{\partial}_{y_3} + \frac{y_4^2}{2} \, \bs{\partial}_{y_1}, \allowbreak y_3 \, \bs{\partial}_{y_4} + \frac{y_3^2}{2} \, \bs{\partial}_{y_1}$ \\ \hline
{[3,3,\{8,9\}]} & $\emptyset$ \\ \hline\hline
{[4,3,\{1,3\}]} & $y_1 \, \bs{\partial}_{y_1}$ \\ \hline
{[4,3,\{2,4,9\}]} & $\emptyset$ \\ \hline
{[4,3,5]} & $y_1 \, \bs{\partial}_{y_1} \allowbreak + \frac{y_3}{2} \, \bs{\partial}_{y_3} \allowbreak + \frac{y_4}{2} \, \bs{\partial}_{y_4}$ \\ \hline
{[4,3,6]} & $y_1 \, \bs{\partial}_{y_1}, \allowbreak y_3 \, \bs{\partial}_{y_3} + y_4 \, \bs{\partial}_{y_4}$ \\ \hline
{[4,3,8]} & $y_4 \, \bs{\partial}_{y_4}$ \\ \hline
{[4,3,10]} & $y_1 \, \bs{\partial}_{y_1} + y_3 \, \bs{\partial}_{y_3}$ \\ \hline
{[4,3,11]} & $y_1 \, \bs{\partial}_{y_1} + y_2 \, \bs{\partial}_{y_2}, \allowbreak y_4 \, \bs{\partial}_{y_4}$ \\ \hline\hline
{[4,4,\{1,2\}]} & $y_2 \, \bs{\partial}_{y_2}$ \\ \hline
{[4,4,9]} & $y_2 \, \bs{\partial}_{y_1}, y_3 \, \bs{\partial}_{y_1},y_4 \, \bs{\partial}_{y_1} - y_2 \, \bs{\partial}_{y_4}, y_3 \, \bs{\partial}_{y_4} - \tfrac{1}{2} y_3^2 \, \bs{\partial}_{y_1}, \allowbreak y_1 \, \bs{\partial}_{y_1} + \tfrac{1}{2} y_3 \, \bs{\partial}_{y_3} + \tfrac{1}{2} y_4 \, \bs{\partial}_{y_4}, \allowbreak y_2 \, \bs{\partial}_{y_2} - \tfrac{1}{2} y_3 \, \bs{\partial}_{y_3} + \tfrac{1}{2} y_4 \, \bs{\partial}_{y_4}, \allowbreak y_3 \, \bs{\partial}_{y_2} + \tfrac{1}{2} y_3^2 \, \bs{\partial}_{y_4} - \tfrac{1}{6} y_3^3 \, \bs{\partial}_{y_1}$ \\ \hline
{[4,4,18]} & $y_2 \, \bs{\partial}_{y_1}, y_1 \, \bs{\partial}_{y_1} + y_3 \, \bs{\partial}_{y_3},y_3 \, \bs{\partial}_{y_3} - y_4 \, \bs{\partial}_{y_4}$ \\ \hline
{[4,4,22]} & $y_2 \, \bs{\partial}_{y_1}, y_1 \, \bs{\partial}_{y_1} + \tfrac12 y_3 \, \bs{\partial}_{y_3} + \tfrac12 y_4 \, \bs{\partial}_{y_4}, y_4 \, \bs{\partial}_{y_3} - y_3 \, \bs{\partial}_{y_4} - \tfrac12 (y_3^2 - y_4^2) \, \bs{\partial}_{y_1}
$ \\ \hline \hline
{[5,4,\{1,2\}]} & $y_2 \, \bs{\partial}_{y_1}, y_2 \, \bs{\partial}_{y_2} $ \\ \hline
{[5,4,3]} & $y_2 \, \bs{\partial}_{y_1}, \, y_2 \, \bs{\partial}_{y_2}, \, y_1 \, \bs{\partial}_{y_1} \, + \, \frac{1}{2}y_3 \, \bs{\partial}_{y_3} \, + \, \frac{1}{2} y_4 \, \bs{\partial}_{y_4}$ \\ \hline
{[5,4,6]} & $y_2 \, \bs{\partial}_{y_1}, y_2 \, \bs{\partial}_{y_2}, y_1 \, \bs{\partial}_{y_1} + y_3 \, \bs{\partial}_{y_3}$ \\ \hline
{[5,4,7]} & $y_4 \, \bs{\partial}_{y_3}, y_4 \, \bs{\partial}_{y_4} $ \\ \hline \hline
{[6,3,\{1,3,4,6\}]} & $\emptyset$ \\ \hline
{[6,3,2]} & $y_2 \, \bs{\partial}_{y_2} + y_3 \bs{\partial}_{y_3} + y_4 \, \bs{\partial}_{y_4}$ \\ \hline  
{[6,3,5]} & $y_1 \, \bs{\partial}_{y_1} + y_2 \bs{\partial}_{y_2} + y_4 \, \bs{\partial}_{y_4}$ \\ \hline \hline 
{[6,4,\{1,2\}]} & $\emptyset$ \\ \hline
{[6,4,3]} & $y_3 \, \bs{\partial}_{y_3} \, + \, y_4 \, \bs{\partial}_{y_4}$ \\ \hline
{[6,4,\{4,5\}]} & $y_1 \, \bs{\partial}_{y_1} \, + \, y_2 \, \bs{\partial}_{y_2}$ \\ \hline \hline
{[7,4,\{1,2\}]} & $y_1 \, \bs{\partial}_{y_1}$ \\ \hline
{[7,4,\{3,4\}]} & $y_4 \, \bs{\partial}_{y_4}$ \\ \hline
\end{longtable}

\begin{longtable}{|l||>{\raggedright\arraybackslash}p{13cm}|}
\caption{Noether identities}\label{tab:Noetherid}
\\ \hline
{\scriptsize Hicks \#} & {\scriptsize Noether identities from infinite-dimensional $\bs{W}$ (i.e., infinite-dimensional $\bs{V}$)} \\\hline\hline
{[3,2,\{1,2,3\}]} & $\underline{E}_1 \partial_{y_1} \phi_1 + \underline{E}_2 \partial_{y_1} \phi_2 + \underline{E}_3 \partial_{y_1} \phi_3 + \underline{E}_4 \partial_{y_1} \phi_4 -  \partial_{y_1} (2 \underline{E}_1 \phi_1 + \underline{E}_2 \phi_2) \allowbreak + \partial_{y_2} (\underline{E}_2 \phi_1 - 2 \underline{E}_3 \phi_2) = 0$, \allowbreak $\underline{E}_1 \partial_{y_2} \phi_1 + \underline{E}_2 \partial_{y_2} \phi_2 + \underline{E}_3 \partial_{y_2} \phi_3 \allowbreak + \underline{E}_4 \partial_{y_2} \phi_4 + \partial_{y_1} (2 \underline{E}_1 \phi_2 -  \underline{E}_2 \phi_3) -  \partial_{y_2} (\underline{E}_2 \phi_2 + 2 \underline{E}_3 \phi_3) = 0$ \\ \hline
{[3,2,\{4,5\}]} & $\underline{E}_1 \partial_{y_3} \phi_1 + \underline{E}_2 \partial_{y_3} \phi_2 + \underline{E}_3 \partial_{y_3} \phi_3 + \underline{E}_4 \partial_{y_3} \phi_4 -  \partial_{y_3} (2 \underline{E}_2 \phi_2 + \underline{E}_3 \phi_3) \allowbreak - \partial_{y_4} (\underline{E}_3 \phi_2 + 2 \underline{E}_4 \phi_3) = 0$, \allowbreak $\underline{E}_1 \partial_{y_4} \phi_1 + \underline{E}_2 \partial_{y_4} \phi_2 + \underline{E}_3 \partial_{y_4} \phi_3 \allowbreak + \underline{E}_4 \partial_{y_4} \phi_4 -  \partial_{y_3} (2 \underline{E}_2 \phi_3 + \underline{E}_3 \phi_4) -  \partial_{y_4} (\underline{E}_3 \phi_3 + 2 \underline{E}_4 \phi_4) = 0$ \\ \hline\hline
{[3,3,2]} & $\partial_{y_3} (2 \underline{E}_2 \phi_7 - \underline{E}_3 \phi_8 + \underline{E}_7 \phi_1 -  \underline{E}_9 \phi_6) = 0$, \allowbreak $\partial_{y_3} (2 \underline{E}_2 \phi_9 + \underline{E}_3 \phi_{10} + \underline{E}_7 \phi_6 + \underline{E}_9 \phi_5) = 0$, \allowbreak $\underline{E}_1 \partial_{y_3} \phi_1 + \underline{E}_2 \partial_{y_3} \phi_2 \allowbreak + \underline{E}_3 \partial_{y_3} \phi_3 \allowbreak + \underline{E}_4 \partial_{y_3} \phi_4 + \underline{E}_5 \partial_{y_3} \phi_5 + \underline{E}_6 \partial_{y_3} \phi_6 + \underline{E}_7 \partial_{y_3} \phi_7 + \underline{E}_8 \partial_{y_3} \phi_8 \allowbreak + \underline{E}_9 \partial_{y_3} \phi_9 + \underline{E}_{10} \partial_{y_3} \phi_{10} -  \partial_{y_3} (2 \underline{E}_2 \phi_2 + \underline{E}_3 \phi_3 + \underline{E}_7 \phi_7 + \underline{E}_9 \phi_9) = 0$, \allowbreak $\partial_{y_3} (2 \underline{E}_2 \phi_3 + \underline{E}_3 \phi_4 + \underline{E}_7 \phi_8 + \underline{E}_9 \phi_{10}) = 0$ \\ \hline
{[3,3,3]} & $\partial_{y_2} (\underline{E}_2 \phi_1 - 2 \underline{E}_3 \phi_2 -  \underline{E}_7 \phi_5 -  \underline{E}_8 \phi_6) = 0$, \allowbreak $\underline{E}_1 \partial_{y_2} \phi_1  + \underline{E}_2 \partial_{y_2} \phi_2 \allowbreak + \underline{E}_3 \partial_{y_2} \phi_3 \allowbreak + \underline{E}_4 \partial_{y_2} \phi_4 \allowbreak + \underline{E}_5 \partial_{y_2} \phi_5 + \underline{E}_6 \partial_{y_2} \phi_6 + \underline{E}_7 \partial_{y_2} \phi_7 + \underline{E}_8 \partial_{y_2} \phi_8 \allowbreak + \underline{E}_9 \partial_{y_2} \phi_9 + \underline{E}_{10} \partial_{y_2} \phi_{10} \allowbreak - \partial_{y_2} (\underline{E}_2 \phi_2 + 2 \underline{E}_3 \phi_3 + \underline{E}_7 \phi_7 + \underline{E}_8 \phi_8) = 0$, \allowbreak $\underline{E}_6 \phi_1 -  \underline{E}_8 \phi_2 -  \underline{E}_9 \phi_5 - 2 \underline{E}_{10} \phi_6 \allowbreak - \partial_{y_2} (\underline{E}_2 \phi_5 + 2 \underline{E}_3 \phi_7 + \underline{E}_7 \phi_4 + \underline{E}_8 \phi_9) = 0$, \allowbreak $2 \underline{E}_4 \phi_5 - \underline{E}_5 \phi_1 + \underline{E}_7 \phi_2 + \underline{E}_9 \phi_6 \allowbreak - \partial_{y_2} (\underline{E}_2 \phi_6 + 2 \underline{E}_3 \phi_8 + \underline{E}_7 \phi_9 + \underline{E}_8 \phi_{10}) = 0$ \\ \hline
{[3,3,8]} & $- 2 \underline{E}_4 \phi_7 + \underline{E}_5 \phi_9 + \underline{E}_6 \phi_8 + \underline{E}_7 (\phi_4 - \phi_{10}) -  \underline{E}_8 \phi_6 -  \underline{E}_9 \phi_5 + 2 \underline{E}_{10} \phi_7 \allowbreak + \partial_{y_2} (\underline{E}_2 \phi_1 \allowbreak - 2 \underline{E}_3 \phi_2 \allowbreak - \underline{E}_5 \phi_6 \allowbreak - \underline{E}_9 \phi_8) = 0$, \allowbreak $\underline{E}_1 \partial_{y_2} \phi_1 + \underline{E}_2 \partial_{y_2} \phi_2 \allowbreak  + \underline{E}_3 \partial_{y_2} \phi_3 + \underline{E}_4 \partial_{y_2} \phi_4 + \underline{E}_5 \partial_{y_2} \phi_5 \allowbreak + \underline{E}_6 \partial_{y_2} \phi_6 \allowbreak + \underline{E}_7 \partial_{y_2} \phi_7 + \underline{E}_8 \partial_{y_2} \phi_8 \allowbreak + \underline{E}_9 \partial_{y_2} \phi_9 + \underline{E}_{10} \partial_{y_2} \phi_{10} -  \partial_{y_2} (\underline{E}_2 \phi_2 + 2 \underline{E}_3 \phi_3 + \underline{E}_5 \phi_5 + \underline{E}_9 \phi_9) = 0$, \allowbreak $- 2 \underline{E}_1 \phi_6 + \underline{E}_2 \phi_5 + \underline{E}_5 \phi_2 - \underline{E}_6 (\phi_1 - \phi_{10}) + \underline{E}_7 \phi_8 + \underline{E}_8 \phi_7 + 2 \underline{E}_{10} \phi_6 \allowbreak - \partial_{y_2} (\underline{E}_2 \phi_8 + 2 \underline{E}_3 \phi_9 \allowbreak + \underline{E}_5 \phi_7 + \underline{E}_9 \phi_4) = 0$, \allowbreak $2 \underline{E}_1 \phi_8 - \underline{E}_2 \phi_9 - 2 \underline{E}_4 \phi_8 - \underline{E}_6 \phi_7 - \underline{E}_7 \phi_6 + \underline{E}_8 (\phi_1 - \phi_4) - \underline{E}_9 \phi_2 \allowbreak - \partial_{y_2} (\underline{E}_2 \phi_6 + 2 \underline{E}_3 \phi_5 + \underline{E}_5 \phi_{10} + \underline{E}_9 \phi_7) = 0$ \\ \hline
{[3,3,9]} & $- 2 \underline{E}_4 \phi_7 + \underline{E}_5 \phi_9 + \underline{E}_6 \phi_8 + \underline{E}_7 (\phi_4 - \phi_{10}) - \underline{E}_8 \phi_6 -  \underline{E}_9 \phi_5 + 2 \underline{E}_{10} \phi_7 \allowbreak + \partial_{y_2} (\underline{E}_2 \phi_1 - 2 \underline{E}_3 \phi_2 \allowbreak - \underline{E}_5 \phi_6 -  \underline{E}_9 \phi_8) = 0$, \allowbreak $\underline{E}_1 \partial_{y_2} \phi_1 + \underline{E}_2 \partial_{y_2} \phi_2 \allowbreak + \underline{E}_3 \partial_{y_2} \phi_3 + \underline{E}_4 \partial_{y_2} \phi_4 + \underline{E}_5 \partial_{y_2} \phi_5 \allowbreak + \underline{E}_6 \partial_{y_2} \phi_6 \allowbreak + \underline{E}_7 \partial_{y_2} \phi_7 + \underline{E}_8 \partial_{y_2} \phi_8 \allowbreak + \underline{E}_9 \partial_{y_2} \phi_9 + \underline{E}_{10} \partial_{y_2} \phi_{10} - \partial_{y_2} (\underline{E}_2 \phi_2 + 2 \underline{E}_3 \phi_3 + \underline{E}_5 \phi_5 + \underline{E}_9 \phi_9) = 0$, \allowbreak $2 \underline{E}_1 \phi_6 - \underline{E}_2 \phi_5 + \underline{E}_5 \phi_2 - \underline{E}_6 (\phi_1 + \phi_{10}) + \underline{E}_7 \phi_8 - \underline{E}_8 \phi_7 + 2 \underline{E}_{10} \phi_6 \allowbreak + \partial_{y_2} (\underline{E}_2 \phi_8 + 2 \underline{E}_3 \phi_9 \allowbreak + \underline{E}_5 \phi_7 + \underline{E}_9 \phi_4) = 0$, \allowbreak $2 \underline{E}_1 \phi_8 - \underline{E}_2 \phi_9 + 2 \underline{E}_4 \phi_8 -  \underline{E}_6 \phi_7 + \underline{E}_7 \phi_6 - \underline{E}_8 (\phi_1 + \phi_4) + \underline{E}_9 \phi_2 \allowbreak - \partial_{y_2} (\underline{E}_2 \phi_6 + 2 \underline{E}_3 \phi_5 + \underline{E}_5 \phi_{10} + \underline{E}_9 \phi_7) = 0$ \\ \hline\hline
{[4,3,\{1,2,3,4,5,6,10\}]} & $\partial_{y_2} (\underline{E}_2 \phi_1 - 2 \underline{E}_3 \phi_2) = 0$, \allowbreak $\underline{E}_1 \partial_{y_2} \phi_1 + \underline{E}_2 \partial_{y_2} \phi_2 + \underline{E}_3 \partial_{y_2} \phi_3 + \underline{E}_4 \partial_{y_2} \phi_4 -  \partial_{y_2} (\underline{E}_2 \phi_2 + 2 \underline{E}_3 \phi_3) = 0$ \\ \hline
{[4,3,\{8,11\}]} & $\underline{E}_1 \partial_{y_3} \phi_1 + \underline{E}_2 \partial_{y_3} \phi_2 + \underline{E}_3 \partial_{y_3} \phi_3 + \underline{E}_4 \partial_{y_3} \phi_4 -  \partial_{y_3} (2 \underline{E}_2 \phi_2 + \underline{E}_3 \phi_3) = 0$, \allowbreak $\partial_{y_3} (2 \underline{E}_2 \phi_3 + \underline{E}_3 \phi_4) = 0$ \\ \hline
{[4,3,9]} & $\partial_{y_4} (\underline{E}_3 \phi_2 + 2 \underline{E}_4 \phi_3) = 0$, \allowbreak $\underline{E}_1 \partial_{y_4} \phi_1 + \underline{E}_2 \partial_{y_4} \phi_2 + \underline{E}_3 \partial_{y_4} \phi_3 + \underline{E}_4 \partial_{y_4} \phi_4 -  \partial_{y_4} (\underline{E}_3 \phi_3 + 2 \underline{E}_4 \phi_4) = 0$ \\ \hline\hline
{[6,3,\{1,2,3\}]} & $\underline{E}_1 \partial_{y_1} \phi_1 + \underline{E}_2 \partial_{y_1} \phi_2 - 2 \partial_{y_1} (\underline{E}_1 \phi_1) = 0$ \\ \hline
{[6,3,\{4,6\}]} & $\underline{E}_1 \partial_{y_4} \phi_1 + \underline{E}_2 \partial_{y_4} \phi_2 - 2 \partial_{y_4} (\underline{E}_2 \phi_2) = 0$ \\ \hline
{[6,3,5]} & $\underline{E}_1 \partial_{y_3} \phi_1 + \underline{E}_2 \partial_{y_3} \phi_2 - 2 \partial_{y_3} (\underline{E}_2 \phi_2) = 0$ \\ \hline
\end{longtable}

\section{Examples}\label{sec:examples}
Let us now analyze several examples with which we can illustrate the methodology of the rigorous symmetry reduction of Lagrangians in various common situations. We will only consider the cases in which PSC is satisfied, i.e., infinitesimal group actions for which the field equations derived from the reduced Lagrangian are fully equivalent to the reduced field equations. (The examples violating PSC1 and/or PSC2 were explored in \cite{Fels:2001rv,Torre:2010xa}.) Moreover, we focus on examples leading to the exact closed-form solutions that are relatively simple and well known (with an exception of possibly new solutions in Sec.~\ref{ssc:541}). 


\subsection{[3,2,3] in general relativity (pure radiation/vacuum) -- Vaidya \& Birkhoff's theorem}

First, we consider the general relativity with fixed $\Gamma$-invariant matter fields, $\mathcal{L}=\frac{1}{2\varkappa} R+\mathcal{L}_{\textrm{m}}$, and the (PSC-compatible) infinitesimal group action [3,2,3] that corresponds to the usual spherical symmetry. The $\Gamma$-invariant metric and $l$-chain are
\begin{equation}
\begin{aligned}
    \hat{\bs{g}} &=- \phi_1(y_1, y_2)\, \bs{\mathrm{d}}{y_1}^2 \allowbreak + \phi_2(y_1, y_2) (\bs{\mathrm{d}}{y_1}\vee\bs{\mathrm{d}}{y_2}) \allowbreak + \phi_3(y_1, y_2)\, \bs{\mathrm{d}}{y_2}^2 \allowbreak + \phi_4(y_1, y_2) (\bs{\mathrm{d}}{y_3}^2 \allowbreak + \sin^2y_3\, \bs{\mathrm{d}}{y_4}^2)\;,
    \\
    \bs{\chi} &=\csc y_3\, \bs{\partial}_{y_3} \wedge \bs{\partial}_{y_4}\;,
\end{aligned}
\end{equation}
where $\phi_i$, ${i=1,2,3,4}$, are $\Gamma$-invariant functions of two variables $y_1$ and $y_2$ that cover the $2$-dimensional reduced spacetime, ${r=2}$. The Lorentzian signature implies that ${\phi_4}$ and ${\phi_1\phi_3+\phi_2^2}$ are positive. By using \eqref{eq:reducedL}, we arrive at the gravitational part of reduced Lagrangian,
\begin{equation}\label{eq:redLag323}
\begin{aligned}
    \hat{\underline{L}}_{\textrm{g}} &=\frac{1}{2\kappa}\frac{1}{\phi_4 \left(\phi_1 \phi_3+\phi_2^2\right)^{3/2}}\bigg[\phi_2^2 \big(2 \phi_4 (\phi_1' \phi_4'-\dot{\phi}_3 \dot{\phi}_4)+\phi_1 (8 \phi_3 \phi_4+\phi_4'^2)-4 \dot{\phi}_2' \phi_4^2-\phi_3 \dot{\phi}_4^2\big)+\phi_2 \Big(\phi_4^2 (-\dot{\phi}_1 \phi_3'+\phi_1' \dot{\phi}_3
    \\
    &\feq+4 \phi_2' \dot{\phi}_2)+2 \phi_4 \big(\phi_3 (\dot{\phi}_1 \phi_4'+\phi_1' \dot{\phi}_4-4 \phi_1 \dot{\phi}_4')+\phi_1 (\dot{\phi}_3 \phi_4'+\phi_3' \dot{\phi}_4)\big)+2 \phi_1 \phi_3 \phi_4' \dot{\phi}_4\Big)-\phi_1 \Big(2 \phi_3 \phi_4 \big(-\phi_4' (\phi_1'-2 \dot{\phi}_2)
    \\
    &\feq +2 \dot{\phi}_2' \phi_4 +\dot{\phi}_4 (2 \phi_2'+\dot{\phi}_3) \big)-2 \dot{\phi}_2 \phi_3' \phi_4^2+\phi_3^2 \dot{\phi}_4^2\Big)+2 \dot{\phi}_1 \phi_2' \phi_3 \phi_4^2+\phi_1^2 \phi_3 (4 \phi_3 \phi_4+\phi_4'^2)+2 \phi_2^3 (\phi_4' \dot{\phi}_4-4 \phi_4 \dot{\phi}_4')
    \\
    &\feq+4 \phi_2^4 \phi_4\bigg]\underline{\mathrm{d}}y_1\wedge \underline{\mathrm{d}}y_2\;,
\end{aligned}
\end{equation}
where $\dot{}:=\partial_{y_1}$ and $':=\partial_{y_2}$. Note that we also added a total derivative in order to simplify the expression by eliminating terms with $\ddot{\phi}_3$, $\ddot{\phi}_4$, $\phi_1''$, and $\phi_4''$. By taking the variation with respect to all $\phi_i$, we obtain the Euler-Lagrange expressions
\begin{equation}
    \begin{aligned}
        \underline{E}_1(\hat{\underline{L}}_{\textrm{g}}) &=\frac{1}{2\varkappa}\frac{4 y_2 \phi_2 \phi_2' }{|\phi_2|^3}\underline{\mathrm{d}}y_1\wedge \underline{\mathrm{d}}y_2\;,
        \\
        \underline{E}_2(\hat{\underline{L}}_{\textrm{g}}) &=-\frac{1}{2\varkappa}\frac{4  }{\phi_2 |\phi_2|}(y_2 \phi_1'+\phi_1-\phi_2^2)\underline{\mathrm{d}}y_1\wedge \underline{\mathrm{d}}y_2\;,
        \\
        \underline{E}_3(\hat{\underline{L}}_{\textrm{g}}) &=\frac{1}{2\varkappa}\frac{2  }{|\phi_2|^3}\big[\phi_1 \big(\phi_2^2-y_2 (\phi_1'+2 \dot{\phi}_2)\big)+y_2 \dot{\phi}_1 \phi_2-\phi_1^2\big] \underline{\mathrm{d}}y_1\wedge \underline{\mathrm{d}}y_2\;,
        \\
        \underline{E}_4(\hat{\underline{L}}_{\textrm{g}}) &=\frac{1}{2\varkappa}\frac{2  }{y_2\phi_2|\phi_2|}\big[\phi_2'\big(2\phi_1+y_2(\phi_1'+2\dot{\phi}_2\big)-\phi_2\big(2\phi_1'+y_2(\phi_1''+2\dot{\phi}_2'\big)\big] \underline{\mathrm{d}}y_1\wedge \underline{\mathrm{d}}y_2\;,
    \end{aligned}
\end{equation}
where we also imposed the gauge condition ${\phi_3=0}$ and ${\phi_4=y_2^2}$ by virtue of the residual diffeomorphisms generated by ${\bs{W}=f_1(y_1,y_2) \, \bs{\partial}_{y_1}}$ and ${\bs{W}=f_2(y_1,y_2) \, \bs{\partial}_{y_2}}$, respectively. We consider the pure-radiation energy-momentum tensor ${\hat{\bs{T}}=\rho\bs{k}^2}$, where $\bs{k}=\bs{\partial}_{y_2}$ is a null vector field, ${\bs{k}^2=0}$, and ${\rho=\rho(y_1,y_2)}$ is a $\Gamma$-invariant scalar function corresponding to the radiation density. After canceling the reduced Levi-Civita tensor  ${\hat{\underline{\epsilon}}=2y_2^2|\phi_2|\underline{\mathrm{d}}y_1\wedge \underline{\mathrm{d}}y_2}$ and recasting $\hat{\bs{T}}$ in terms of the $\Gamma$-invariant $\binom{(2)}{0}$ tensor field base ${\bs{p}_1=-\bs{\partial}_{y_1}^2}$, ${\bs{p}_2=\frac{1}{2}(\bs{\partial}_{y_1}\vee\bs{\partial}_{y_2})}$,  ${\bs{p}_3=\bs{\partial}_{y_2}^2}$, ${\bs{p}_4=\frac{1}{2}\left(\bs{\partial}_{y_3}^2 + (\sin^2 y_3)^{-1}\, \bs{\partial}_{y_4}^2\right)}$ [see \eqref{eq:dual}], i.e., ${\hat{\bs{T}}=\rho\bs{p}_3}$, we can write the field equations \eqref{eq:feqEE} as the following set of partial differential equations:
\begin{equation}\label{eq:323eqns}
    \begin{aligned}
        -2\mathcal{E}_1(\hat{\underline{L}}_{\textrm{g}}) &=-\frac{1}{\varkappa}\frac{2  \phi_2' }{y_2\phi_2^3}=0\;,
        \\
        -2\mathcal{E}_2(\hat{\underline{L}}_{\textrm{g}}) &=\frac{1}{\varkappa}\frac{2  }{y_2^2\phi_2^3}(y_2 \phi_1'+\phi_1-\phi_2^2)=0\;,
        \\
        -2\mathcal{E}_3(\hat{\underline{L}}_{\textrm{g}}) &=-\frac{1}{\varkappa}\frac{1  }{y_2^2\phi_2^4}\big[\phi_1 \big(\phi_2^2-y_2 (\phi_1'+2 \dot{\phi}_2)\big)+y_2 \dot{\phi}_1 \phi_2-\phi_1^2\big] =\rho\;,
        \\
        -2\mathcal{E}_4(\hat{\underline{L}}_{\textrm{g}}) &=-\frac{1}{\varkappa}\frac{1  }{y_2^3\phi_2^3}\big[\phi_2'\big(2\phi_1+y_2(\phi_1'+2\dot{\phi}_2\big)-\phi_{2}\big(2\phi_1'+y_2(\phi_1''+2\dot{\phi}_2'\big)\big]=0\;.
    \end{aligned}
\end{equation}
The general solution of these equations is given by ${\phi_1=\phi_2^2-2m/y_2}$ with ${\rho={(2\dot{m}\phi_2-m\dot{\phi}_2)}{/(\varkappa y_2^2)}}$. Here, ${\phi_2=\phi_2(y_1)}$ and ${m=m(y_1)}$ are arbitrary functions of a single variable $y_1$. One can further gauge fix ${\phi_2=-1}$ by means of ${\bs{W}=\tilde{f}_1(y_1) \, \bs{\partial}_{y_1}}$, and we arrive at ${\phi_1=1-2m/y_2}$ with ${\rho={-2\dot{m}}{/(\varkappa y_2^2)}}$, in which we easily recognize the Vaidya spacetime \eqref{eq:Vaidya} with ${\epsilon=1}$.

Let us repeat the derivation for the vacuum case, ${\rho=0}$, which allows us to make an interesting simplification. Provided that ${\partial_{y_2}\phi_4\neq0}$, the second Noether identity leads to the following implication: 
\begin{equation}
    \underline{E}_1=\underline{E}_2=\underline{E}_3=0 \implies \underline{E}_4=0\;.
\end{equation}
This tells us that varying $\phi_4$ only provides a redundant equation, so we can set ${\phi_4=y_2^2}$ already in the reduced Lagrangian,
\begin{equation}\label{eq:323Lagrvac}
\begin{aligned}
    \hat{\underline{L}} &=\frac{1}{2\varkappa}\frac{1}{\left(\phi_1 \phi_3+\phi_2^2\right)^{3/2}}\Big[y_2 \phi_2 (-y_2 \dot{\phi}_1 \phi_3'+y_2 \phi_1' \dot{\phi}_3+4 \dot{\phi}_1 \phi_3+4 y_2 \phi_2' \dot{\phi}_2)+2 y_2^2 \dot{\phi}_1 \phi_2' \phi_3-4 y_2 \phi_2^2 (\phi_1'+y_2 \dot{\phi}_2')+4 \phi_2^4
    \\
    &\feq+2 \phi_1 \Big(2 y_2 \phi_2 (2 \phi_2'+\dot{\phi}_3)+y_2 \big(\dot{\phi}_2 (y_2 \phi_3'-4 \phi_3)-2 y_2 \dot{\phi}_2' \phi_3\big)+\phi_2^2 (4 \phi_3-2)\Big)+4 \phi_1^2 \big(y_2 \phi_3'+(\phi_3-1) \phi_3\big)\Big] \underline{\mathrm{d}}y_1\wedge \underline{\mathrm{d}}y_2\;.
\end{aligned}
\end{equation}
On the other hand, the gauge choice ${\phi_3=0}$ is only possible after the variation; this fact was noticed before, e.g., in \cite{Deser:2004gi} but was not sufficiently justified there. By taking the variation of \eqref{eq:323Lagrvac} with respect to $\phi_1$, $\phi_2$, and $\phi_3$ we eventually obtain the first three equations in \eqref{eq:323eqns} with ${\rho=0}$, whose general solution, upon the same gauge fixing ${\phi_2=-1}$, is again ${\phi_1=1-2m/y_2}$, but with $m$ being a constant this time. Therefore, we recover the Schwarzschild spacetime \eqref{eq:schw} with ${\epsilon=1}$ (in Eddington-Finkelstein coordinates) as the only spherically symmetric vacuum solution of general relativity, which is in agreement with the Birkhoff theorem.


\subsection{[4,3,3] in general relativity (vacuum) -- Schwarzschild}\label{scc:433}
It is also instructive to go through the seminal example of symmetry reduction by [4,3,3] (first performed in \cite{Weyl:1917gp}), where, in addition to the spherical symmetry, the stationarity is also assumed from the outset. We will restrict ourselves to the general relativity in vacuum, ${\mathcal{L}=\frac{1}{2\varkappa} R}$. The $\Gamma$-invariant metric and $l$-chain is now given by
\begin{equation}
\begin{aligned}
    \hat{\bs{g}} &=- \phi_1(y_2)\, \bs{\mathrm{d}}{y_1}^2 \allowbreak + \phi_2(y_2) (\bs{\mathrm{d}}{y_1}\vee\bs{\mathrm{d}}{y_2}) \allowbreak + \phi_3(y_2)\, \bs{\mathrm{d}}{y_2}^2 \allowbreak + \phi_4(y_2) (\bs{\mathrm{d}}{y_3}^2 \allowbreak + \sin^2y_3\, \bs{\mathrm{d}}{y_4}^2)\;,
    \\
    \bs{\chi} &=\csc y_3\, \bs{\partial}_{y_1} \wedge \bs{\partial}_{y_3} \wedge \bs{\partial}_{y_4}\;,
\end{aligned}
\end{equation}
where $\phi_i$, ${i=1,2,3,4}$, are $\Gamma$-invariant functions of a single variable $y_2$ that cover the $1$-dimensional reduced spacetime, and again the Lorentzian signature implies that ${\phi_4}$ and ${\phi_1\phi_3+\phi_2^2}$ are positive. The dimension of the reduced spacetime, ${r=1}$, is a significant simplification (when compared to [3,2,3]) because the field equations ${\underline{E}_i=0}$ now become a set of ordinary differential equations (rather than partial differential equations). The residual diffeomorphism generators ${\bs{W}=f_1(y_2)\bs{\partial}_{y_1}}$ and ${\bs{W}=f_2(y_2)\bs{\partial}_{y_2}}$ allow us to make the standard gauge choice ${\phi_2=0}$ and ${\phi_4=y_2^2}$ (with ${y_2>0}$), respectively. Nevertheless, the question remains whether we can do that already at the level of the reduced Lagrangian. The first Noether identity implies that the component of the 1-form ${\underline{E}_2 \phi_1 -  2 \underline{E}_3  \phi_2}$ (without the coordinate volume form $\underline{\mathrm{d}}y_2$) is a constant, which we denote by $\kappa$, that only depends on the theory (the Lagrangian $\underline{L}$) but it is independent of ${\mathrm{x}\in M}$ as well as of $\phi_i$.\footnote{This follows directly from the chain-rule expansion of the Noether identity, which holds for every $\phi_k$, and the fact that it the depends on $y_2$ (also) through $\phi_k$ and its derivatives. Since they all can be specified arbitrarily, the coefficients must vanish separately.} The diffeomorphism generator ${\bs{W}=y_1 \bs{\partial}_{y_1}}$ (i.e., constant rescaling of $y_1$) induces a nontrivial scaling transformation of $\kappa$, which is impossible unless $\kappa$ vanishes identically, therefore this Noether identity simplifies to ${\underline{E}_2 \phi_1 -  2 \underline{E}_3 \phi_2=0}$. Together with the one remaining Noether identity, it gives us, for example, the implication (alternative implications are also possible)
\begin{equation}
    \underline{E}_1=\underline{E}_3=0 \implies \underline{E}_2=\underline{E}_4=0\;,
\end{equation}
provided that ${\phi_1\neq0}$ and ${\partial_{y_2}\phi_4\neq0}$. This means that the field equations ${\underline{E}_2=\underline{E}_4=0}$ are redundant and this is true irrespective of the theory. Therefore, we do not need to vary ${\phi_2}$ and ${\phi_4}$, so we can simply set ${\phi_2=0}$ and ${\phi_4=y_2^2}$ already in the reduced Lagrangian,
\begin{equation}
    \hat{\underline{L}}=-\frac{1}{2\varkappa}\frac{12 \phi_1^2 \left(y_2 \phi_3'+(\phi_3-1) \phi_3\right)}{\left( \phi_1 \phi_3\right)^{3/2}}\underline{\mathrm{d}}y_2\;,
\end{equation}
where $':=\partial_{y_2}$ and we also added a total derivative to eliminate $\phi_1''$. A similar line of arguments for gauge fixing (in this specific case) appeared in some form already in \cite{Lovelock1973SphericallySM,Anderson:1999cn}. All that is left for us to calculate is the variation with respect to $\phi_1$ and $\phi_3$, which gives us the field equations,
\begin{equation}
\begin{aligned}
    \underline{E}_1 &=-\frac{1}{2\varkappa}\frac{6 \phi_1 \left(y_2 \phi_3'+(\phi_3-1) \phi_3\right)}{\left( \phi_1 \phi_3\right)^{3/2}}\underline{\mathrm{d}}y_2=0\;,
    \\
    \underline{E}_3 &=\frac{1}{2\varkappa}\frac{6 \phi_1 \left(y_2 \phi_1'-(\phi_3-1)\phi_1 \right)}{\left( \phi_1 \phi_3\right)^{3/2}}\underline{\mathrm{d}}y_2=0\;.
\end{aligned}
\end{equation}
After solving the first equation, ${\phi_3=(1-2m/y_2)^{-1}}$, with $m$ being a constant, we substitute into the second one and find ${\phi_1=1-2m/y_2}$, where we already fixed the overall constant by constant rescaling of $y_1$ via ${\bs{W}=y_1 \, \bs{\partial}_{y_1}}$. This provides one of the fastest (but still rigorous) derivations of the Schwarzschild spacetime \eqref{eq:schw} with ${\epsilon=1}$.


\subsection{[4,3,4] in general relativity (vacuum) -- Taub-NUT}

Although [4,3,4] is the same abstract algebra as [4,3,3], i.e., $\mathrm{so(3)\oplus n(1,1)}$, its action on $M$ is very different. Already its ``so(3) part'' acts on the 3-dimensional orbits rather then the 2-dimensional ones, cf. [3,3,9] and [3,2,3]. To better compare these two cases, we again focus on the general relativity in vacuum, ${\mathcal{L}=\frac{1}{2\varkappa} R}$. The $\Gamma$-invariant metric and $l$-chain read
\begin{equation}
\begin{aligned}
    \hat{\bs{g}} &=-\phi_1(y_2) (\bs{\mathrm{d}}{y_1} \allowbreak - \cos y_3\, \bs{\mathrm{d}}{y_4})^2 \allowbreak + \phi_2(y_2) \bigl(\bs{\mathrm{d}}{y_2}\vee(\bs{\mathrm{d}}{y_1} \allowbreak - \cos y_3\, \bs{\mathrm{d}}{y_4})\bigr) \allowbreak + \phi_3(y_2)\, \bs{\mathrm{d}}{y_2}^2 \allowbreak + \phi_4(y_2) (\bs{\mathrm{d}}{y_3}^2 \allowbreak + \sin^2y_3\, \bs{\mathrm{d}}{y_4}^2)\;,
    \\
    \bs{\chi} &=\csc y_3\, \bs{\partial}_{y_1} \wedge \bs{\partial}_{y_3} \wedge \bs{\partial}_{y_4}\;,
\end{aligned}
\end{equation}
where $\phi_i$, ${i=1,2,3,4}$, are $\Gamma$-invariant functions of $y_2$ again, and the Lorentzian signature leads to positive ${\phi_4}$ and ${\phi_1\phi_3+\phi_2^2}$. The Noether identities remain unchanged for this case, but the important difference from [4,3,4] is the absence of any residual diffeomorphism generator corresponding to scaling transformation, see Tab.~\ref{tab:residualgenerators}. This means that we cannot remove the derivative in the first Noether identity in a complete generality  (like we did in [4,3,3]), since the constant $\kappa$ may not be zero; furthermore, its value is theory dependent. However, $\kappa$ is independent of $\phi_k$. Therefore, if there exists any solution of a given theory within [4,3,4], ${\underline{E}_i=0}$, then ${\kappa=0}$ irrespective of $\phi_k$ and then again ${\underline{E}_2 \phi_1  -  2 \underline{E}_3 \phi_2=0}$ for that theory. In our case, this is somewhat trivial because we already know from the literature that the Taub-NUT spacetime is a vacuum solution of general relativity. Nevertheless, all we need to know is the mere \textit{existence} of a solution (not the specific form of $\phi_k$) and only of a \textit{single} representative (not the general solution; massless Taub-NUT with a specific value of the NUT parameter suffices).\footnote{Note that neither the Schwarzschild nor the Minkowski spacetimes can be used in this argument as they do not belong to [4,3,4]. The metric $\hat{\bs{g}}$ becomes flat for ${\phi_1=-\phi_4}$, ${\phi_3=(\phi_2^2+\phi_4'^2)/\phi_4}$, but it has Euclidean signature.} The improved Noether identity together with the remaining one then imply, for example,
\begin{equation}
    \underline{E}_3=\underline{E}_4=0 \implies \underline{E}_1=\underline{E}_2=0\;,
\end{equation}
provided that ${\phi_1\neq0}$ and ${\partial_{y_2}\phi_1\neq0}$. (For convenience, we have chosen a different implication compared to [4,3,3], but both are possible in both cases.) In contrast to the theory-independent arguments we made in [4,3,3], however, the above statements are only valid for theories that admit vacuum solutions (within [4,3,4]).  After fixing ${\phi_2=0}$ and ${\phi_1=(2l)^2/\phi_3}$ (for some constant ${l>0}$ and assuming $\phi_3$ to be monotonic) by means of ${\bs{W}=f_1(y_2)\bs{\partial}_{y_1}}$ and ${\bs{W}=f_2(y_2)\bs{\partial}_{y_2}}$, respectively, in the reduced Lagrangian, we get
\begin{equation}
    \hat{\underline{L}}=\frac{1}{2\varkappa}\frac{6 l  }{\phi_3^2 \phi_4}\big[4\phi_3 \left( l^2 + \phi_3 \phi_4\right) -2\phi_4 \phi_3'\phi_4'+ \phi_3 \phi_4'^2 \big]\underline{\mathrm{d}}y_2\;,
\end{equation}
where again ${':=\partial_{y_2}}$; this time we added a total derivative to eliminate $\phi_3''$ and $\phi_4''$. Upon taking the variation with respect to $\phi_3$ and $\phi_4$ we get the field equations
\begin{equation}
    \begin{aligned}
        \underline{E}_3 &=-\frac{1}{2\varkappa}\frac{6 l  }{\phi_3^2 \phi_4}\big[4 l^2-2 \phi_4 \phi_4''+\phi_4'^2\big]\underline{\mathrm{d}}y_2 = 0\;,
        \\
        \underline{E}_4 &=\frac{1}{2\varkappa}\frac{6 l  }{\phi_3^3 \phi_4^2}\big[\phi_3^2 \left(-4 l^2-2 \phi_4 \phi_4''+\phi_4'^2\right)-4 \phi_4^2 \phi_3'^2+2 \phi_3 \phi_4 \left(\phi_4 \phi_3''+\phi_3' \phi_4'\right)\big]\underline{\mathrm{d}}y_2 = 0\;.
    \end{aligned}
\end{equation}
Finding the general solution is straightforward; its special case is the Taub-NUT spacetime \eqref{eq:TaubNUT} with ${\epsilon=1}$, for which ${\phi_4=y_2^2+l^2}$, ${\phi_3=(y_2^2+l^2)/(y_2^2-l^2-2m y_2)}$.


\subsection{[4,3,6] in conformal gravity (vacuum) -- conformally Einstein}\label{scc:436}

To demonstrate the usefulness of symmetry reduction in higher-derivative theories, let us take the conformal gravity in vacuum, ${\mathcal{L}=\bs{C}^2}$, with $\bs{C}$ being the Weyl tensor. We will consider the infinitesimal group action [4,3,6], which combines the planar symmetry with stationarity (i.e., the planar analogue of [4,3,3]). The $\Gamma$-invariant metric and the $l$-chain are
\begin{equation}
\begin{aligned}
    \hat{\bs{g}} &= - \phi_1(y_2)\, \bs{\mathrm{d}}{y_1}^2 \allowbreak + \phi_2(y_2) (\bs{\mathrm{d}}{y_1}\vee\bs{\mathrm{d}}{y_2}) \allowbreak + \phi_3(y_2)\, \bs{\mathrm{d}}{y_2}^2 \allowbreak + \phi_4(y_2) (\bs{\mathrm{d}}{y_3}^2 \allowbreak +\, \bs{\mathrm{d}}{y_4}^2)\;,
    \\
    \bs{\chi} &= \bs{\partial}_{y_1} \wedge\, \bs{\partial}_{y_3} \wedge \bs{\partial}_{y_4}\;,
\end{aligned}
\end{equation}
with $\phi_i$, ${i=1,2,3,4}$, being $\Gamma$-invariant functions of $y_2$ again; the Lorentzian signature implies that ${\phi_4}$ and ${\phi_1\phi_3+\phi_2^2}$ are positive. Exactly along the lines of [4,3,3], the diffeomorphism generators ${\bs{W}=f_1(y_2)\bs{\partial}_{y_1}}$ and ${\bs{W}=f_2(y_2)\bs{\partial}_{y_2}}$ together with their Noether identities improved by the scaling transformation generated by ${\bs{W}=y_1 \bs{\partial}_{y_1}}$ allow us to fix ${\phi_3=0}$ and ${\phi_4=y_2^2}$, ${y_2>0}$ in the reduced Lagrangian (or any other theory), which now reads
\begin{equation}
    \hat{\underline{L}} = \frac{2}{y_2^2 |\phi_2|\phi_2^4 }\big[y_2 \left(y_2 \phi_2 \phi_1''-\phi_1' \left(y_2 \phi_2'+2 \phi_2\right)\right)+2 \phi_1 \left(y_2 \phi_2'+\phi_2\right)\big]^2\underline{\mathrm{d}}y_2\;.
\end{equation}
By taking the variation with respect to $\phi_1$ and $\phi_2$, we obtain the field equations
\begin{equation}
    \begin{aligned}
        \underline{E}_1 &=\frac{4 }{ y_2|\phi_2|\phi_2^5} \Big[y_2 \Big(-15 y_2^2 \phi_1' \phi_2'^3+y_2 \left(y_2 \phi_1''''+4 \phi_1'''\right) \phi_2^3+5 y_2 \phi_2 \phi_2' \left(3 y_2 \phi_1'' \phi_2'-2 \phi_1' \left(\phi_2'-y_2 \phi_2''\right)\right)
        \\
        &\feq+\phi_2^2 \big(-6 y_2^2 \phi_1''' \phi_2'-4 y_2 \phi_1'' \left(y_2 \phi_2''+2 \phi_2'\right)+\phi_1' (-y_2^2\phi_2'''+2 y_2 \phi_2''+10 \phi_2')\big)\Big)
        \\
        &\feq+2 \phi_1 \big(15 y_2^2 \phi_2'^3-5 y_2 \phi_2 \phi_2' \left(2 y_2 \phi_2''+\phi_2'\right)+\phi_2^2 (y_2^2 \phi_2'''+2 y_2 \phi_2''-2 \phi_2')\big)\Big] \underline{\mathrm{d}}y_2 = 0\;,
        \\
        \underline{E}_2 &=-\frac{2}{ y_2^4 |\phi_2|\phi_2^5} \Big[y_2^2 \Big(-5 y_2^2 \phi_1'^2 \phi_2'^2+2 y_2 \phi_2 \phi_1' \left(y_2 \phi_1' \phi_2''-2 \phi_2' \left(\phi_1'-y_2 \phi_1''\right)\right)+\phi_2^2 \big(\left(y_2 \phi_1''-2 \phi_1'\right)^2-2 y_2^2 \phi_1''' \phi_1'\big)\Big)
        \\
        &\feq+4 y_2 \phi_1 \Big(5 y_2^2 \phi_1' \phi_2'^2-2 y_2 \phi_2 \big(y_2 \phi_1' \phi_2''+\phi_2' (y_2 \phi_1''-\phi_1')\big)+\phi_2^2 \big(y_2 (y_2 \phi_1'''+\phi_1'')-2 \phi_1'\big)\Big)
        \\
        &\feq+4 \phi_1^2 (2 y_2^2 \phi_2 \phi_2''-5 y_2^2 \phi_2'^2+\phi_2^2)\Big] \underline{\mathrm{d}}y_2 = 0\;.
    \end{aligned}
\end{equation}
Finding a general solution seems difficult, but we can confirm that, for example, the planar Schwarzschild ${\phi_1=-2m/y_2}$, ${\phi_2=-1}$, see \eqref{eq:schw} with ${\epsilon=0}$ (rewritten in Eddington-Finkelstein coordinates), as well as the metric with ${\phi_1=y_2^8}$, ${\phi_2=-4y_2^3/b}$ from \cite{Liu:2013fna} are solutions of the above field equations. (Note that both examples are rather simple in a sense that every conformally Einstein metric is known to be a vacuum solution of the conformal gravity; the latter example, however, is not a vacuum solution of the general relativity.)


\subsection{[4,3,9] in general relativity (vacuum) -- NHEK}
Now, we wish to explore an interesting case of the enhanced so(2,1) symmetry [4,3,9] in the general relativity in vacuum, ${\mathcal{L}=\frac{1}{2\varkappa}R}$. Here, the $\Gamma$-invariant metric and $l$-chain are given by
\begin{equation}
\begin{aligned}
    \hat{\bs{g}} &=\phi_1(y_4) (-\cosh^2y_2\, \bs{\mathrm{d}}{y_1}^2 \allowbreak + \, \bs{\mathrm{d}}{y_2}^2) \allowbreak + \phi_2(y_4) (\sinh y_2\, \bs{\mathrm{d}}{y_1} \allowbreak +\, \bs{\mathrm{d}}{y_3})^2 \allowbreak + \phi_3(y_4) \bigl(\bs{\mathrm{d}}{y_4}\vee(\sinh y_2\, \bs{\mathrm{d}}{y_1} \allowbreak +\, \bs{\mathrm{d}}{y_3})\bigr) \allowbreak + \phi_4(y_4)\, \bs{\mathrm{d}}{y_4}^2\;,
    \\
    \bs{\chi} &=\sech y_2\, \bs{\partial}_{y_1} \wedge \bs{\partial}_{y_2} \wedge \bs{\partial}_{y_3}\;,
\end{aligned}
\end{equation}
where $\phi_i$, ${i=1,2,3,4}$, are $\Gamma$-invariant functions of $y_4$. Here, the Lorentzian signature leads to positiveness of $\phi_2$, $\phi_4$, and ${\phi_2\phi_4-\phi_3^2}$. The problem of gauge fixing is somewhat similar to that of [4,3,4]. In analogy to [4,3,\{3,4,6\}] discussed above, we denote the constant coming from the first Noether identity, i.e., the component of the 1-form ${\underline{E}_3 \phi_2 + 2 \underline{E}_4 \phi_3}$, by $\lambda$. Similar to [4,3,4], the Noether identities do not allow us to gauge fix at the level of the Lagrangian in general because $\lambda$ may not be zero for every theory. However, the mere existence of the NHEK spacetime as a vacuum solution of general relativity implies ${\lambda=0}$.\footnote{Again, the Minkowski spacetime cannot be used here because it is does not belong to [4,3,9]. The metric $\hat{\bs{g}}$ becomes flat for ${\phi_3=0}$, ${\phi_1=\phi_2}$, ${\phi_4=-(\phi_2')^2/\phi_2}$, which has the Euclidean signature.} As a consequence of ${\underline{E}_3 \phi_2 + 2 \underline{E}_4 \phi_3=0}$ and the remaining Noether identity, we obtain, for example,
\begin{equation}
    \underline{E}_1=\underline{E}_4=0 \implies \underline{E}_2=\underline{E}_3=0\;,
\end{equation}
if ${\phi_2\neq0}$ and ${\partial_{y_4}\phi_2\neq0}$. Due to redundancy of ${\underline{E}_2=\underline{E}_3=0}$, the gauge fixing of $\phi_2$ and  $\phi_3$ is possible at the level of the reduced Lagrangian. As before this holds true only in theories that admit vacuum solutions within [4,3,9]. Upon setting ${\phi_3=0}$ and ${\phi_2=(2a)^2/\phi_4}$ (for some constant ${a>0}$ and assuming $\phi_4$ to be monotonic) by means of ${\bs{W}=f_1(y_4)\, \bs{\partial}_{y_3}}$ and  ${\bs{W}=f_2(y_4)\, \bs{\partial}_{y_4}}$, respectively, and adding a total derivative to remove $\phi_1''$ and $\phi_4''$, the reduced Lagrangian reads
\begin{equation}
    \hat{\underline{L}}=\frac{1}{2\varkappa}\frac{12 a  }{|\phi_1| \phi_4^2}\left[\phi_4 \left(4 a^2-4 \phi_1 \phi_4+\phi_1'^2\right)-2 \phi_1 \phi_1' \phi_4'\right]\underline{\mathrm{d}}y_4\;,
\end{equation}
where ${':=\partial_{y_4}}$. By calculating the variations with respect to $\phi_1$ and $\phi_4$, we obtain the field equations
\begin{equation}
\begin{aligned}
    \underline{E}_1 &=\frac{1}{2\varkappa}\frac{12 a }{|\phi_1|\phi_1 \phi_4^3}\left[\phi_4^2 \left(-4 a^2-2 \phi_1 \phi_1''+\phi_1'^2\right)+2 \phi_1 \phi_4 \left(\phi_1' \phi_4'+\phi_1 \phi_4''\right)-4 \phi_1^2 \phi_4'^2\right]\underline{\mathrm{d}}y_4=0\;,
    \\
    \underline{E}_4 &=-\frac{1}{2\varkappa}\frac{12 a }{|\phi_1| \phi_4^2}\left(4 a^2-2 \phi_1 \phi_1''+\phi_1'^2\right)\underline{\mathrm{d}}y_4=0\;.
\end{aligned}
\end{equation}
The general solution can be found straightforwardly; it contains ${\phi_1=a^2+y_4^2}$, ${\phi_4=(a^2+y_4^2)/(a^2-y_4^2)}$, as a special case, which reproduces the NHEK spacetime \eqref{eq:NHEK}.


\subsection{[5,4,1] in quadratic gravity (vacuum) -- G\"odel-like}\label{ssc:541}

Illustrating the symmetry reduction in the homogeneous spacetimes (${l=4}$), we will consider [5,4,1], which stands out as the symmetries of the G\"odel spacetime \eqref{eq:Godel}. However, we will search for vacuum solutions in a generic class of quadratic curvature theories, ${\mathcal{L}=\alpha(R-2\Lambda)+\beta R^2 + \gamma \bs{C}^2}$, with arbitrary constants $\alpha$, $\beta$, and $\gamma$. The $\Gamma$-invariant metric and $l$-chain are
\begin{equation}
\begin{aligned}
    \hat{\bs{g}} &=-\phi_1 (\bs{\mathrm{d}}{y_1} \allowbreak - \cosh y_3\, \bs{\mathrm{d}}{y_4})^2 \allowbreak + \phi_2 \bigl(\bs{\mathrm{d}}{y_2}\vee(\bs{\mathrm{d}}{y_1} \allowbreak - \cosh y_3\, \bs{\mathrm{d}}{y_4})\bigr) \allowbreak + \phi_3\, \bs{\mathrm{d}}{y_2}^2 \allowbreak + \phi_4 (\bs{\mathrm{d}}{y_3}^2 \allowbreak + \sinh^2y_3\, \bs{\mathrm{d}}{y_4}^2)\;,
    \\
    \bs{\chi} &= \csch y_3\, \bs{\partial}_{y_1} \wedge \bs{\partial}_{y_2} \wedge \bs{\partial}_{y_3} \wedge \bs{\partial}_{y_4}\;,
\end{aligned}
\end{equation}
where $\phi_i$, ${i=1,2,3,4}$, are just constants since the reduced spacetime is $0$-dimensional (a point), ${r=0}$. Despite this degeneracy, the symmetry reduction makes a perfect sense and becomes very tractable. The reduced Lagrangian is no longer a form that is a function of fields and their derivatives, but only a scalar (0-form) function of ordinary variables~$\phi_i$,
 \begin{equation}
     \hat{L}=\frac{4   \sqrt{\phi_1 \phi_3+\phi_2^2} }{\phi_4^3}\left[6 \alpha  \phi_4^2 (\phi_1-4 \phi_4 (\Lambda  \phi_4+1))+3 \beta  (\phi_1-4 \phi_4)^2+16 \gamma  (\phi_1-\phi_4)^2\right]\;.
 \end{equation}
The corresponding equations can be obtained by ordinary differentiation with respect to $\phi_1$, $\phi_2$, $\phi_3$, and $\phi_4$; they are no longer differential equations for fields but only algebraic equations for the values $\phi_i$. Specifically, we get
\begin{equation}
\begin{aligned}
        E_1 &=\frac{2}{ \phi_4^3\sqrt{\phi_1}}\Big[6 \alpha  \phi_4^2 \left(3 \phi_1-4 \phi_4 (\Lambda  \phi_4+1)\right)+3 \beta  (\phi_1-4 \phi_4) \left(5 \phi_1-4 \phi_4\right)+16 \gamma  (\phi_1-\phi_4)(5\phi_1 - \phi_4)\Big]=0\;,
        \\
        E_3 &=\frac{2 \phi_1 }{ \phi_4^3\sqrt{\phi_1}}\left[6 \alpha  \phi_4^2 (\phi_1-4 \phi_4 (\Lambda  \phi_4+1))+3 \beta  (\phi_1-4 \phi_4)^2+16 \gamma  (\phi_1-\phi_4)^2\right]=0\;,
        \\
        E_4 &=-4\frac{\sqrt{\phi_1}}{\phi_4^4} \left[6 \alpha  \phi_4^2 \left(4 \Lambda  \phi_4^2+\phi_1\right)+3 \beta  (\phi_1-4 \phi_4) (3 \phi_1-4 \phi_4)+16 \gamma  \left(\phi_1-\phi_4\right)\left(3\phi_1-\phi_4\right)\right]=0\;,
\end{aligned}
\end{equation}
where we set ${\phi_2=0}$ and ${\phi_3=1}$ by means of ${\bs{W} = y_2 \bs{\partial}_{y_1}}$ and ${\bs{W} = y_2 \bs{\partial}_{y_2}}$, respectively. We omitted the equation ${E_2=0}$ as it was identically satisfied by this gauge choice. The form of $E_2$ and $E_3$ before our gauge fixing hint at the existence of algebraic identity ${-E_2\phi_1+2E_3\phi_2=0}$, which is a consequence of invariance of $\hat{L}$ under the variation $\delta\phi_i$ that is induced by ${\bs{W} = y_2 \bs{\partial}_{y_1}}$. By solving this set of equations, one may found the following vacuum solutions:
i) ${\phi_1=\phi_4}$ (a conformally flat non-Einstein spacetime) either for a theory with ${\alpha=\beta=0}$, ${\gamma\neq0}$ or with ${\alpha+8\beta\Lambda=0}$, but then ${\phi_4=-3/(8\Lambda)}$  ii) ${\phi_1=4\phi_4}$ (a Ricci-scalar flat spacetime) for a theory with ${\alpha=\gamma=0}$, ${\beta\neq0}$ iii) ${\phi_1=4\phi_4(1+2\Lambda\phi_4)}$ for a theory with ${\gamma=0}$, ${\alpha+8\beta\Lambda=0}$.


\subsection{[6,3,2] in general relativity (perfect fluid/vacuum) -- flat FLRW \& de Sitter}
Since many applications of symmetry reduction of Lagrangian have been considered in the context of cosmology (e.g. \cite{Hawking:1968zw,Maccallum:1972er}), it is worth revisiting the simplest case of the flat spatial curvature [6,3,2] in the general relativity with a cosmological constant, ${\mathcal{L}=\frac{1}{2\varkappa}(R-2\Lambda)+\mathcal{L}_{\textrm{m}}}$. The $\Gamma$-invariant metric and $l$-chain are given by
\begin{equation}
\begin{aligned}
    \hat{\bs{g}} &=- \phi_1(y_1)\, \bs{\mathrm{d}}{y_1}^2 \allowbreak + \phi_2(y_1) (\bs{\mathrm{d}}{y_2}^2 \allowbreak +\, \bs{\mathrm{d}}{y_3}^2 \allowbreak +\, \bs{\mathrm{d}}{y_4}^2)\;,
    \\
    \bs{\chi} &=\bs{\partial}_{y_2} \wedge\, \bs{\partial}_{y_3} \wedge \bs{\partial}_{y_4}\;.
\end{aligned}
\end{equation}
The fact that there are more Killing vector fields, ${d=6}$, has a pleasant consequence that $\hat{\bs{g}}$ contains only two $\Gamma$-invariant functions of $y_4$, namely $\phi_1$ and $\phi_2$. Furthermore, the Lorentzian signature implies that ${\phi_1}$ and $\phi_2$ are positive. The gravitational part of the reduced Lagrangian is
\begin{equation}
    \hat{\underline{L}}_{\textrm{g}}=-\frac{1}{2\varkappa}\frac{3(3\dot{\phi}_2^2+4\Lambda \phi_1\phi_2^2) }{\sqrt{\phi_1 \phi_2}}\underline{\mathrm{d}}y_1\;,
\end{equation}
where we denoted ${\dot{}:=\partial_{y_1}}$ and added a total derivative to remove $\ddot{\phi}_2$. By taking variation with respect to $\phi_1$ and $\phi_2$, we obtain the Euler-Lagrange expressions,
\begin{equation}
\begin{aligned}
    \underline{E}_1(\hat{\underline{L}}_{\textrm{g}}) &=\frac{1}{2\varkappa}\frac{3(3 \dot{\phi}_2^2-4\Lambda\phi_2^2) }{2 \sqrt{\phi_2}}\underline{\mathrm{d}}y_1\;,
    \\
    \underline{E}_2(\hat{\underline{L}}_{\textrm{g}}) &=-\frac{1}{2\varkappa}\frac{9   }{2 \phi_2^{3/2}}\big(-4 \phi_2   \ddot{\phi}_2+\dot{\phi}_2^2+4\Lambda\phi_2^2\big)\underline{\mathrm{d}}y_1\;,
\end{aligned}
\end{equation}
where we imposed ${\phi_1=1}$ by means of ${\bs{W}=f_1(y_1)\, \bs{\partial}_{y_1}}$, which corresponds to reparametrization of $y_1$. The energy-momentum tensor of the perfect fluid is ${\hat{\bs{T}}=(\rho+p)\bs{u}\bs{u}+p\hat{\bs{g}}^{-1}}$, where ${\bs{u}=\bs{\partial}_t}$ is a timelike vector field, ${\bs{u}^2=-1}$, and $\rho$ and $p$ are $\Gamma$-invariant functions characterizing the energy density and pressure. As before, we cancel the reduced Levi-Civita tensor, ${\hat{\underline{\epsilon}}= 6 \phi_2^{3/2} \underline{\mathrm{d}}y_1}$, and rewrite $\hat{\bs{T}}$ in terms of $\Gamma$-invariant $\binom{(2)}{0}$ tensor field base $\bs{p}_1=-\bs{\partial}_{y_1}^2$, $\bs{p}_2=\frac{1}{3}(\bs{\partial}_{y_2}^2+\bs{\partial}_{y_3}^2+\bs{\partial}_{y_4}^2)$, i.e., ${\hat{\bs{T}}=-\rho\bs{p}_1+\frac{3p}{\phi_2}\bs{p}_2}$. We arrive at the field equations 
\begin{equation}\label{eq:feqFLRW}
\begin{aligned}
    -2\mathcal{E}_1(\hat{\underline{L}}_{\textrm{g}}) &=-\frac{1}{\varkappa}\frac{3 \dot{\phi}_2^2-4\Lambda\phi_2^2}{4 \phi_2^2}=-\rho\;,
    \\
    -2\mathcal{E}_2(\hat{\underline{L}}_{\textrm{g}}) &=\frac{1}{\varkappa}\frac{3   }{4 \phi_2^3}\big(-4 \phi_2\ddot{\phi}_2+ \dot{\phi}_2^2+4\Lambda\phi_2^2\big)=\frac{3p}{\phi_2}\;,
\end{aligned}
\end{equation}
which are nothing but the \textit{Friedmann equations} after denoting ${\phi_2=a^2}$ (see, e.g., \cite{Griffiths:2009dfa}). Notice the two equations are related via the reduced conservation law, ${\hat{\bs{\nabla}}\cdot\hat{\bs{T}}=0}$, which corresponds to ${3\dot{\phi}_2(\rho+p)+2\phi_2\dot{\rho}=0}$\;.

Let us also consider the vacuum case, ${\rho=p=0}$, in which we can further simplify the process of symmetry reduction. The Noether identity gives us the implication 
\begin{equation}
    \underline{E}_1=0 \implies \underline{E}_2=0
\end{equation} 
as long as ${\partial_{y_1}\phi_2\neq0}$. Hence, the equation ${\underline{E}_2=0}$ is redundant and we may freely set $\phi_2$ to any non-constant function by means of ${f_1(y_1)\, \bs{\partial}_{y_1}}$ already in the reduced Lagrangian. We will set ${\phi_2=\exp({2 \sqrt{|\Lambda| /3} y_1})}$ as it gives us the reduced Lagrangian in a particularly simple form, 
\begin{equation}
    \hat{\underline{L}}_{\textrm{g}}=-\frac{1}{2\varkappa}\frac{12 e^{\sqrt{3|\Lambda|}y_1}(|\Lambda|+\Lambda\phi_1)}{\sqrt{\phi_1}}\underline{\mathrm{d}}y_1\;.
\end{equation}
By variation with respect to $\phi_1$ we obtain the field equation
\begin{equation}
    \underline{E}_1(\hat{\underline{L}}_{\textrm{g}}) =\frac{1}{2\varkappa}\frac{6 e^{\sqrt{3|\Lambda|}y_1}(|\Lambda|-\Lambda\phi_1)}{\phi_1^{3/2}}\underline{\mathrm{d}}y_1=0\;.
\end{equation}
Since ${\phi_1}$ has to be positive, the solution only exists for ${\Lambda>0}$ and it is given by ${\phi_1=1}$, in which we easily recognize the de~Sitter spacetime.

\section{Summary}\label{sec:concl}

In the present paper, we provided a comprehensive cookbook on symmetry reduction of Lagrangians for any 4-dimensional metric theory of gravity. Building on the Hicks classification, we identified and catalogued all 44 possible infinitesimal group actions that respect PSC, ensuring the equivalence of the field equations of the reduced Lagrangian with the reduced field equations. We calculated all necessary ingredients for a successful and rigorous symmetry reduction of Lagrangians, such as the $\Gamma$-invariant metrics and $l$-chains. We also clarified the issue of imposing gauge conditions (through the residual diffeomorphisms) at the level of the Lagrangian, which is only feasible if the Noether identities render some field equations redundant. The infinitesimal group actions have been recast into new coordinates and vector-field bases that highlight well-known structures such as the mutual relations among the symmetries and the well-known spacetimes that possess them, see Fig.~\ref{fig:actionspider}; this aids in navigating all possible symmetries allowing for PSC. On top of that, we demonstrated the appropriate treatment of symmetry reduction of Lagrangians on a series of examples while deriving known (and possibly some new) solutions in a fast but rigorous way. Finally, we build a database with all necessary ingredients for the rigorous symmetry reduction of gravitational Lagrangians in the \texttt{xAct} package of \textsc{Mathematica}\footnote{In contrast to the Hicks database (based on \texttt{DifferentialGeometry}/\textsc{Maple}) \cite{Hicks:thesis}, our database focuses solely on PSC-compatible cases but contains many extra geometrical objects that are necessary for the symmetry reduction such as the $\Gamma$-invariant $l$-chains and residual diffeomorphism generators (all of which are implemented in more suitable adapted coordinates and vector-field bases).} and developed a first code that fully automate the symmetry reduction process, which is attached as supplementary material to this work.

Hopefully, this paper will serve as a remedy for all the unjustified but correct symmetry reductions of Lagrangians, as well as a warning against those that are truly incorrect, both of which are common in the literature. All allowed cases are explicitly listed in Tab.~\ref{tab:PSCcompgroupactions} with all necessities for the rigorous symmetry reduction in Tab.~\ref{tab:invmetrforPSCcomp} and Tab.~\ref{tab:lchainsPSCcomp}. The possible gauge fixing has to be always justifiable through the residual diffeomorphisms Tab.~\ref{tab:invvectfieldsPSCcomp} and Tab.~\ref{tab:residualgenerators} by means of the Noether identities \ref{tab:Noetherid}. Other practices, e.g., the reduction by symmetries that violate PSC or by ansatzes not (fully) captured by symmetries, incorrect reduction of the volume element, unjustified gauge choices, may (and often will) result either in wrong or insufficient set of field equations. 

With our results one can finally make the most out of PSC in all sorts of applications. Naturally, the symmetry-reduced gravitational Lagrangians with unchanged dynamics (fully equivalent field equations) are of their own interest as they define interesting mini/midi-superspace models (e.g., \cite{DeWitt:1967,Misner:1969,Kuchar:1971,Torre:1998dy}), many of which have not been analyzed yet. Furthermore, the Weyl trick can now be used for scanning gravitational theories for solutions like it was suggested e.g. in \cite{Deser:2005gr} but for any PSC-compatible symmetry and without the need to verify the obtained solutions against the reduced field equations. The fact that we can bypass derivation and reduction of the field equations becomes especially useful in higher-derivative theories whose Lagrangians are significantly simpler then their field equations. Last but not least, it may be possible to define theories with unique properties (e.g., the reduced order of derivatives) when restricted to certain symmetries by following the lines of \cite{Oliva:2010eb, Myers:2010ru,
Dehghani:2011vu, Cisterna:2017umf, Bueno:2019ycr, Bueno:2022res,Moreno:2023arp} but beyond the (stationary) spherically symmetric metrics and spatially homogeneous isotropic cosmologies. A small problem that remains open for future works is the derivation of algebraic identities corresponding to the finite-dimensional residual diffeomorphism generators that may help fixing the constants $\phi_k$ in homogeneous spacetimes also in the reduced Lagrangian.

\section*{Acknowledgements}
We thank Pavel Krtou\v{s} (Prague, Czech Republic) for the helpful comments on our work. I.K. acknowledges financial support by Primus grant PRIMUS/23/SCI/005 from Charles University and the support from the Charles University Research Center Grant No. UNCE24/SCI/016. T.M. acknowledges the support of the Czech Academy of Sciences (RVO 67985840) and the Czech Science Foundation GAČR grant no. 25-15544S.

\appendix

\appsection{Transformations of coordinates and vector-field bases}\label{apx:transfcoord}

\begin{longtable}{|l||>{\raggedright\arraybackslash}p{15cm}|l}
\caption{Transformations between Hicks and adapted coordinates}\label{tab:transf}\\
\hline
{\scriptsize Hicks \#} & {\scriptsize Transformation of coordinates: $x_i \leftrightarrow y_i$} \\\hline\hline
{[3,2,1]} & $x_1=y_3$, $x_2=y_4$, $x_3=y_1$, $x_4=y_2$ \\ \hline
{[3,2,2]} & $x_1 = \sinh^{-1} (\sinh y_3 \sin y_4)$, $x_2 = \tanh^{-1} (\tanh y_3 \cos y_4)$, $x_3 = y_1$, $x_4 = y_2$ \\ \hline
{[3,2,3]} & $x_1 = y_3 - \tfrac{\pi}{2}$, $x_2 = y_4$, $x_3 = y_1$, $x_4 = y_2$ \\ \hline
{[3,2,4]} & $x_1=(y_1+y_2)/\sqrt{2}$, $x_2=(-y_1+y_2)/\sqrt{2}$, $x_3=y_3$, $x_4=y_4$ \\ \hline
{[3,2,5]} & $x_1=y_2$, $x_2=y_1$, $x_3=y_3$, $x_4=y_4$ \\ \hline\hline
{[3,3,2]} & $x_1=y_1$, $x_2=y_2$, $x_3=y_4$, $x_4=y_3$ \\ \hline
{[3,3,3]} & $x_1=y_3$, $x_2=y_1$, $x_3=-y_4$, $x_4=y_2$ \\ \hline
{[3,3,8]} & $x_1= \tan \left(\frac{y_1+y_3}{2}+\tan ^{-1}\left(\cot y_4-\tanh \frac{y_3}{2} \csc y_4\right)\right)$, $x_2 = \frac{\tan \left(\frac{y_1+y_4}{2}+\tan ^{-1}\left(\cot y_4-\tanh \frac{y_3}{2} \csc y_4\right)\right)}{\cosh y_3-\sinh y_3 \cos y_4}+\frac{\sin y_4}{\coth y_3-\cos y_4}$, $x_3= \log \frac{\sec ^2\left(\frac{y_1+y_4}{2}+\tan ^{-1}\left(\cot y_4-\tanh \frac{y_3}{2} \csc y_4\right)\right)}{\cosh y_3-\sinh y_3 \cos y_4}$, $x_4= y_2$ \\ \hline
{[3,3,9]} & $x_1=y_3$, $x_2=y_4$, $x_3=-y_1$, $x_4=y_2$ \\ \hline\hline
{[4,3,1]} & $x_1 = y_1$, $x_2 = y_3$, $x_3 = y_4$, $x_4 = y_2$ \\ \hline
{[4,3,2]} & $x_1=-\tfrac{y_1+y_4}{2} - \tan ^{-1}\left(\cot y_4-\tanh \tfrac{y_3}{2}\csc y_4\right)$, $x_2 = \tfrac{\sin y_4}{\coth y_3-\cos y_4}$, $x_3 = -\log \left(\cosh y_3-\sinh y_3 \cos y_4\right)$, $x_4=y_2$ \\ \hline
{[4,3,3]} & $x_1=y_1$, $x_2=y_3$, $x_3=y_4$, $x_4=y_2$ \\ \hline
{[4,3,4]} & $x_1=y_1$, $x_2=y_4$, $x_3=y_3-\pi/2$, $x_4=y_2$ \\ \hline
{[4,3,5]} & $x_1=y_3$, $x_2=y_1$, $x_3=-y_4$, $x_4=y_2$ \\ \hline
{[4,3,6]} & $x_1=-y_1$, $x_2=y_3$, $x_3=y_4$, $x_4=y_2$ \\ \hline
{[4,3,8]} & $x_1=y_4$, $x_2=\frac{\sin y_1}{\cos y_1- \tanh y_2}$, $x_3 = \log\frac{\sech y_2}{\cos y_1-\tanh y_2}$, $x_4=y_3$ \\ \hline
{[4,3,9]} & $x_1=\tfrac{y_3}{2} - \tanh^{-1}\left(\tan \tfrac{y_1 }{2}\right) + \tanh^{-1}\big(\csc y_1-\cot y_1 \tanh \tfrac{y_2}{2}\big)$, $x_2=-\frac{\sin y_1}{\cos y_1 - \tanh y_2}$, $x_3 = \log\frac{\sech y_2}{\cos y_1-\tanh y_2}$, $x_4=y_4$ \\ \hline
{[4,3,10]} & $x_1=y_3$, $x_2=y_1$, $x_3=-y_4$, $x_4=y_2$ \\ \hline
{[4,3,11]} & $x_1=-y_4$, $x_2=y_2$, $x_3=y_1$, $x_4=y_3$ \\ \hline\hline
{[4,4,1]} & $x_1=y_3$, $x_2=y_4$, $x_3=-y_1$, $x_4=y_2$ \\ \hline
{[4,4,2]} & $x_1 = - \frac{\cos\tfrac{y_1 - y_4}{2} \cosh \tfrac{y_3}{2} - \cos\tfrac{y_1 + y_4}{2} \sinh\tfrac{y_3}{2}}{\cosh\tfrac{y_3}{2} \sin\tfrac{y_1 - y_4}{2} - \sin\tfrac{y_1 + y_4}{2} \sinh\tfrac{y_3}{2}} + y_2$, $x_2 = \tfrac{1}{2} \bigl( \sin y_4 (\cos y_1 \cosh y_3 + \sinh y_3) - \cos y_4 \sin y_1 \bigr)$, $x_3 = 2\log \frac{1 -  \cos y_1 \cos y_4 -  \cosh y_3 \sin y_1 \sin y_4}{2 \cosh\tfrac{y_3}{2} \sin\tfrac{y_1 - y_4}{2} + 2 \sin\tfrac{y_1 + y_4}{2} \sinh\tfrac{y_3}{2}}$, $x_4 = - y_2$ \\ \hline
{[4,4,9]} & $x_1=y_1$, $x_2=y_2$, $x_3=y_4$, $x_4=y_3$ \\ \hline
{[4,4,18]} & $x_1=y_3$, $x_2=y_1$, $x_3=-y_4$, $x_4=y_2$ \\ \hline 
{[4,4,22]} & $x_1=y_3$, $x_2=y_1$, $x_3=-y_4$, $x_4=y_2$ \\ \hline\hline
{[5,4,1]} & $x_1=-\tfrac{y_1+y_4}{2} - \tan ^{-1}\left(\cot y_4-\tanh \tfrac{y_3}{2}\csc y_4\right)$, $x_2 = \tfrac{\sin y_4}{\coth y_3-\cos y_4}$, $x_3 = -\log \left(\cosh y_3-\sinh y_3 \cos y_4\right)$, $x_4=y_2$ \\ \hline
{[5,4,2]} & $x_1=y_3-\pi$, $x_2=y_4$, $x_3=y_1$, $x_4=y_2$ \\ \hline
{[5,4,3]} & $x_1=y_3$, $x_2=y_1$, $x_3=-y_4$, $x_4=y_2$ \\ \hline
{[5,4,6]} & $x_1=y_3$, $x_2=y_1$, $x_3=-y_4$, $x_4=y_2$ \\ \hline
{[5,4,7]} & $x_1=\tfrac{y_3}{2} - \tanh^{-1}\left(\tan \tfrac{y_1 }{2}\right) + \tanh^{-1}\big(\csc y_1-\cot y_1 \tanh \tfrac{y_2}{2}\big)$, $x_2=-\frac{\sin y_1}{\cos y_1 - \tanh y_2}$, $x_3 = \log\frac{\sech y_2}{\cos y_1-\tanh y_2}$, $x_4=y_4$ \\ \hline\hline
{[6,3,1]} &   $x_1=y_1$, $x_2=y_2$, $x_3=y_3$, $x_4=y_4$\\ \hline
{[6,3,2]} &   $x_1=y_1$, $x_2=y_2$, $x_3=y_3$, $x_4=y_4$\\ \hline
{[6,3,3]} &   $x_1= \sqrt{-y_1} \sinh y_2 \cos y_3$, $x_2= \sqrt{-y_1} \sinh y_2 \sin y_3  \cos y_4$, $x_3= \sqrt{-y_1} \sinh y_2 \sin y_3 \sin y_4$, $x_4= \sqrt{-y_1} \cosh y_2$\\ \hline
{[6,3,4]} & $x_1= \tfrac{\sin y_1 - \tanh y_2\sin y_3}{\cos y_1-\tanh y_2\cos y_3}$, $x_2 = \tfrac{\sin y_1+\tanh y_2\sin y_3}{\cos y_1-\tanh y_2\cos y_3}$, $x_3= -2\log(\cos y_1\cosh y_2-\cos y_3 \sinh y_2)$, $x_4=y_4$ \\ \hline
{[6,3,5]} & $x_1=y_3$, $x_2=y_2$, $x_3=y_4$, $x_4=y_1$ \\ \hline
{[6,3,6]} &   $x_1= \sqrt{y_4} \cosh y_1 \cos y_2$, $x_2= \sqrt{y_4} \cosh y_1 \sin y_2 \cos y_3$, $x_3= \sqrt{y_4} \cosh y_1 \sin y_2 \sin y_3$, $x_4= \sqrt{y_4} \sinh y_1$ \\ \hline\hline
{[6,4,1]} & $x_1=y_2$, $x_2=y_1$, $x_3=y_3 + \frac{\pi}{2}$, $x_4=y_4$ \\ \hline
{[6,4,2]} & $x_1 = y_2$, $x_2 = y_1$, $x_3 = \sinh^{-1} (\sinh y_3 \sin y_4)$, $x_4 = \tanh^{-1} (\tanh y_3 \cos y_4)$ \\ \hline
{[6,4,3]} & $x_1=y_3$, $x_2=y_4$, $x_3=y_2$, $x_4=y_1$ \\ \hline
{[6,4,4]} & $x_1=y_1$, $x_2=y_2$, $x_3=y_3 + \tfrac{\pi}{2}$, $x_4=y_4$ \\ \hline
{[6,4,5]} & $x_1 = y_1$, $x_2 = y_2$, $x_3 = \sinh^{-1} (\sinh y_3 \sin y_4)$, $x_4 = \tanh^{-1} (\tanh y_3 \cos y_4)$ \\ \hline\hline
{[7,4,1]} &  $x_1=y_1$, $x_2=y_2$, $x_3=y_3$, $x_4=y_4$ \\ \hline
{[7,4,2]} &  $x_1= e^{y_1} \sinh y_2 \cos y_3$, $x_2= e^{y_1}  \sinh y_2 \sin y_3 \cos y_4$, $x_3= e^{y_1}  \sinh y_2 \sin y_3 \sin y_4$, $x_4= e^{y_1} \cosh y_2$\\ \hline
[7,4,3] &  $x_1= \tfrac{1}{\sqrt{2y_4}} \cosh y_1 \cos y_2$, $x_2= \tfrac{1}{\sqrt{2y_4}}  \cosh y_1 \sin y_2 \cos y_3$, $x_3= \tfrac{1}{\sqrt{2y_4}}  \cosh y_1 \sin y_2 \sin y_3$, $x_4= \tfrac{1}{\sqrt{2y_4}} \sinh y_1$\\ \hline
{[7,4,4]} &  $x_1= \tfrac{\sin y_1-\tanh y_2\sin y_3}{\cos y_1-\tanh y_2\cos y_3}$, $x_2 = \tfrac{\sin y_1+\tanh y_2\sin y_3}{\cos y_1-\tanh y_2\cos y_3}$, $x_3= -2\log(\cos y_1 \cosh y_2 - \cos y_3 \sinh y_2)$, $x_4=y_4$\\ \hline
\end{longtable}

\begin{longtable}{|l||>{\raggedright\arraybackslash}p{15cm}|l}
\caption{Redefinition of the Hicks and adapted vector-field bases}\label{tab:redefKV}\\\hline
{\scriptsize Hicks \#} & {\scriptsize Redefinition of $\Gamma$ base: $\bs{X}_i \leftrightarrow \bs{Y}_i$} \\\hline\hline
{[3,2,1]} & no change \\ \hline
{[3,2,2]} & $\bs{Y}_1 = \bs{X}_2$, $\bs{Y}_2 = - \bs{X}_1$, $\bs{Y}_3 = \bs{X}_3$ (analogous to rotations around $x$, $y$ and $z$ axis in $\mathrm{E_3}$) \\ \hline
{[3,2,3]} & $\bs{Y}_1 = \bs{X}_1$, $\bs{Y}_2 = - \bs{X}_3$, $\bs{Y}_3 = \bs{X}_2$ (rotations around $x$, $y$ and $z$ axis of $\mathrm{E_3}$ in spherical coordinates) \\ \hline
{[3,2,4]} & $\bs{Y}_1 = \frac{\bs{X}_1 - \bs{X}_2}{\sqrt{2}} = \bs\partial_{y_1}$, $\bs{Y}_2 = \frac{\bs{X}_1 + \bs{X}_2}{\sqrt{2}} = \bs\partial_{y_2}$, $\bs{Y}_3 = \bs{X}_3 = y_2 \bs\partial_{y_1} + y_1 \bs\partial_{y_2}$ \\ \hline
{[3,2,5]} & $\bs{Y}_1 = \bs{X}_1$, $\bs{Y}_2 = - \bs{X}_3$, $\bs{Y}_3 = \bs{X}_2$ \\ \hline\hline
{[3,3,2]} & no change \\ \hline
{[3,3,3]} & $\bs{Y}_1 = \bs{X}_1$, $\bs{Y}_2 = - \bs{X}_2$, $\bs{Y}_3 = -\bs{X}_3$ \\ \hline
{[3,3,8]} & $\bs{Y}_1 = \bs{X}_2$, $\bs{Y}_2 = \frac{\bs{X}_3 - \bs{X}_1}{2}$, $\bs{Y}_3 = -\frac{\bs{X}_1 + \bs{X}_3}{2}$  \\ \hline
{[3,3,9]} & $\bs{Y}_1 = \bs{X}_2$, $\bs{Y}_2 = \bs{X}_3$, $\bs{Y}_3 = \bs{X}_1$ \\ \hline\hline
{[4,3,1]} & $\bs{Y}_1 = \frac{\bs{X}_1 + \bs{X}_3}{2}$, $\bs{Y}_2 = - \bs{X}_2$, $\bs{Y}_3 = \frac{\bs{X}_1 - \bs{X}_3}{2}$, $\bs{Y}_4 = \bs{X}_4$ ($\bs{Y}_1$, $\bs{Y}_2$, $\bs{Y}_3$ identical to {[3,2,2]}) \\ \hline
{[4,3,2]} & $\bs{Y}_1 = \bs{X}_2$, $\bs{Y}_2 = \frac{\bs{X}_3 - \bs{X}_1}{2}$, $\bs{Y}_3 = -\frac{\bs{X}_1 + \bs{X}_3}{2}$, $\bs{Y}_4 = -\bs{X}_4$ ($\bs{Y}_1$, $\bs{Y}_2$, $\bs{Y}_3$ identical to {[3,3,8]}) \\ \hline
{[4,3,3]} & $\bs{Y}_1 = \bs{X}_3$, $\bs{Y}_2 = - \bs{X}_2$, $\bs{Y}_3 = \bs{X}_1$, $\bs{Y}_4 = \bs{X}_4$ ($\bs{Y}_1$, $\bs{Y}_2$, $\bs{Y}_3$ identical to {[3,2,3]}) \\ \hline
{[4,3,4]} & $\bs{Y}_1 = \bs{X}_3$, $\bs{Y}_2 = - \bs{X}_2$, $\bs{Y}_3 = \bs{X}_1$, $\bs{Y}_4 = \bs{X}_4$ ($\bs{Y}_1$, $\bs{Y}_2$, $\bs{Y}_3$ identical to {[3,3,9]}) \\ \hline
{[4,3,5]} & $\bs{Y}_1 = \bs{X}_1$, $\bs{Y}_2 = - \bs{X}_2$, $\bs{Y}_3 = -\bs{X}_3$, $\bs{Y}_4 = \bs{X}_4$ ($\bs{Y}_1$, $\bs{Y}_2$, $\bs{Y}_3$ identical to {[3,3,3]}) \\ \hline
{[4,3,6]} & $\bs{Y}_1 = \bs{X}_1 = \bs\partial_{y_3}$, $\bs{Y}_2 = \bs{X}_2 = \bs\partial_{y_4}$, $\bs{Y}_3 = \bs{X}_4 = -y_4 \bs\partial_{y_3} + y_3 \bs\partial_{y_4}$, $\bs{Y}_4 = \bs{X}_3 = \bs\partial_{y_1}$ \\ \hline
{[4,3,8]} & $\bs{Y}_1 = \bs{X}_2$, $\bs{Y}_2 = \frac{\bs{X}_1 - \bs{X}_3}{2}$, $\bs{Y}_3 = \frac{\bs{X}_1 + \bs{X}_3}{2}$, $\bs{Y}_4 = \bs{X}_4$ ($\bs{Y}_1$, $\bs{Y}_2$, $\bs{Y}_3$ identical to {[3,2,5]}) \\ \hline
{[4,3,9]} & $\bs{Y}_1 = \bs{X}_2$, $\bs{Y}_2 = \frac{\bs{X}_3 - \bs{X}_1}{2}$, $\bs{Y}_3 = -\frac{\bs{X}_1 + \bs{X}_3}{2}$, $\bs{Y}_4 = \bs{X}_4$ ($\bs{Y}_1$, $\bs{Y}_2$, $\bs{Y}_3$ identical to {[3,3,8]}) \\ \hline
{[4,3,10]} & $\bs{Y}_1 = \bs{X}_1$, $\bs{Y}_2 = - \bs{X}_2$, $\bs{Y}_3 = -\bs{X}_3$, $\bs{Y}_4 = -\bs{X}_4$ ($\bs{Y}_1$, $\bs{Y}_2$, $\bs{Y}_3$ corresponds to {[3,3,3]}) \\ \hline
{[4,3,11]} & $\bs{Y}_1 = \bs{X}_2 = \bs\partial_{y_1}$, $\bs{Y}_2 = \bs{X}_1 = \bs\partial_{y_2}$, $\bs{Y}_3 = \bs{X}_3 = \bs\partial_{y_4}$, $\bs{Y}_4 = \bs{X}_4 = y_2 \bs\partial_{y_1} + y_1 \bs\partial_{y_2}$
($\bs{Y}_1$, $\bs{Y}_2$, $\bs{Y}_4$ corresponds to {[3,2,4]}, $\bs{Y}_1$, $\bs{Y}_2$, $\bs{Y}_3$ corresponds to {[3,3,2]}) \\ \hline\hline
{[4,4,1]} & $\bs{Y}_1 = \bs{X}_2$, $\bs{Y}_2 = \bs{X}_3$, $\bs{Y}_3 = \bs{X}_1$, $\bs{Y}_4 = \bs{X}_4$ ($\bs{Y}_1$, $\bs{Y}_2$, $\bs{Y}_3$ identical to {[3,3,9]}) \\ \hline
{[4,4,2]} & $\bs{Y}_1 = \bs{X}_2$, $\bs{Y}_2 = \frac{\bs{X}_3 - \bs{X}_1}{2}$, $\bs{Y}_3 = -\frac{\bs{X}_1 + \bs{X}_3}{2}$, $\bs{Y}_4 = \bs{X}_4$ ($\bs{Y}_1$, $\bs{Y}_2$, $\bs{Y}_3$ identical to {[3,3,8]}) \\ \hline
{[4,4,9]} & $\bs{Y}_1 = - \bs{X}_3 = \bs\partial_{y_1}$, $\bs{Y}_2 = \bs{X}_1 = \bs\partial_{y_2}$, $\bs{Y}_3 = \bs{X}_2 = \bs\partial_{y_4}$, $\bs{Y}_4 = \bs{X}_4 = - y_4 \bs\partial_{y_1} + \bs\partial_{y_3} + y_2 \bs\partial_{y_4}$ ($\bs{Y}_1$, $\bs{Y}_2$, $\bs{Y}_3$ identical to {[3,3,2]}) \\ \hline
{[4,4,18]} & $\bs{Y}_1 = \bs{X}_1$, $\bs{Y}_2 = - \bs{X}_2$, $\bs{Y}_3 = -\bs{X}_3$, $\bs{Y}_4 = \bs{X}_4$ ($\bs{Y}_1$, $\bs{Y}_2$, $\bs{Y}_3$ identical to {[3,3,3]}) \\ \hline
{[4,4,22]} & $\bs{Y}_1 = \bs{X}_1$, $\bs{Y}_2 = - \bs{X}_2$, $\bs{Y}_3 = -\bs{X}_3$, $\bs{Y}_4 = \bs{X}_4$ ($\bs{Y}_1$, $\bs{Y}_2$, $\bs{Y}_3$ identical to {[3,3,3]}) \\ \hline\hline
{[5,4,1]} & $\bs{Y}_1 = \bs{X}_2$, $\bs{Y}_2 = \frac{\bs{X}_3 - \bs{X}_1}{2}$, $\bs{Y}_3 = -\frac{\bs{X}_1 + \bs{X}_3}{2}$, $\bs{Y}_4 = -\bs{X}_4$, $\bs{Y}_5 = \bs{X}_5$ ($\bs{Y}_1$, $\bs{Y}_2$, $\bs{Y}_3$, $\bs{Y}_4$ identical to {[4,3,2]}) \\ \hline
{[5,4,2]} & $\bs{Y}_1 = \bs{X}_2$, $\bs{Y}_2 = \bs{X}_3$, $\bs{Y}_3 = \bs{X}_1$, $\bs{Y}_4 = \bs{X}_4$, $\bs{Y}_5 = \bs{X}_5$ ($\bs{Y}_1$, $\bs{Y}_2$, $\bs{Y}_3$, $\bs{Y}_4$ identical to {[4,3,4]}, $\bs{Y}_1$, $\bs{Y}_2$, $\bs{Y}_3$, $\bs{Y}_5$ identical to {[4,4,1]}) \\ \hline
{[5,4,3]} & $\bs{Y}_1 = \bs{X}_1$, $\bs{Y}_2 = - \bs{X}_2$, $\bs{Y}_3 = -\bs{X}_3$, $\bs{Y}_4 = \bs{X}_4$, $\bs{Y}_5 = \bs{X}_5$ ($\bs{Y}_1$, $\bs{Y}_2$, $\bs{Y}_3$, $\bs{Y}_4$ identical to {[4,3,5]}) \\ \hline
{[5,4,6]} & $\bs{Y}_1 = \bs{X}_1$, $\bs{Y}_2 = - \bs{X}_2$, $\bs{Y}_3 = -\bs{X}_3$, $\bs{Y}_4 = -\bs{X}_4$, $\bs{Y}_5 = \bs{X}_5$ ($\bs{Y}_1$, $\bs{Y}_2$, $\bs{Y}_3$, $\bs{Y}_4$ identical to {[4,3,10]}) \\ \hline
{[5,4,7]} & $\bs{Y}_1 = \bs{X}_2$, $\bs{Y}_2 = \frac{\bs{X}_3 - \bs{X}_1}{2}$, $\bs{Y}_3 = -\frac{\bs{X}_1 + \bs{X}_3}{2}$, $\bs{Y}_4 = \bs{X}_4$, $\bs{Y}_5 = \bs{X}_5$ ($\bs{Y}_1$, $\bs{Y}_2$, $\bs{Y}_3$, $\bs{Y}_4$ identical to {[4,3,9]}) \\ \hline\hline
{[6,3,1]} & $\bs{Y}_1 = -\bs{X}_3$, $\bs{Y}_2 = -\bs{X}_4$, $\bs{Y}_3 = \bs{X}_5$, $\bs{Y}_4 = -\bs{X}_1$, $\bs{Y}_5 = \bs{X}_2$, $\bs{Y}_6 = -\bs{X}_6$ ($\bs{Y}_1$, $\bs{Y}_2$, $\bs{Y}_3$ identical to {[3,2,3]}) \\ \hline
{[6,3,2]} & $\bs{Y}_1 = -\bs{X}_5$, $\bs{Y}_2 = -\bs{X}_3$, $\bs{Y}_3 = -\bs{X}_2$, $\bs{Y}_4 = \bs{X}_1$, $\bs{Y}_5 = \bs{X}_4$, $\bs{Y}_6 = \bs{X}_6$ ($\bs{Y}_1$, $\bs{Y}_2$, $\bs{Y}_3$ identical to {[3,2,1]}, $\bs{Y}_3$, $\bs{Y}_5$, $\bs{Y}_6$ identical to {[3,3,2]} with $y_1 \longleftrightarrow y_4$, $\bs{Y}_1$, $\bs{Y}_2$, $\bs{Y}_4$ identical to {[3,2,3]} in Cartesian coordinates) \\ \hline
{[6,3,3]} & $\bs{Y}_1 = -\bs{X}_6$, $\bs{Y}_2 = \bs{X}_5$, $\bs{Y}_3 = -\bs{X}_4$, $\bs{Y}_4 = \bs{X}_1$, $\bs{Y}_5 = \bs{X}_2$, $\bs{Y}_6 = \bs{X}_3$ ($\bs{Y}_1$, $\bs{Y}_2$, $\bs{Y}_3$ identical to {[3,2,3]}) \\ \hline
{[6,3,4]} & no change \\ \hline
{[6,3,5]} & $\bs{Y}_1 = \bs{X}_3$, $\bs{Y}_2 = -\bs{X}_6$, $\bs{Y}_3 = -\bs{X}_5$, $\bs{Y}_4 = \bs{X}_1$, $\bs{Y}_5 = \bs{X}_2$, $\bs{Y}_6 = \bs{X}_4$ ($\bs{Y}_1$, $\bs{Y}_2$, $\bs{Y}_3$, -$\bs{Y}_4$ identical to {[4,3,11]}, $\bs{Y}_2$, $\bs{Y}_3$, $\bs{Y}_6$, $\bs{Y}_1$ after $y_2 \leftrightarrow y_3$ identical to {[4,3,6]}) \\ \hline
{[6,3,6]} & $\bs{Y}_1 = \bs{X}_6$, $\bs{Y}_2 = -\bs{X}_5$, $\bs{Y}_3 = \bs{X}_4$, $\bs{Y}_4 = \bs{X}_1$, $\bs{Y}_5 = \bs{X}_2$, $\bs{Y}_6 = \bs{X}_3$ ($\bs{Y}_1$, $\bs{Y}_2$, $\bs{Y}_3$ identical to {[3,2,3]} with $y_2 \to y_4$, $y_3 \to y_2$, $y_4 \to y_3$) \\ \hline\hline
{[6,4,1]} & $\bs{Y}_1 = \bs{X}_2$, $\bs{Y}_2 = \bs{X}_3$, $\bs{Y}_3 = \bs{X}_1$, $\bs{Y}_4 = \bs{X}_5$, $\bs{Y}_5 = \bs{X}_6$, $\bs{Y}_6 = -\bs{X}_4$ ($\bs{Y}_4$, $\bs{Y}_5$, $\bs{Y}_6$, $\bs{Y}_3$ identical to {[4,3,8]}, $\bs{Y}_1$, $\bs{Y}_2$, $\bs{Y}_3$, $\bs{Y}_6$ identical to {[4,3,3]}) \\ \hline
{[6,4,2]} & $\bs{Y}_1 = \bs{X}_1$, $\bs{Y}_2 = - \frac{\bs{X}_2 - \bs{X}_3}{2}$, $\bs{Y}_3 = - \frac{\bs{X}_2 + \bs{X}_3}{2}$, $\bs{Y}_4 = \bs{X}_5$, $\bs{Y}_5 = \bs{X}_6$, $\bs{Y}_6 = -\bs{X}_4$ ($\bs{Y}_4$, $\bs{Y}_5$, $\bs{Y}_6$, $\bs{Y}_3$ identical to {[4,3,8]}, $\bs{Y}_1$, $\bs{Y}_2$, $\bs{Y}_3$, $\bs{Y}_6$ identical to {[4,3,1]}) \\ \hline
{[6,4,3]} & $\bs{Y}_1 = \bs{X}_2$, $\bs{Y}_2 = \bs{X}_3$, $\bs{Y}_3 = -\bs{X}_1$, $\bs{Y}_4 = \bs{X}_6$, $\bs{Y}_5 = \bs{X}_5$, $\bs{Y}_6 = \bs{X}_4$ ($\bs{Y}_1$, $\bs{Y}_2$, $\bs{Y}_3$, $\bs{Y}_5$ identical to {[4,3,8]}, $\bs{Y}_4$, $\bs{Y}_5$, $\bs{Y}_6$, $\bs{Y}_3$ identical to {[4,3,6]}) \\ \hline
{[6,4,4]} & $\bs{Y}_1 = \bs{X}_2$, $\bs{Y}_2 = \bs{X}_3$, $\bs{Y}_3 = \bs{X}_1$, $\bs{Y}_4 = \bs{X}_6$, $\bs{Y}_5 = -\bs{X}_5$, $\bs{Y}_6 = -\bs{X}_4$ ($\bs{Y}_1$, $\bs{Y}_2$, $\bs{Y}_3$, $\bs{Y}_4$ identical to {[4,3,3]}, $\bs{Y}_4$, $\bs{Y}_5$, $\bs{Y}_3$, $\bs{Y}_6$ identical to {[4,3,11]}) \\ \hline
{[6,4,5]} & $\bs{Y}_1 = \bs{X}_1$, $\bs{Y}_2 = \frac{\bs{X}_3 - \bs{X}_2}{2}$, $\bs{Y}_3 = -\frac{\bs{X}_2 + \bs{X}_3}{2}$, $\bs{Y}_4 = -\bs{X}_4$, $\bs{Y}_5 = \bs{X}_6$, $\bs{Y}_6 = -\bs{X}_5$ ($\bs{Y}_1$, $\bs{Y}_2$, $\bs{Y}_3$, $\bs{Y}_5$ identical to {[4,3,1]}, $\bs{Y}_5$, $\bs{Y}_6$, $\bs{Y}_3$, $\bs{Y}_4$ identical to {[4,3,11]}) \\ \hline\hline
{[7,4,2]} & $\bs{Y}_1 = -\bs{X}_7$, $\bs{Y}_2 = \bs{X}_6$, $\bs{Y}_3 = -\bs{X}_5$, $\bs{Y}_4 = \bs{X}_2$, $\bs{Y}_5 = \bs{X}_3$, $\bs{Y}_6 = \bs{X}_4$, $\bs{Y}_7 = \bs{X}_1$ ($\bs{Y}_1$, $\bs{Y}_2$, $\bs{Y}_3$, $\bs{Y}_4$, $\bs{Y}_5$, $\bs{Y}_6$ identical to {[6,3,3]}, $\bs{Y}_1$, $\bs{Y}_2$, $\bs{Y}_3$, $\bs{Y}_7$ identical to {[4,3,3]}) \\ \hline
{[7,4,3]} & $\bs{Y}_1 = \bs{X}_6$, $\bs{Y}_2 = -\bs{X}_5$, $\bs{Y}_3 = \bs{X}_4$, $\bs{Y}_4 = \bs{X}_1$, $\bs{Y}_5 = \bs{X}_2$, $\bs{Y}_6 = \bs{X}_3$, $\bs{Y}_7 = \bs{X}_7$ ($\bs{Y}_1$, $\bs{Y}_2$, $\bs{Y}_3$, $\bs{Y}_4$, $\bs{Y}_5$, $\bs{Y}_6$ identical to {[6,3,6]}, $\bs{Y}_1$, $\bs{Y}_2$, $\bs{Y}_3$, $\bs{Y}_7$ identical to {[4,3,3]} with with $y_1 \to y_4$, $y_2 \to y_1$, $y_3 \to y_2$, $y_4 \to y_3$) \\ \hline
\end{longtable}

\vspace{1cm}

\appsection{Relations among infinitesimal group actions}\label{apx:relinfgra}

\begin{longtable}{|l||>{\raggedright\arraybackslash}p{7cm}|>{\raggedright\arraybackslash}p{7cm}|}
\caption{Details of relations among infinitesimal group actions}\label{tab:arrows}\\\hline
{\scriptsize Relation} & {\scriptsize Extra Killing vectors $\bs{Y}_i$} & {\scriptsize Specialization of $\Gamma$-invariant metrics $\hat{\bs{g}}$} \\\hline\hline
{[3,2,1]$\to$[4,3,6]} & $\bs{\partial}_{y_1}$ & $\phi_i(y_1,y_2) \to \phi_i(y_2)$ \\ \hline
{[3,2,2]$\to$[4,3,1]} & $\bs{\partial}_{y_1}$ & $\phi_i(y_1,y_2) \to \phi_i(y_2)$ \\ \hline
{[3,2,3]$\to$[4,3,3]} & $\bs{\partial}_{y_1}$ & $\phi_i(y_1,y_2) \to \phi_i(y_2)$ \\ \hline
{[3,2,3]$\to$[6,3,1]} & $\cos y_4 (\sin y_3 \, \bs{\partial}_{y_2} + \cos y_3 \cot y_2 \, \bs{\partial}_{y_3}) - \cot y_2 \csc y_3 \sin y_4 \, \bs{\partial}_{y_4}$, $\sin y_4 (\sin y_3 \, \bs{\partial}_{y_2} + \cos y_3 \cot y_2 \, \bs{\partial}_{y_3}) + \cos y_4 \cot y_2 \csc y_3 \, \bs{\partial}_{y_4}$, $\cos y_3 \, \bs{\partial}_{y_2} -  \cot y_2 \sin y_3 \, \bs{\partial}_{y_3}$ & $\phi_1(y_1, y_2) \to \phi_1(y_1)$, $\phi_2(y_1, y_2) \to 0$, $\phi_3(y_1, y_2) \to \phi_2(y_1)$, $\phi_4(y_1, y_2) \to \sin^2y_2 \, \phi_2(y_1)$ \\ \hline
{[3,2,3]\footnote{Upon coordinate transformation $y_2 \to \sqrt{y_2^2 + y_3^2 + y_4^2}$, $\cos y_3 \to \frac{y_4}{\sqrt{y_2^2 + y_3^2 + y_4^2}}$, $\tan y_4 \to \frac{y_3}{y_2}$.}$\to$[6,3,2]} & $\bs{\partial}_{y_3}$, $\bs{\partial}_{y_4}$, $ y_4 \bs{\partial}_{y_3} - y_3 \bs{\partial}_{y_4}$ & $\phi_1(y_1, r) \to \phi_1(y_1)$, $\phi_1(y_1, r) \to 0$, $\phi_3(y_1, r) \to \phi_2(y_1)$, $\phi_4(y_1, r) \to r^2 \phi_2(y_1)$, where $r = \sqrt{y_2^2 + y_3^2 + y_4^2}$ \\ \hline
{[3,2,3]$\to$[6,3,3]} & $\cos y_4 (\sin y_3 \bs{\partial}_{y_2} + \cos y_3 \coth y_2 \bs{\partial}_{y_3}) - \coth y_2 \csc y_3 \sin y_4 \bs{\partial}_{y_4}$, $\sin y_4 (\sin y_3 \bs{\partial}_{y_2} + \cos y_3 \coth y_2 \bs{\partial}_{y_3}) + \cos y_4 \coth y_2 \csc y_3 \bs{\partial}_{y_4}$, $\cos y_3 \bs{\partial}_{y_2} - \coth y_2 \sin y_3 \bs{\partial}_{y_3}$ & $\phi_1(y_1, y_2) \to \phi_1(y_1)$, $\phi_2(y_1, y_2) \to 0$, $\phi_3(y_1, y_2) \to \phi_2(y_1)$, $\phi_4(y_1, y_2) \to \sinh^2y_2 \, \phi_2(y_1)$ \\ \hline
{[3,2,3]\footnote{Upon coordinate transformation $y_2 \to y_4$, $y_3 \to y_2$, $y_4 \to y_3$.}$\to$[6,3,6]} & $\cos y_3 (\sin y_2 \bs{\partial}_{y_1} + \cos y_2 \tanh y_1 \bs{\partial}_{y_2}) -  \csc y_2 \sin y_3 \tanh y_1 \bs{\partial}_{y_3}$, $\sin y_3 (\sin y_2 \bs{\partial}_{y_1} + \cos y_2 \tanh y_1 \bs{\partial}_{y_2}) + \cos y_3 \csc y_2 \tanh y_1 \bs{\partial}_{y_3}$, $\cos y_2 \bs{\partial}_{y_1} -  \sin y_2 \tanh y_1 \bs{\partial}_{y_2}$ & $\phi_1(y_1, y_4) \to \phi_1(y_4)$, $\phi_2(y_1, y_4) \to 0$, $\phi_3(y_1, y_4) \to \phi_2(y_4)$, $\phi_4(y_1, y_4) \to \cosh^2y_1 \, \phi_1(y_4)$ \\ \hline
{[3,2,4]$\to$[4,3,11]} & $\bs{\partial}_{y_4}$ & $\phi_i(y_3,y_4) \to \phi_i(y_3)$ \\ \hline
{[3,2,5]$\to$[4,3,8]} & $\bs{\partial}_{y_4}$ & $\phi_i(y_3,y_4) \to \phi_i(y_3)$ \\ \hline\hline
{[3,3,2]$\to$[4,3,11]} & $y_2 \bs{\partial}_{y_1} + y_1 \bs{\partial}_{y_2}$ & $\phi_i(y_2) \to \phi_i(y_2)$ for $1\leq i \leq 4$, $\phi_5(y_2) \to \phi_1(y_2)$, $\phi_i(y_2) \to 0$ for $6 \leq i \leq 10$ \\ \hline
{[3,3,2]$\to$[4,4,9]} & $\bs{\partial}_{y_3} - y_4 \bs{\partial}_{y_1} + y_2 \bs{\partial}_{y_4}$ & $\phi_1(y_3) \to \phi_1$, $\phi_2(y_3) \to \phi_2$, $\phi_3(y_3) \to \phi_3 + \phi_7 y_3$, $\phi_4(y_3) \to \phi_4 + 2 \phi_8 y_3 - \phi_1 y_3^2$, $\phi_5(y_3) \to \phi_5 - 2 \phi_{10} y_3 + (\phi_4 - \phi_6)y_3^2 + y_3^3 \phi_8 - \tfrac{1}{4} \phi_1 y_3^4$, $\phi_6(y_3) \to \phi_6 - \phi_8 y_3 + \tfrac{1}{2}  \phi_1 y_3^2$, $\phi_7(y_3) \to \phi_7$, $\phi_8(y_3) \to \phi_8 - \phi_1 y_3$, $\phi_9(y_3) \to \phi_9 - \phi_3 y_3 - \tfrac{1}{2}  \phi_7 y_3^2$, $\phi_{10}(y_3) \to \phi_{10} - (\phi_4 - \phi_6) y_3 - \tfrac{3}{2}  \phi_8 y_3^2 + \tfrac{1}{2}  \phi_1 y_3^3$ \\ \hline
{[3,3,2]\footnote{Upon coordinate transformation $y_1 \to y_3$, $y_3 \to y_1$.}$\to$[6,3,2]} & $y_3 \bs{\partial}_{y_2} - y_2 \bs{\partial}_{y_3}$, $y_4 \bs{\partial}_{y_2} -  y_2 \bs{\partial}_{y_4}$, $y_4 \bs{\partial}_{y_3} - y_3 \bs{\partial}_{y_4}$ & $\phi_1(y_1) \to - \phi_2(y_1)$, $\phi_2(y_1) \to - \phi_1(y_1)$, $\phi_3(y_1) \to 0$, $\phi_4(y_1) \to \phi_2(y_1)$, $\phi_5(y_1) \to \phi_2(y_1)$, $\phi_i(y_1) \to 0$ for $6 \leq i \leq 10$ \\ \hline
{[3,3,3]$\to$[4,3,5]} & $y_4 \bs{\partial}_{y_3} - y_3 \bs{\partial}_{y_4} - \tfrac12 (y_3^2 - y_4^2) \bs{\partial}_{y_1}$ & $\phi_i(y_2) \to \phi_i(y_2)$ for $1\leq i \leq 4$, $\phi_i(y_2) \to 0$ for $5 \leq i \leq 9$, $\phi_{10}(y_2) \to \phi_4(y_2)$ \\ \hline
{[3,3,3]$\to$[4,3,10]} & $y_3 \bs{\partial}_{y_3} -  y_4 \bs{\partial}_{y_4}$ & $\phi_i(y_2) \to \phi_i(y_2)$ for $1\leq i \leq 3$, $\phi_i(y_2) \to 0$ for $4 \leq i \leq 8$ and $i = 10$, $\phi_9(y_2) \to \phi_4(y_2)$ \\ \hline
{[3,3,3]$\to$[4,4,18]} & $\bs{\partial}_{y_2} - y_3 \bs{\partial}_{y_3} + y_4 \bs{\partial}_{y_4}$ & $\phi_i(y_2) \to \phi_i$ for $i \in \{1,2,3,9\}$, $\phi_i(y_2) \to e^{y_2} \phi_i$ for $i \in \{5,7\}$, $\phi_i(y_2) \to e^{-y_2} \phi_i$ for $i \in \{6,8\}$, $\phi_4(y_2) \to e^{2y_2} \phi_4$, $\phi_{10}(y_2) \to e^{-2y_2} \phi_{10}$  \\ \hline
{[3,3,3]$\to$[4,4,22]} & $\bs{\partial}_{y_2} + y_4 \bs{\partial}_{y_3} - y_3 \bs{\partial}_{y_4} - \tfrac12 (y_3^2 - y_4^2) \bs{\partial}_{y_1}$ & $\phi_1(y_2) \to \phi_1$, $\phi_2(y_2) \to \phi_2$, $\phi_3(y_2) \to \phi_3$, $\phi_4(y_2) \to \phi_4 \cos^2y_2 + \phi_{10} \sin^2y_2 + \phi_9 \sin(2 y_2)$, $\phi_5(y_2) \to \phi_5 \cos y_2  + \phi_6 \sin y_2$, $\phi_6(y_2) \to \phi_6 \cos y_2 - \phi_5 \sin y_2$, $\phi_7(y_2) \to \phi_7 \cos y_2 + \phi_8 \sin y_2$, $\phi_8(y_2) \to \phi_8 \cos y_2 - \phi_7 \sin y_2$, $\phi_9(y_2) \to \phi_9 \cos(2 y_2) - (\phi_4 - \phi_{10}) \cos y_2 \sin y_2$, $\phi_{10}(y_2) \to \phi_{10} \cos^2y_2 + \phi_4 \sin^2y_2 - \phi_9 \sin(2 y_2)$ \\ \hline
{[3,3,8]$\to$[4,3,2]} & $\bs{\partial}_{y_1}$ & $\phi_i(y_2) \to \phi_i(y_2)$ for $1\leq i \leq 4$, $\phi_i(y_2) \to 0$ for $5 \leq i \leq 9$, $\phi_{10}(y_2) \to \phi_4(y_2)$  \\ \hline
{[3,3,8]\footnote{Upon coordinate transformation $\tan y_1 \to \coth y_2 \sinh y_3$, $y_2 \to y_4$, $\sech y_3 \to \sech y_2 \sech y_3$, $\cot y_4 \to \frac{1 - \tan y_1 \csch y_2 \tanh y_3}{\tan y_1 + \csch y_2 \tanh y_3}$.}}$\to$[4,3,9] & $\bs{\partial}_{y_3} $ & $\phi_1(y_4) \to \phi_1(y_4)$, $\phi_i(y_4) \to 0$ for $i \in \{2,5,6,7,8\}$, $\phi_3(y_4) \to \phi_4(y_4)$, $\phi_4(y_4) \to \phi_2(y_4)$,
 $\phi_9(y_4) \to \phi_3(y_4)$, $\phi_{10}(y_4) \to \phi_1(y_4)$ \\ \hline
{[3,3,8]$\to$[4,4,2]} & $\bs{\partial}_{y_2}$ & $\phi_i(y_2) \to \phi_i$ \\ \hline
{[3,3,9]$\to$[4,3,4]} & $\bs{\partial}_{y_1}$ & $\phi_i(y_2) \to \phi_i(y_2)$ for $1\leq i \leq 4$, $\phi_i(y_2) \to 0$ for $5 \leq i \leq 9$, $\phi_{10}(y_2) \to \phi_4(y_2)$ \\ \hline
{[3,3,9]$\to$[4,4,1]} & $\bs{\partial}_{y_2}$ & $\phi_i(y_2) \to \phi_i$ \\ \hline\hline
{[4,3,1]$\to$[6,4,2]} & $\sin y_1 \tanh y_2 \bs{\partial}_{y_1} - \cos y_1 \bs{\partial}_{y_2}$, $\cos y_1 \tanh y_2 \bs{\partial}_{y_1} + \sin y_1 \bs{\partial}_{y_2}$ & $\phi_1(y_2) \to \cosh^2 y_2 \, \phi_1$, $\phi_2(y_2) \to 0$, $\phi_3(y_2) \to \phi_1$, $\phi_4(y_2) \to \phi_2$ \\ \hline
{[4,3,1]$\to$[6,4,5]} & $\bs{\partial}_{y_2}$, $y_2 \bs{\partial}_{y_1} + y_1 \bs{\partial}_{y_2}$ & $\phi_1(y_2) \to \phi_1$, $\phi_2(y_2) \to 0$, $\phi_3(y_2) \to \phi_1$, $\phi_4(y_2) \to \phi_2$ \\ \hline
{[4,3,2]$\to$[5,4,1]} & $\bs{\partial}_{y_2}$ & $\phi_i(y_2) \to \phi_i$ \\ \hline
{[4,3,3]$\to$[6,4,1]} & $\sin y_1 \tanh y_2 \bs{\partial}_{y_1} - \cos y_1 \bs{\partial}_{y_2}$, $\cos y_1 \tanh y_2 \bs{\partial}_{y_1} + \sin y_1 \bs{\partial}_{y_2}$ & $\phi_1(y_2) \to \cosh^2 y_2 \, \phi_1$, $\phi_2(y_2) \to 0$, $\phi_3(y_2) \to \phi_1$, $\phi_4(y_2) \to \phi_2$ \\ \hline
{[4,3,3]$\to$[6,4,4]} & $\bs{\partial}_{y_2}$, $y_2 \bs{\partial}_{y_1} + y_1 \bs{\partial}_{y_2}$ & $\phi_1(y_2) \to \phi_1$, $\phi_2(y_2) \to 0$, $\phi_3(y_2) \to \phi_1$, $\phi_4(y_2) \to \phi_2$ \\ \hline
{[4,3,3]$\to$[7,4,1]} & $\cos y_4 (\sin y_3 \bs{\partial}_{y_2} + \cos y_3 \cot y_2 \bs{\partial}_{y_3}) -  \cot y_2 \csc y_3 \sin y_4 \bs{\partial}_{y_4}$, $\sin y_4 (\sin y_3 \bs{\partial}_{y_2} + \cos y_3 \cot y_2 \bs{\partial}_{y_3}) + \cos y_4 \cot y_2 \csc y_3 \bs{\partial}_{y_4}$, $\cos y_3 \bs{\partial}_{y_2} -  \cot y_2 \sin y_3 \bs{\partial}_{y_3}$ & $\phi_1(y_2) \to \phi_1$, $\phi_2(y_2) \to 0$, $\phi_3(y_2) \to \phi_2$, $\phi_4(y_2) \to \sin^2y_2 \, \phi_2$ \\ \hline
{[4,3,3]$\to$[7,4,2]} & $\cos y_4 (\sin y_3 \bs{\partial}_{y_2} + \cos y_3 \coth y_2 \bs{\partial}_{y_3}) -  \coth y_2 \csc y_3 \sin y_4 \bs{\partial}_{y_4}$, $\sin y_4 (\sin y_3 \bs{\partial}_{y_2} + \cos y_3 \coth y_2 \bs{\partial}_{y_3}) + \cos y_4 \coth y_2 \csc y_3 \bs{\partial}_{y_4}$, $\cos y_3 \bs{\partial}_{y_2} -  \coth y_2 \sin y_3 \bs{\partial}_{y_3}$ & $\phi_1(y_2) \to \phi_1$, $\phi_2(y_2) \to 0$, $\phi_3(y_2) \to \phi_2$, $\phi_4(y_2) \to \sinh^2y_2 \, \phi_2$ \\ \hline
{[4,3,3]\footnote{Upon coordinate transformation $y_1 \to y_4$, $y_2 \to y_1$, $y_3 \to y_2$, $y_4 \to y_3$}$\to$[7,4,3]} & $\cos y_3 (\sin y_2 \bs{\partial}_{y_1} + \cos y_2 \tanh y_1 \bs{\partial}_{y_2}) -  \tanh y_1 \csc y_2 \sin y_3 \bs{\partial}_{y_3}$, $\sin y_3 (\sin y_2 \bs{\partial}_{y_1} + \cos y_2 \tanh y_1 \bs{\partial}_{y_2}) + \cos y_3 \tanh y_1 \csc y_2 \bs{\partial}_{y_3}$, $\cos y_2 \bs{\partial}_{y_1} -  \tanh y_1 \sin y_2 \bs{\partial}_{y_2}$ & $\phi_1(y_1) \to -\phi_2$, $\phi_2(y_1) \to 0$, $\phi_3(y_1) \to -\phi_1$, $\phi_4(y_1) \to \cosh^2y_1 \, \phi_1$ \\ \hline
{[4,3,4]$\to$[5,4,2]} & $\bs{\partial}_{y_2}$ & $\phi_i(y_2) \to \phi_i$ \\ \hline
{[4,3,5]$\to$[5,4,3]} & $\bs{\partial}_{y_2}$ & $\phi_i(y_2) \to \phi_i$ \\ \hline
{[4,3,6]\footnote{Upon coordinate transformation $y_1 \to y_2$, $y_2 \to y_1$.}$\to$[6,3,2]} & $y_3 \bs{\partial}_{y_2} -  y_2 \bs{\partial}_{y_3}$, $y_4 \bs{\partial}_{y_2} -  y_2 \bs{\partial}_{y_4}$ & $\phi_1(y_1) \to - \phi_2(y_1)$, $\phi_2(y_1) \to 0$, $\phi_3(y_1) \to - \phi_1(y_1)$, $\phi_4(y_1) \to \phi_2(y_1)$ \\ \hline
{[4,3,6]\footnote{Upon coordinate transformation $y_2 \to y_3$, $y_3 \to y_2$.}$\to$[6,3,5]} & $y_2 \bs{\partial}_{y_1} + y_1 \bs{\partial}_{y_2}$, $y_4 \bs{\partial}_{y_1} + y_1 \bs{\partial}_{y_4}$ & $\phi_1(y_3) \to \phi_1(y_3)$, $\phi_2(y_3) \to 0$, $\phi_3(y_3) \to \phi_2(y_3)$, $\phi_4(y_3) \to \phi_1(y_3)$ \\ \hline
{[4,3,6]$\to$[6,4,3]} &
$\sin y_1 \tanh y_2 \bs{\partial}_{y_1} - \cos y_1 \bs{\partial}_{y_2}$, $\cos y_1 \tanh y_2 \bs{\partial}_{y_1} + \sin y_1 \bs{\partial}_{y_2}$ & $\phi_1(y_2) \to \cosh^2 y_2 \, \phi_1$, $\phi_2(y_2) \to 0$, $\phi_3(y_2) \to \phi_1$, $\phi_4(y_2) \to \phi_2$ \\ \hline
{[4,3,8]$\to$[6,4,1]} & $\cos y_4 \bs{\partial}_{y_3} -  \cot y_3 \sin y_4 \bs{\partial}_{y_4}$, $\sin y_4 \bs{\partial}_{y_3} + \cos y_4 \cot y_3 \bs{\partial}_{y_4}$ & $\phi_1(y_3) \to \phi_1$, $\phi_2(y_3) \to \phi_2$, $\phi_3(y_3) \to 0$, $\phi_4(y_3) \to \sin^2y_3 \, \phi_2$ \\ \hline
{[4,3,8]$\to$[6,4,2]} & $\cos y_4 \bs{\partial}_{y_3} - \coth y_3 \sin y_4 \bs{\partial}_{y_4}$, $\sin y_4 \bs{\partial}_{y_3} + \cos y_4 \coth y_3 \bs{\partial}_{y_4}$ & $\phi_1(y_3) \to \phi_1$, $\phi_2(y_3) \to \phi_2$, $\phi_3(y_3) \to 0$, $\phi_4(y_3) \to \sinh^2y_3 \, \phi_2$ \\ \hline
{[4,3,8]$\to$[6,4,3]} & $\bs{\partial}_{y_3}$, $y_4 \bs{\partial}_{y_3} - y_3 \bs{\partial}_{y_4}$ & $\phi_1(y_3) \to \phi_1$, $\phi_2(y_3) \to \phi_2$, $\phi_3(y_3) \to 0$, $\phi_4(y_3) \to \phi_2$ \\ \hline
{[4,3,9]$\to$[5,4,7]} & $\bs{\partial}_{y_4}$ & $\phi_i(y_4) \to \phi_i$ \\ \hline
{[4,3,10]$\to$[5,4,6]} & $\bs{\partial}_{y_2}$ & $\phi_i(y_2) \to \phi_i$ \\ \hline
{[4,3,11]$\to$[6,3,5]} & $y_4 \bs{\partial}_{y_1} + y_1 \bs{\partial}_{y_4}$, $y_4 \bs{\partial}_{y_2} - y_2 \bs{\partial}_{y_4}$ & $\phi_1(y_3) \to \phi_1(y_3)$, $\phi_2(y_3) \to \phi_2(y_3)$, $\phi_3(y_3) \to 0$, $\phi_4(y_3) \to \phi_1(y_3)$ \\ \hline
{[4,3,11]$\to$[6,4,4]} & $\cos y_4 \bs{\partial}_{y_3} -  \cot y_3 \sin y_4 \bs{\partial}_{y_4}$, $\sin y_4 \bs{\partial}_{y_3} + \cos y_4 \cot y_3 \bs{\partial}_{y_4}$ & $\phi_1(y_3) \to \phi_1$, $\phi_2(y_3) \to \phi_2$, $\phi_3(y_3) \to 0$, $\phi_4(y_3) \to \sin^2y_3 \, \phi_2$ \\ \hline
{[4,3,11]$\to$[6,4,5]} & $\cos y_4 \bs{\partial}_{y_3} -  \coth y_3 \sin y_4 \bs{\partial}_{y_4}$, $\sin y_4 \bs{\partial}_{y_3} + \cos y_4 \coth y_3 \bs{\partial}_{y_4}$ & $\phi_1(y_3) \to \phi_1$, $\phi_2(y_3) \to \phi_2$, $\phi_3(y_3) \to 0$, $\phi_4(y_3) \to \sinh^2y_3 \, \phi_2$ \\ \hline\hline
{[4,4,1]$\to$[5,4,2]} & $\bs{\partial}_{y_1}$ & $\phi_i \to \phi_i$ for $1\leq i \leq 4$, $\phi_i \to 0$ for $5 \leq i \leq 9$, $\phi_{10} \to \phi_4$ \\ \hline
{[4,4,2]$\to$[5,4,1]} & $\bs{\partial}_{y_1}$ & $\phi_i \to \phi_i$ for $1\leq i \leq 4$, $\phi_i \to 0$ for $5 \leq i \leq 9$, $\phi_{10} \to \phi_4$ \\ \hline
{[4,4,18]$\to$[5,4,6]} & $\bs{\partial}_{y_2}$ & $\phi_i \to \phi_i$ for $1\leq i \leq 3$, $\phi_i \to 0$ for $4 \leq i \leq 8$, $\phi_9 \to \phi_4$, $\phi_{10} \to 0$ \\ \hline
{[4,4,22]$\to$[5,4,3]} & $\bs{\partial}_{y_2}$ & $\phi_i \to \phi_i$ for $1\leq i \leq 4$, $\phi_i \to 0$ for $5 \leq i \leq 9$, $\phi_{10} \to \phi_4$ \\ \hline \hline
{[6,3,1]$\to$[7,4,1]} & $\bs{\partial}_{y_1}$ & $\phi_i(y_1) \to \phi_i$ \\ \hline
{[6,3,3]$\to$[7,4,2]} & $\bs{\partial}_{y_1}$ & $\phi_i(y_1) \to \phi_i$ \\ \hline
{[6,3,4]$\to$[7,4,4]} & $\bs{\partial}_{y_4}$ & $\phi_i(y_4) \to \phi_i$ \\ \hline
{[6,3,6]$\to$[7,4,3]} & $\bs{\partial}_{y_4}$ & $\phi_i(y_4) \to \phi_i$ \\ \hline
\end{longtable}




\begin{thebibliography}{45}%
\makeatletter
\providecommand \@ifxundefined [1]{%
 \@ifx{#1\undefined}
}%
\providecommand \@ifnum [1]{%
 \ifnum #1\expandafter \@firstoftwo
 \else \expandafter \@secondoftwo
 \fi
}%
\providecommand \@ifx [1]{%
 \ifx #1\expandafter \@firstoftwo
 \else \expandafter \@secondoftwo
 \fi
}%
\providecommand \natexlab [1]{#1}%
\providecommand \enquote  [1]{``#1''}%
\providecommand \bibnamefont  [1]{#1}%
\providecommand \bibfnamefont [1]{#1}%
\providecommand \citenamefont [1]{#1}%
\providecommand \href@noop [0]{\@secondoftwo}%
\providecommand \href [0]{\begingroup \@sanitize@url \@href}%
\providecommand \@href[1]{\@@startlink{#1}\@@href}%
\providecommand \@@href[1]{\endgroup#1\@@endlink}%
\providecommand \@sanitize@url [0]{\catcode `\\12\catcode `\$12\catcode `\&12\catcode `\#12\catcode `\^12\catcode `\_12\catcode `\%12\relax}%
\providecommand \@@startlink[1]{}%
\providecommand \@@endlink[0]{}%
\providecommand \url  [0]{\begingroup\@sanitize@url \@url }%
\providecommand \@url [1]{\endgroup\@href {#1}{\urlprefix }}%
\providecommand \urlprefix  [0]{URL }%
\providecommand \Eprint [0]{\href }%
\providecommand \doibase [0]{https://doi.org/}%
\providecommand \selectlanguage [0]{\@gobble}%
\providecommand \bibinfo  [0]{\@secondoftwo}%
\providecommand \bibfield  [0]{\@secondoftwo}%
\providecommand \translation [1]{[#1]}%
\providecommand \BibitemOpen [0]{}%
\providecommand \bibitemStop [0]{}%
\providecommand \bibitemNoStop [0]{.\EOS\space}%
\providecommand \EOS [0]{\spacefactor3000\relax}%
\providecommand \BibitemShut  [1]{\csname bibitem#1\endcsname}%
\let\auto@bib@innerbib\@empty
\bibitem [{\citenamefont {Stephani}\ \emph {et~al.}(2009)\citenamefont {Stephani}, \citenamefont {Kramer}, \citenamefont {MacCallum}, \citenamefont {Hoenselaers},\ and\ \citenamefont {Herlt}}]{Stephani2009-dt}%
  \BibitemOpen
  \bibfield  {author} {\bibinfo {author} {\bibfnamefont {H.}~\bibnamefont {Stephani}}, \bibinfo {author} {\bibfnamefont {D.}~\bibnamefont {Kramer}}, \bibinfo {author} {\bibfnamefont {M.}~\bibnamefont {MacCallum}}, \bibinfo {author} {\bibfnamefont {C.}~\bibnamefont {Hoenselaers}},\ and\ \bibinfo {author} {\bibfnamefont {E.}~\bibnamefont {Herlt}},\ }\href@noop {} {\emph {\bibinfo {title} {Cambridge monographs on mathematical physics: Exact solutions of Einstein's field equations}}},\ \bibinfo {edition} {2nd}\ ed.\ (\bibinfo  {publisher} {Cambridge University Press},\ \bibinfo {address} {Cambridge, England},\ \bibinfo {year} {2009})\BibitemShut {NoStop}%
\bibitem [{\citenamefont {Griffiths}\ and\ \citenamefont {Podolsk\'y}(2009)}]{Griffiths:2009dfa}%
  \BibitemOpen
  \bibfield  {author} {\bibinfo {author} {\bibfnamefont {J.~B.}\ \bibnamefont {Griffiths}}\ and\ \bibinfo {author} {\bibfnamefont {J.}~\bibnamefont {Podolsk\'y}},\ }\href {https://doi.org/10.1017/CBO9780511635397} {\emph {\bibinfo {title} {{Exact Space-Times in Einstein's General Relativity}}}},\ Cambridge Monographs on Mathematical Physics\ (\bibinfo  {publisher} {Cambridge University Press},\ \bibinfo {address} {Cambridge},\ \bibinfo {year} {2009})\BibitemShut {NoStop}%
\bibitem [{\citenamefont {Weyl}(1917)}]{Weyl:1917gp}%
  \BibitemOpen
  \bibfield  {author} {\bibinfo {author} {\bibfnamefont {H.}~\bibnamefont {Weyl}},\ }\bibfield  {title} {\bibinfo {title} {{The theory of gravitation}},\ }\href {https://doi.org/10.1007/s10714-011-1310-7} {\bibfield  {journal} {\bibinfo  {journal} {Annalen Phys.}\ }\textbf {\bibinfo {volume} {54}},\ \bibinfo {pages} {117} (\bibinfo {year} {1917})}\BibitemShut {NoStop}%
\bibitem [{\citenamefont {Lovelock}(1973)}]{Lovelock1973SphericallySM}%
  \BibitemOpen
  \bibfield  {author} {\bibinfo {author} {\bibfnamefont {D.}~\bibnamefont {Lovelock}},\ }\bibfield  {title} {\bibinfo {title} {Spherically symmetric metrics and field equations in four-dimensional space},\ }\href {https://api.semanticscholar.org/CorpusID:117848886} {\bibfield  {journal} {\bibinfo  {journal} {Il Nuovo Cimento B (1971-1996)}\ }\textbf {\bibinfo {volume} {14}},\ \bibinfo {pages} {260} (\bibinfo {year} {1973})}\BibitemShut {NoStop}%
\bibitem [{\citenamefont {Deser}\ and\ \citenamefont {Tekin}(2003)}]{Deser:2003up}%
  \BibitemOpen
  \bibfield  {author} {\bibinfo {author} {\bibfnamefont {S.}~\bibnamefont {Deser}}\ and\ \bibinfo {author} {\bibfnamefont {B.}~\bibnamefont {Tekin}},\ }\bibfield  {title} {\bibinfo {title} {{Shortcuts to high symmetry solutions in gravitational theories}},\ }\href {https://doi.org/10.1088/0264-9381/20/22/011} {\bibfield  {journal} {\bibinfo  {journal} {Class. Quant. Grav.}\ }\textbf {\bibinfo {volume} {20}},\ \bibinfo {pages} {4877} (\bibinfo {year} {2003})},\ \Eprint {https://arxiv.org/abs/gr-qc/0306114} {arXiv:gr-qc/0306114} \BibitemShut {NoStop}%
\bibitem [{\citenamefont {Deser}\ and\ \citenamefont {Franklin}(2005{\natexlab{a}})}]{Deser:2004gi}%
  \BibitemOpen
  \bibfield  {author} {\bibinfo {author} {\bibfnamefont {S.}~\bibnamefont {Deser}}\ and\ \bibinfo {author} {\bibfnamefont {J.}~\bibnamefont {Franklin}},\ }\bibfield  {title} {\bibinfo {title} {{Schwarzschild and Birkhoff a la Weyl}},\ }\href {https://doi.org/10.1119/1.1830505} {\bibfield  {journal} {\bibinfo  {journal} {Am. J. Phys.}\ }\textbf {\bibinfo {volume} {73}},\ \bibinfo {pages} {261} (\bibinfo {year} {2005}{\natexlab{a}})},\ \Eprint {https://arxiv.org/abs/gr-qc/0408067} {arXiv:gr-qc/0408067} \BibitemShut {NoStop}%
\bibitem [{\citenamefont {Deser}\ and\ \citenamefont {Franklin}(2005{\natexlab{b}})}]{Deser:2005gr}%
  \BibitemOpen
  \bibfield  {author} {\bibinfo {author} {\bibfnamefont {S.}~\bibnamefont {Deser}}\ and\ \bibinfo {author} {\bibfnamefont {J.}~\bibnamefont {Franklin}},\ }\bibfield  {title} {\bibinfo {title} {{Birkhoff for Lovelock redux}},\ }\href {https://doi.org/10.1088/0264-9381/22/16/L03} {\bibfield  {journal} {\bibinfo  {journal} {Class. Quant. Grav.}\ }\textbf {\bibinfo {volume} {22}},\ \bibinfo {pages} {L103} (\bibinfo {year} {2005}{\natexlab{b}})},\ \Eprint {https://arxiv.org/abs/gr-qc/0506014} {arXiv:gr-qc/0506014} \BibitemShut {NoStop}%
\bibitem [{\citenamefont {Franklin}(2019)}]{sym11070845}%
  \BibitemOpen
  \bibfield  {author} {\bibinfo {author} {\bibfnamefont {J.}~\bibnamefont {Franklin}},\ }\bibfield  {title} {\bibinfo {title} {Symmetric criticality and magnetic monopoles in general relativity},\ }\href {https://www.mdpi.com/2073-8994/11/7/845} {\bibfield  {journal} {\bibinfo  {journal} {Symmetry}\ }\textbf {\bibinfo {volume} {11}} (\bibinfo {year} {2019})}\BibitemShut {NoStop}%
\bibitem [{\citenamefont {Hawking}(1969)}]{Hawking:1968zw}%
  \BibitemOpen
  \bibfield  {author} {\bibinfo {author} {\bibfnamefont {S.~W.}\ \bibnamefont {Hawking}},\ }\bibfield  {title} {\bibinfo {title} {{On the Rotation of the universe}},\ }\href@noop {} {\bibfield  {journal} {\bibinfo  {journal} {Mon. Not. Roy. Astron. Soc.}\ }\textbf {\bibinfo {volume} {142}},\ \bibinfo {pages} {129} (\bibinfo {year} {1969})}\BibitemShut {NoStop}%
\bibitem [{\citenamefont {Maccallum}\ and\ \citenamefont {Taub}(1972)}]{Maccallum:1972er}%
  \BibitemOpen
  \bibfield  {author} {\bibinfo {author} {\bibfnamefont {M.~A.~H.}\ \bibnamefont {Maccallum}}\ and\ \bibinfo {author} {\bibfnamefont {A.~H.}\ \bibnamefont {Taub}},\ }\bibfield  {title} {\bibinfo {title} {{Variational principles and spatially-homogeneous universes, including rotation}},\ }\href {https://doi.org/10.1007/BF01877589} {\bibfield  {journal} {\bibinfo  {journal} {Commun. Math. Phys.}\ }\textbf {\bibinfo {volume} {25}},\ \bibinfo {pages} {173} (\bibinfo {year} {1972})}\BibitemShut {NoStop}%
\bibitem [{\citenamefont {DeWitt}(1967)}]{DeWitt:1967}%
  \BibitemOpen
  \bibfield  {author} {\bibinfo {author} {\bibfnamefont {B.~S.}\ \bibnamefont {DeWitt}},\ }\bibfield  {title} {\bibinfo {title} {Quantum theory of gravity. i. the canonical theory},\ }\href {https://doi.org/10.1103/PhysRev.160.1113} {\bibfield  {journal} {\bibinfo  {journal} {Phys. Rev.}\ }\textbf {\bibinfo {volume} {160}},\ \bibinfo {pages} {1113} (\bibinfo {year} {1967})}\BibitemShut {NoStop}%
\bibitem [{\citenamefont {Misner}(1969)}]{Misner:1969}%
  \BibitemOpen
  \bibfield  {author} {\bibinfo {author} {\bibfnamefont {C.~W.}\ \bibnamefont {Misner}},\ }\bibfield  {title} {\bibinfo {title} {Mixmaster universe},\ }\href {https://doi.org/10.1103/PhysRevLett.22.1071} {\bibfield  {journal} {\bibinfo  {journal} {Phys. Rev. Lett.}\ }\textbf {\bibinfo {volume} {22}},\ \bibinfo {pages} {1071} (\bibinfo {year} {1969})}\BibitemShut {NoStop}%
\bibitem [{\citenamefont {Kucha\ifmmode~\check{r}\else \v{r}\fi{}}(1971)}]{Kuchar:1971}%
  \BibitemOpen
  \bibfield  {author} {\bibinfo {author} {\bibfnamefont {K.}~\bibnamefont {Kucha\ifmmode~\check{r}\else \v{r}\fi{}}},\ }\bibfield  {title} {\bibinfo {title} {Canonical quantization of cylindrical gravitational waves},\ }\href {https://doi.org/10.1103/PhysRevD.4.955} {\bibfield  {journal} {\bibinfo  {journal} {Phys. Rev. D}\ }\textbf {\bibinfo {volume} {4}},\ \bibinfo {pages} {955} (\bibinfo {year} {1971})}\BibitemShut {NoStop}%
\bibitem [{\citenamefont {Torre}(1999)}]{Torre:1998dy}%
  \BibitemOpen
  \bibfield  {author} {\bibinfo {author} {\bibfnamefont {C.~G.}\ \bibnamefont {Torre}},\ }\bibfield  {title} {\bibinfo {title} {{Midisuperspace models of canonical quantum gravity}},\ }\href {https://doi.org/10.1023/A:1026650212053} {\bibfield  {journal} {\bibinfo  {journal} {Int. J. Theor. Phys.}\ }\textbf {\bibinfo {volume} {38}},\ \bibinfo {pages} {1081} (\bibinfo {year} {1999})},\ \Eprint {https://arxiv.org/abs/gr-qc/9806122} {arXiv:gr-qc/9806122} \BibitemShut {NoStop}%
\bibitem [{\citenamefont {Oliva}\ and\ \citenamefont {Ray}(2010)}]{Oliva:2010eb}%
  \BibitemOpen
  \bibfield  {author} {\bibinfo {author} {\bibfnamefont {J.}~\bibnamefont {Oliva}}\ and\ \bibinfo {author} {\bibfnamefont {S.}~\bibnamefont {Ray}},\ }\bibfield  {title} {\bibinfo {title} {{A new cubic theory of gravity in five dimensions: Black hole, Birkhoff's theorem and C-function}},\ }\href {https://doi.org/10.1088/0264-9381/27/22/225002} {\bibfield  {journal} {\bibinfo  {journal} {Class. Quant. Grav.}\ }\textbf {\bibinfo {volume} {27}},\ \bibinfo {pages} {225002} (\bibinfo {year} {2010})},\ \Eprint {https://arxiv.org/abs/1003.4773} {arXiv:1003.4773 [gr-qc]} \BibitemShut {NoStop}%
\bibitem [{\citenamefont {Myers}\ and\ \citenamefont {Robinson}(2010)}]{Myers:2010ru}%
  \BibitemOpen
  \bibfield  {author} {\bibinfo {author} {\bibfnamefont {R.~C.}\ \bibnamefont {Myers}}\ and\ \bibinfo {author} {\bibfnamefont {B.}~\bibnamefont {Robinson}},\ }\bibfield  {title} {\bibinfo {title} {{Black Holes in Quasi-topological Gravity}},\ }\href {https://doi.org/10.1007/JHEP08(2010)067} {\bibfield  {journal} {\bibinfo  {journal} {JHEP}\ }\textbf {\bibinfo {volume} {2010}}\bibfield  {number} {\bibinfo  {number} { (8)},\ \bibinfo {pages} {067}},\ }\Eprint {https://arxiv.org/abs/1003.5357} {arXiv:1003.5357 [gr-qc]} \BibitemShut {NoStop}%
\bibitem [{\citenamefont {Dehghani}\ \emph {et~al.}(2012)\citenamefont {Dehghani}, \citenamefont {Bazrafshan}, \citenamefont {Mann}, \citenamefont {Mehdizadeh}, \citenamefont {Ghanaatian},\ and\ \citenamefont {Vahidinia}}]{Dehghani:2011vu}%
  \BibitemOpen
  \bibfield  {author} {\bibinfo {author} {\bibfnamefont {M.~H.}\ \bibnamefont {Dehghani}}, \bibinfo {author} {\bibfnamefont {A.}~\bibnamefont {Bazrafshan}}, \bibinfo {author} {\bibfnamefont {R.~B.}\ \bibnamefont {Mann}}, \bibinfo {author} {\bibfnamefont {M.~R.}\ \bibnamefont {Mehdizadeh}}, \bibinfo {author} {\bibfnamefont {M.}~\bibnamefont {Ghanaatian}},\ and\ \bibinfo {author} {\bibfnamefont {M.~H.}\ \bibnamefont {Vahidinia}},\ }\bibfield  {title} {\bibinfo {title} {{Black Holes in Quartic Quasitopological Gravity}},\ }\href {https://doi.org/10.1103/PhysRevD.85.104009} {\bibfield  {journal} {\bibinfo  {journal} {Phys. Rev. D}\ }\textbf {\bibinfo {volume} {85}},\ \bibinfo {pages} {104009} (\bibinfo {year} {2012})},\ \Eprint {https://arxiv.org/abs/1109.4708} {arXiv:1109.4708 [hep-th]} \BibitemShut {NoStop}%
\bibitem [{\citenamefont {Cisterna}\ \emph {et~al.}(2017)\citenamefont {Cisterna}, \citenamefont {Guajardo}, \citenamefont {Hassaine},\ and\ \citenamefont {Oliva}}]{Cisterna:2017umf}%
  \BibitemOpen
  \bibfield  {author} {\bibinfo {author} {\bibfnamefont {A.}~\bibnamefont {Cisterna}}, \bibinfo {author} {\bibfnamefont {L.}~\bibnamefont {Guajardo}}, \bibinfo {author} {\bibfnamefont {M.}~\bibnamefont {Hassaine}},\ and\ \bibinfo {author} {\bibfnamefont {J.}~\bibnamefont {Oliva}},\ }\bibfield  {title} {\bibinfo {title} {{Quintic quasi-topological gravity}},\ }\href {https://doi.org/10.1007/JHEP04(2017)066} {\bibfield  {journal} {\bibinfo  {journal} {JHEP}\ }\textbf {\bibinfo {volume} {2017}}\bibfield  {number} {\bibinfo  {number} { (4)},\ \bibinfo {pages} {066}},\ }\Eprint {https://arxiv.org/abs/1702.04676} {arXiv:1702.04676 [hep-th]} \BibitemShut {NoStop}%
\bibitem [{\citenamefont {Bueno}\ \emph {et~al.}(2020)\citenamefont {Bueno}, \citenamefont {Cano},\ and\ \citenamefont {Hennigar}}]{Bueno:2019ycr}%
  \BibitemOpen
  \bibfield  {author} {\bibinfo {author} {\bibfnamefont {P.}~\bibnamefont {Bueno}}, \bibinfo {author} {\bibfnamefont {P.~A.}\ \bibnamefont {Cano}},\ and\ \bibinfo {author} {\bibfnamefont {R.~A.}\ \bibnamefont {Hennigar}},\ }\bibfield  {title} {\bibinfo {title} {{(Generalized) quasi-topological gravities at all orders}},\ }\href {https://doi.org/10.1088/1361-6382/ab5410} {\bibfield  {journal} {\bibinfo  {journal} {Class. Quant. Grav.}\ }\textbf {\bibinfo {volume} {37}},\ \bibinfo {pages} {015002} (\bibinfo {year} {2020})},\ \Eprint {https://arxiv.org/abs/1909.07983} {arXiv:1909.07983 [hep-th]} \BibitemShut {NoStop}%
\bibitem [{\citenamefont {Bueno}\ \emph {et~al.}(2023)\citenamefont {Bueno}, \citenamefont {Cano}, \citenamefont {Hennigar}, \citenamefont {Lu},\ and\ \citenamefont {Moreno}}]{Bueno:2022res}%
  \BibitemOpen
  \bibfield  {author} {\bibinfo {author} {\bibfnamefont {P.}~\bibnamefont {Bueno}}, \bibinfo {author} {\bibfnamefont {P.~A.}\ \bibnamefont {Cano}}, \bibinfo {author} {\bibfnamefont {R.~A.}\ \bibnamefont {Hennigar}}, \bibinfo {author} {\bibfnamefont {M.}~\bibnamefont {Lu}},\ and\ \bibinfo {author} {\bibfnamefont {J.}~\bibnamefont {Moreno}},\ }\bibfield  {title} {\bibinfo {title} {{Generalized quasi-topological gravities: the whole shebang}},\ }\href {https://doi.org/10.1088/1361-6382/aca236} {\bibfield  {journal} {\bibinfo  {journal} {Class. Quant. Grav.}\ }\textbf {\bibinfo {volume} {40}},\ \bibinfo {pages} {015004} (\bibinfo {year} {2023})},\ \Eprint {https://arxiv.org/abs/2203.05589} {arXiv:2203.05589 [hep-th]} \BibitemShut {NoStop}%
\bibitem [{\citenamefont {Moreno}\ and\ \citenamefont {Murcia}(2023)}]{Moreno:2023arp}%
  \BibitemOpen
  \bibfield  {author} {\bibinfo {author} {\bibfnamefont {J.}~\bibnamefont {Moreno}}\ and\ \bibinfo {author} {\bibfnamefont {A.~J.}\ \bibnamefont {Murcia}},\ }\href@noop {} {\bibinfo {title} {{Cosmological higher-curvature gravities}}} (\bibinfo {year} {2023}),\ \Eprint {https://arxiv.org/abs/2311.12104} {arXiv:2311.12104 [gr-qc]} \BibitemShut {NoStop}%
\bibitem [{\citenamefont {Fels}\ and\ \citenamefont {Torre}(2002)}]{Fels:2001rv}%
  \BibitemOpen
  \bibfield  {author} {\bibinfo {author} {\bibfnamefont {M.~E.}\ \bibnamefont {Fels}}\ and\ \bibinfo {author} {\bibfnamefont {C.~G.}\ \bibnamefont {Torre}},\ }\bibfield  {title} {\bibinfo {title} {{The Principle of symmetric criticality in general relativity}},\ }\href {https://doi.org/10.1088/0264-9381/19/4/303} {\bibfield  {journal} {\bibinfo  {journal} {Class. Quant. Grav.}\ }\textbf {\bibinfo {volume} {19}},\ \bibinfo {pages} {641} (\bibinfo {year} {2002})},\ \Eprint {https://arxiv.org/abs/gr-qc/0108033} {arXiv:gr-qc/0108033} \BibitemShut {NoStop}%
\bibitem [{\citenamefont {Anderson}\ and\ \citenamefont {Fels}(1997)}]{Anderson1997}%
  \BibitemOpen
  \bibfield  {author} {\bibinfo {author} {\bibfnamefont {I.~M.}\ \bibnamefont {Anderson}}\ and\ \bibinfo {author} {\bibfnamefont {M.~E.}\ \bibnamefont {Fels}},\ }\bibfield  {title} {\bibinfo {title} {Symmetry reduction of variational bicomplexes and the principle of symmetric criticality},\ }\href {https://doi.org/10.1353/ajm.1997.0015} {\bibfield  {journal} {\bibinfo  {journal} {American Journal of Mathematics}\ }\textbf {\bibinfo {volume} {119}},\ \bibinfo {pages} {609} (\bibinfo {year} {1997})}\BibitemShut {NoStop}%
\bibitem [{\citenamefont {Anderson}\ \emph {et~al.}(2000)\citenamefont {Anderson}, \citenamefont {Fels},\ and\ \citenamefont {Torre}}]{Anderson:1999cn}%
  \BibitemOpen
  \bibfield  {author} {\bibinfo {author} {\bibfnamefont {I.~M.}\ \bibnamefont {Anderson}}, \bibinfo {author} {\bibfnamefont {M.~E.}\ \bibnamefont {Fels}},\ and\ \bibinfo {author} {\bibfnamefont {C.~G.}\ \bibnamefont {Torre}},\ }\bibfield  {title} {\bibinfo {title} {{Group Invariant Solutions Without Transversality}},\ }\href {https://doi.org/10.1007/s002200000215} {\bibfield  {journal} {\bibinfo  {journal} {Commun. Math. Phys.}\ }\textbf {\bibinfo {volume} {212}},\ \bibinfo {pages} {653} (\bibinfo {year} {2000})},\ \Eprint {https://arxiv.org/abs/math-ph/9910015} {arXiv:math-ph/9910015} \BibitemShut {NoStop}%
\bibitem [{\citenamefont {Anderson}\ \emph {et~al.}(1999)\citenamefont {Anderson}, \citenamefont {Fels},\ and\ \citenamefont {Torre}}]{Anderson:1999cm}%
  \BibitemOpen
  \bibfield  {author} {\bibinfo {author} {\bibfnamefont {I.~M.}\ \bibnamefont {Anderson}}, \bibinfo {author} {\bibfnamefont {M.~E.}\ \bibnamefont {Fels}},\ and\ \bibinfo {author} {\bibfnamefont {C.~G.}\ \bibnamefont {Torre}},\ }\href@noop {} {\bibinfo {title} {{Group invariant solutions without transversality and the principle of symmetric criticality}}} (\bibinfo {year} {1999}),\ \Eprint {https://arxiv.org/abs/math-ph/9910014} {arXiv:math-ph/9910014} \BibitemShut {NoStop}%
\bibitem [{\citenamefont {Torre}(2011)}]{Torre:2010xa}%
  \BibitemOpen
  \bibfield  {author} {\bibinfo {author} {\bibfnamefont {C.~G.}\ \bibnamefont {Torre}},\ }\bibfield  {title} {\bibinfo {title} {{Symmetric Criticality in Classical Field Theory}},\ }\href {https://doi.org/10.1063/1.3599128} {\bibfield  {journal} {\bibinfo  {journal} {AIP Conf. Proc.}\ }\textbf {\bibinfo {volume} {1360}},\ \bibinfo {pages} {63} (\bibinfo {year} {2011})},\ \Eprint {https://arxiv.org/abs/1011.3429} {arXiv:1011.3429 [math-ph]} \BibitemShut {NoStop}%
\bibitem [{\citenamefont {Petrov}(1969)}]{Petrov}%
  \BibitemOpen
  \bibfield  {author} {\bibinfo {author} {\bibfnamefont {A.~Z.}\ \bibnamefont {Petrov}},\ }\href@noop {} {\emph {\bibinfo {title} {Einstein Spaces}}}\ (\bibinfo  {publisher} {Pergamon},\ \bibinfo {year} {1969})\BibitemShut {NoStop}%
\bibitem [{\citenamefont {Hicks}(2016)}]{Hicks:thesis}%
  \BibitemOpen
  \bibfield  {author} {\bibinfo {author} {\bibfnamefont {J.~W.}\ \bibnamefont {Hicks}},\ }\emph {\bibinfo {title} {Classification of Spacetimes with Symmetry}},\ \href {https://doi.org/10.26076/1A08-57A2} {Ph.D. thesis},\ \bibinfo  {school} {Utah State University} (\bibinfo {year} {2016})\BibitemShut {NoStop}%
\bibitem [{\citenamefont {Fels}\ and\ \citenamefont {Renner}(2006)}]{Fels_Renner_2006}%
  \BibitemOpen
  \bibfield  {author} {\bibinfo {author} {\bibfnamefont {M.~E.}\ \bibnamefont {Fels}}\ and\ \bibinfo {author} {\bibfnamefont {A.~G.}\ \bibnamefont {Renner}},\ }\bibfield  {title} {\bibinfo {title} {Non-reductive homogeneous pseudo-{Riemannian} manifolds of dimension four},\ }\href {https://doi.org/10.4153/CJM-2006-012-1} {\bibfield  {journal} {\bibinfo  {journal} {Canadian Journal of Mathematics}\ }\textbf {\bibinfo {volume} {58}},\ \bibinfo {pages} {282–311} (\bibinfo {year} {2006})}\BibitemShut {NoStop}%
\bibitem [{\citenamefont {Bowers}(2012)}]{Bowers:2012}%
  \BibitemOpen
  \bibfield  {author} {\bibinfo {author} {\bibfnamefont {A.}~\bibnamefont {Bowers}},\ }\bibfield  {title} {\bibinfo {title} {An algebraic construction of lorentz homogeneous spaces of low dimension},\ }\href@noop {} {\bibfield  {journal} {\bibinfo  {journal} {Journal of Lie Theory}\ }\textbf {\bibinfo {volume} {3}} (\bibinfo {year} {2012})}\BibitemShut {NoStop}%
\bibitem [{\citenamefont {{\v S}nobl}\ and\ \citenamefont {Winternitz}(2014)}]{Snobl2014-te}%
  \BibitemOpen
  \bibfield  {author} {\bibinfo {author} {\bibfnamefont {L.}~\bibnamefont {{\v S}nobl}}\ and\ \bibinfo {author} {\bibfnamefont {P.}~\bibnamefont {Winternitz}},\ }\href@noop {} {\emph {\bibinfo {title} {Classification and Identification of Lie Algebras}}},\ CRM monograph series\ (\bibinfo  {publisher} {American Mathematical Society},\ \bibinfo {address} {Providence, Rhode Island},\ \bibinfo {year} {2014})\BibitemShut {NoStop}%
\bibitem [{\citenamefont {Rozum}(2015)}]{Rozum2015-lp}%
  \BibitemOpen
  \bibfield  {author} {\bibinfo {author} {\bibfnamefont {J.}~\bibnamefont {Rozum}},\ }\emph {\bibinfo {title} {Classification of five-dimensional Lie algebras with one-dimensional subalgebras acting as subalgebras of the Lorentz algebra}},\ \href@noop {} {Master's thesis},\ \bibinfo  {school} {Utah State University} (\bibinfo {year} {2015})\BibitemShut {NoStop}%
\bibitem [{\citenamefont {Anderson}\ and\ \citenamefont {Torre}(2022)}]{diffgeo}%
  \BibitemOpen
  \bibfield  {author} {\bibinfo {author} {\bibfnamefont {I.~M.}\ \bibnamefont {Anderson}}\ and\ \bibinfo {author} {\bibfnamefont {C.~G.}\ \bibnamefont {Torre}},\ }\href {https://digitalcommons.usu.edu/dg_downloads/4} {\bibinfo {title} {{The DifferentialGeometry Package}}} (\bibinfo {year} {2022})\BibitemShut {NoStop}%
\bibitem [{\citenamefont {{Maplesoft, a division of Waterloo Maple Inc..}}(2019)}]{maple}%
  \BibitemOpen
  \bibfield  {author} {\bibinfo {author} {\bibnamefont {{Maplesoft, a division of Waterloo Maple Inc..}}},\ }\href {https://hadoop.apache.org} {\bibinfo {title} {Maple}} (\bibinfo {year} {2019})\BibitemShut {NoStop}%
\bibitem [{\citenamefont {Hwang}(2019)}]{Hwang:thesis}%
  \BibitemOpen
  \bibfield  {author} {\bibinfo {author} {\bibfnamefont {E.}~\bibnamefont {Hwang}},\ }\emph {\bibinfo {title} {Classification of isometry algebras of solutions of Einstein's field equations}},\ \href {https://doi.org/10.26076/21e0-b84b} {Master's thesis},\ \bibinfo  {school} {Utah State University} (\bibinfo {year} {2019})\BibitemShut {NoStop}%
\bibitem [{\citenamefont {Mart\'in-Garc\'ia}(2022)}]{xact}%
  \BibitemOpen
  \bibfield  {author} {\bibinfo {author} {\bibfnamefont {J.~M.}\ \bibnamefont {Mart\'in-Garc\'ia}},\ }\href {http://www.xact.es/} {\bibinfo {title} {{xAct: Efficien tensor computer algebra for the Wolfram Language}}} (\bibinfo {year} {2022})\BibitemShut {NoStop}%
\bibitem [{\citenamefont {Inc.}(2023)}]{Mathematica}%
  \BibitemOpen
  \bibfield  {author} {\bibinfo {author} {\bibfnamefont {W.~R.}\ \bibnamefont {Inc.}},\ }\href {https://www.wolfram.com/mathematica} {\bibinfo {title} {{Mathematica, {V}ersion 13.3}}} (\bibinfo {year} {2023}),\ \bibinfo {note} {champaign, IL, 2023}\BibitemShut {NoStop}%
\bibitem [{\citenamefont {Patera}\ \emph {et~al.}(1975)\citenamefont {Patera}, \citenamefont {Winternitz},\ and\ \citenamefont {Zassenhaus}}]{PateraWinternitzZassenhaus}%
  \BibitemOpen
  \bibfield  {author} {\bibinfo {author} {\bibfnamefont {J.}~\bibnamefont {Patera}}, \bibinfo {author} {\bibfnamefont {P.}~\bibnamefont {Winternitz}},\ and\ \bibinfo {author} {\bibfnamefont {H.}~\bibnamefont {Zassenhaus}},\ }\bibfield  {title} {\bibinfo {title} {{Continuous subgroups of the fundamental groups of physics. I. General method and the Poincaré group}},\ }\href {https://doi.org/10.1063/1.522729} {\bibfield  {journal} {\bibinfo  {journal} {Journal of Mathematical Physics}\ }\textbf {\bibinfo {volume} {16}},\ \bibinfo {pages} {1597} (\bibinfo {year} {1975})},\ \Eprint {https://arxiv.org/abs/https://pubs.aip.org/aip/jmp/article-pdf/16/8/1597/19174337/1597\_1\_online.pdf} {https://pubs.aip.org/aip/jmp/article-pdf/16/8/1597/19174337/1597\_1\_online.pdf} \BibitemShut {NoStop}%
\bibitem [{\citenamefont {Iyer}\ and\ \citenamefont {Wald}(1994{\natexlab{a}})}]{Iyer:1994ys}%
  \BibitemOpen
  \bibfield  {author} {\bibinfo {author} {\bibfnamefont {V.}~\bibnamefont {Iyer}}\ and\ \bibinfo {author} {\bibfnamefont {R.~M.}\ \bibnamefont {Wald}},\ }\bibfield  {title} {\bibinfo {title} {{Some properties of Noether charge and a proposal for dynamical black hole entropy}},\ }\href {https://doi.org/10.1103/PhysRevD.50.846} {\bibfield  {journal} {\bibinfo  {journal} {Phys. Rev. D}\ }\textbf {\bibinfo {volume} {50}},\ \bibinfo {pages} {846} (\bibinfo {year} {1994}{\natexlab{a}})},\ \Eprint {https://arxiv.org/abs/gr-qc/9403028} {arXiv:gr-qc/9403028} \BibitemShut {NoStop}%
\bibitem [{\citenamefont {Palais}(1979)}]{Palais:1979rca}%
  \BibitemOpen
  \bibfield  {author} {\bibinfo {author} {\bibfnamefont {R.~S.}\ \bibnamefont {Palais}},\ }\bibfield  {title} {\bibinfo {title} {{The principle of symmetric criticality}},\ }\href {https://doi.org/10.1007/BF01941322} {\bibfield  {journal} {\bibinfo  {journal} {Commun. Math. Phys.}\ }\textbf {\bibinfo {volume} {69}},\ \bibinfo {pages} {19} (\bibinfo {year} {1979})}\BibitemShut {NoStop}%
\bibitem [{\citenamefont {Bardeen}\ and\ \citenamefont {Horowitz}(1999)}]{Bardeen:1999px}%
  \BibitemOpen
  \bibfield  {author} {\bibinfo {author} {\bibfnamefont {J.~M.}\ \bibnamefont {Bardeen}}\ and\ \bibinfo {author} {\bibfnamefont {G.~T.}\ \bibnamefont {Horowitz}},\ }\bibfield  {title} {\bibinfo {title} {{The Extreme Kerr throat geometry: A Vacuum analog of ${AdS_2 \times S^2}$}},\ }\href {https://doi.org/10.1103/PhysRevD.60.104030} {\bibfield  {journal} {\bibinfo  {journal} {Phys. Rev. D}\ }\textbf {\bibinfo {volume} {60}},\ \bibinfo {pages} {104030} (\bibinfo {year} {1999})},\ \Eprint {https://arxiv.org/abs/hep-th/9905099} {arXiv:hep-th/9905099} \BibitemShut {NoStop}%
\bibitem [{\citenamefont {Kunduri}\ \emph {et~al.}(2007)\citenamefont {Kunduri}, \citenamefont {Lucietti},\ and\ \citenamefont {Reall}}]{Kunduri:2007vf}%
  \BibitemOpen
  \bibfield  {author} {\bibinfo {author} {\bibfnamefont {H.~K.}\ \bibnamefont {Kunduri}}, \bibinfo {author} {\bibfnamefont {J.}~\bibnamefont {Lucietti}},\ and\ \bibinfo {author} {\bibfnamefont {H.~S.}\ \bibnamefont {Reall}},\ }\bibfield  {title} {\bibinfo {title} {{Near-horizon symmetries of extremal black holes}},\ }\href {https://doi.org/10.1088/0264-9381/24/16/012} {\bibfield  {journal} {\bibinfo  {journal} {Class. Quant. Grav.}\ }\textbf {\bibinfo {volume} {24}},\ \bibinfo {pages} {4169} (\bibinfo {year} {2007})},\ \Eprint {https://arxiv.org/abs/0705.4214} {arXiv:0705.4214 [hep-th]} \BibitemShut {NoStop}%
\bibitem [{\citenamefont {Astorino}\ \emph {et~al.}(2022)\citenamefont {Astorino}, \citenamefont {Martelli},\ and\ \citenamefont {Vigan\`o}}]{Astorino:2022aam}%
  \BibitemOpen
  \bibfield  {author} {\bibinfo {author} {\bibfnamefont {M.}~\bibnamefont {Astorino}}, \bibinfo {author} {\bibfnamefont {R.}~\bibnamefont {Martelli}},\ and\ \bibinfo {author} {\bibfnamefont {A.}~\bibnamefont {Vigan\`o}},\ }\bibfield  {title} {\bibinfo {title} {{Black holes in a swirling universe}},\ }\href {https://doi.org/10.1103/PhysRevD.106.064014} {\bibfield  {journal} {\bibinfo  {journal} {Phys. Rev. D}\ }\textbf {\bibinfo {volume} {106}},\ \bibinfo {pages} {064014} (\bibinfo {year} {2022})},\ \Eprint {https://arxiv.org/abs/2205.13548} {arXiv:2205.13548 [gr-qc]} \BibitemShut {NoStop}%
\bibitem [{\citenamefont {Liu}\ \emph {et~al.}(2013)\citenamefont {Liu}, \citenamefont {L\"u}, \citenamefont {Pope},\ and\ \citenamefont {V\'azquez-Poritz}}]{Liu:2013fna}%
  \BibitemOpen
  \bibfield  {author} {\bibinfo {author} {\bibfnamefont {H.-S.}\ \bibnamefont {Liu}}, \bibinfo {author} {\bibfnamefont {H.}~\bibnamefont {L\"u}}, \bibinfo {author} {\bibfnamefont {C.~N.}\ \bibnamefont {Pope}},\ and\ \bibinfo {author} {\bibfnamefont {J.~F.}\ \bibnamefont {V\'azquez-Poritz}},\ }\bibfield  {title} {\bibinfo {title} {{Not Conformally-Einstein Metrics in Conformal Gravity}},\ }\href {https://doi.org/10.1088/0264-9381/30/16/165015} {\bibfield  {journal} {\bibinfo  {journal} {Class. Quant. Grav.}\ }\textbf {\bibinfo {volume} {30}},\ \bibinfo {pages} {165015} (\bibinfo {year} {2013})},\ \Eprint {https://arxiv.org/abs/1303.5781} {arXiv:1303.5781 [hep-th]} \BibitemShut {NoStop}%
\end{thebibliography}

%

\end{document}